\documentclass[twocolumn]{aastex61}
\usepackage{amsmath}
\usepackage{color}
\usepackage{comment}
\usepackage{url}

\received{August 9, 2017}
\revised{January 20, 2018}
\accepted{January 22, 2018, in The Astrophysical Journal}

\begin{document}
\title{Candidate Water Vapor Lines to Locate the $\mathrm{H_2O}$ Snowline through High-dispersion Spectroscopic Observations. III. Submillimeter $\mathrm{H_2}$$^{16}\mathrm{O}$ and $\mathrm{H_2}$$^{18}\mathrm{O}$ Lines}
\author[0000-0003-2493-912X]{Shota Notsu}
\altaffiliation{Research Fellow of Japan Society for the Promotion of Science (DC1)}
\affiliation{Department of Astronomy, Graduate School of Science, Kyoto University, Kitashirakawa-Oiwake-cho, Sakyo-ku, Kyoto 606-8502, Japan; snotsu@kusastro.kyoto-u.ac.jp}

\author{Hideko Nomura}
\affiliation{Department of Earth and Planetary Science, Tokyo Institute of Technology, 2-12-1 Ookayama, Meguro-ku, Tokyo 152-8551, Japan}

\author{Catherine Walsh}
\affiliation{School of Physics and Astronomy, University of Leeds, Leeds, LS2 9JT, UK}

\author{Mitsuhiko Honda}
\affiliation{Department of Physics, School of Medicine, Kurume University, 67 Asahi-machi, Kurume, Fukuoka 830-0011, Japan}

\author{Tomoya Hirota}
\affiliation{National Astronomical Observatory of Japan, 2-21-1 Osawa, Mitaka, Tokyo 181-8588, Japan}

\author{Eiji Akiyama}
\affiliation{National Astronomical Observatory of Japan, 2-21-1 Osawa, Mitaka, Tokyo 181-8588, Japan}

\author{T. J. Millar}
\affiliation{Astrophysics Research Centre, School of Mathematics and Physics, Queen's University Belfast, University Road, Belfast, BT7 1NN, UK}

\begin{abstract}
\noindent
In this paper, we extend the results presented in our former papers \citep{Notsu2016, Notsu2017a} on using ortho-$\mathrm{H_2}$$^{16}\mathrm{O}$ line profiles to constrain the location of the $\mathrm{H_2O}$ snowline in T Tauri and Herbig Ae disks, to include sub-millimeter para-$\mathrm{H_2}$$^{16}\mathrm{O}$ and ortho- and para-$\mathrm{H_2}$$^{18}\mathrm{O}$ lines. Since the number densities of the ortho- and para-H$_{2}$$^{18}$O molecules are about 560 times smaller than their $^{16}$O analogues, they trace deeper into the disk than the ortho-H$_{2}$$^{16}$O lines (down to $z=0$, i.e., the midplane). Thus these H$_{2}$$^{18}$O lines are potentially better probes of the position of the H$_{2}$O snowline at the disk midplane, depending on the dust optical depth.
The values of the Einstein $A$ coefficients of sub-millimeter candidate water lines tend to be lower (typically $<$$10^{-4}$ s$^{-1}$) than infrared candidate water lines \citep{ Notsu2017a}. Thus in the sub-millimeter candidate water line cases, the local intensity from the outer optically thin region in the disk is around $10^{4}$ times smaller than that in the infrared candidate water line cases.
Therefore, in the sub-millimeter lines, especially H$_{2}$$^{18}$O and para-H$_{2}$$^{16}$O lines with relatively lower upper state energies ($\sim$ a few 100K) can also locate the position of the $\mathrm{H_2O}$ snowline.
We also investigate the possibility of future observations with ALMA to identify the position of the water snowline.
There are several candidate water lines that trace the hot water vapor inside the $\mathrm{H_2O}$ snowline in ALMA Bands $5-10$.
\end{abstract}


\section{Introduction}
\noindent Measuring the position of the water snowline (which corresponds to the sublimation front of water molecules, 
e.g., \citealt{Hayashi1981,Hayashi1985}) by observations in protoplanetary disks is crucial because it will constrain the chemical and physical structures of disks
(e.g., \citealt{Oka2011, Banzatti2015, Piso2015, Piso2016, Cieza2016, Krijt2016, Pinilla2017, Schoonenberg2017}), the current planetesimal and planet formation theories (e.g., \citealt{Oberg2011, Okuzumi2012, Ros2013}), and the origin of water on terrestrial planets (e.g., \citealt{Morbidelli2000, Morbidelli2012, Morbidelli2016, Walsh2011, IdaGuillot2016, Sato2016, Raymond2017}).
\\ \\
It has been difficult to locate the H$_{2}$O snowline directly in protoplanetary disks through imaging observations of H$_{2}$O ice (e.g., \citealt{Inoue2008, Honda2009, Honda2016}).
This is because the spatial resolution of these observations is insufficient. Water lines from disks have been detected through recent space infrared spectroscopic observations, such as {\em Spitzer}/IRS and {\em Herschel}/PACS, HIFI (for more details, see e.g., \citealt{Carr2008, Carr2011, Pontoppidan2010a, Hogerheijde2011, Salyk2011, Fedele2012, Fedele2013, Riviere-Marichalar2012, Kamp2013, Najita2013, Podio2013, Zhang2013}; \citealt{vanDishoeck2014}; \citealt{Antonellini2015, Antonellini2016, Antonellini2017, Blevins2016, Banzatti2017, Du2017, Notsu2016, Notsu2017a}). 
However, these lines mainly trace the disk surface and the cold water vapor outside the H$_{2}$O snowline.
\\ \\
Water line profiles were detected by ground-based near- and mid-infrared spectroscopic observations using the Keck and VLT for some bright T Tauri disks \footnote[1]{In the remainder of this paper, we define the protoplanetary disks around T Tauri/Herbig Ae stars as ``T Tauri/Herbig Ae disks".} (e.g., \citealt{Salyk2008, Pontoppidan2010b, Mandell2012}).
Those observations suggested that the hot water vapor resides in the inner part of the disks; however, the spatial and spectral resolution was not sufficient to investigate detailed structures, such as the position of the H$_{2}$O snowline.
In addition, the observed lines, with large Einstein A coefficients, are sensitive to the water vapor in the disk surface and are potentially polluted by slow disk winds.
\\ \\
In our former papers \citep{Notsu2016, Notsu2017a},
we proposed a means to identify the location of the H$_{2}$O snowline more directly by analyzing the Keplerian profiles of H$_{2}$O lines which can be obtained by high dispersion spectroscopic observations and selected based on specific criteria.
We concluded that lines which have small Einstein $A$ coefficients (A$_{\mathrm{ul}}$=$10^{-6} \sim10^{-3}$ s$^{-1}$) and relatively high upper state energies (E$_{\mathrm{up}}$$\sim$ 1000K) trace the hot water reservoir inside the $\mathrm{H_2O}$ snowline, and can locate the position of the H$_{2}$O snowline. In these candidate lines, the contribution of the optically thick hot midplane inside the H$_{2}$O snowline is large compared with that of the outer optically thin surface layer. This is because the intensities of lines from the optically thin region are proportional to the Einstein $A$ coefficient. Moreover, the contribution of the cold water reservoir outside the H$_{2}$O snowline is also small, because lines with high excitation energies are not emitted from the regions at low temperature.
The position of the H$_{2}$O snowline of a Herbig Ae disk exists to a larger radius from the star compared with that around less massive and cooler T Tauri stars. 
Therefore, it is expected to be easier to observe the candidate H$_{2}$O lines, and thus identify the location of the H$_{2}$O snowline, in Herbig Ae disks than in T Tauri disks.
\\ \\
In this paper, we extend our water line calculations beyond ortho-$\mathrm{H_2}$$^{16}\mathrm{O}$ lines only to sub-millimeter para-$\mathrm{H_2}$$^{16}\mathrm{O}$ and ortho- and para-$\mathrm{H_2}$$^{18}\mathrm{O}$ lines.
We discuss the possibility of detecting such candidate water lines to locate the position of the $\mathrm{H_2O}$ snowline with future observations with the Atacama Large Millimeter/Submillimeter Array (ALMA).
Note that we discuss for the first time the properties of protoplanetary disk water line profiles in ALMA Band 5 \citep{Immer2016, Humphreys2017}, which is available from 
the second half of the observing Cycle 5 (which commenced in March 2018).
We also investigate the effects of dust emission on water line profiles.
Section 2 outlines the methods. The results and discussion are described in Sections 3 and 4, respectively. In Section 5, the conclusions are listed.
\section{Methods}
\subsection{The disk physical structures and molecular abundances}
\begin{deluxetable*}{rrrrrrrrr}
\tablewidth{0pt}
\tablecaption{{The central star and disk parameters of our adopted physical models}}\label{tab:T1}
\tablehead{
\colhead{Star}&
\colhead{$M_{\mathrm{*}}$}&
\colhead{$R_{\mathrm{*}}$}&
\colhead{$T_{\mathrm{*}}$} &
\colhead{$\dot{M}$}&
\colhead{$\alpha$}&
\colhead{$g/d$}&
\colhead{$M_{\mathrm{disk}}$} 
& \colhead{UV excess?} \\
\colhead{}&
\colhead{[$M_{\bigodot}$]}&\colhead{[$R_{\bigodot}$]}&
\colhead{[K]}& \colhead{[$M_{\bigodot}$ yr$^{-1}$]}&
\colhead{}&
\colhead{}&
\colhead{[$M_{\bigodot}$]}& \colhead{}}
\startdata      
       T Tauri & 0.5 & 2.0 & 4000 & 10$^{-8}$ & $10^{-2}$ & 100 &$2.4\times 10^{-2}$& Yes \\
       Herbig Ae & 2.5 & 2.0 & 10,000 & 10$^{-8}$ & $10^{-2}$ & 100 & $2.5\times 10^{-2}$& No \\
\enddata
\tablenotetext{}{Notes:  $M_{\mathrm{*}}$: central star mass, $R_{\mathrm{*}}$: star radius, $T_{\mathrm{*}}$: star effective temperature,
$\dot{M}$: mass accretion rate, $\alpha$: viscous parameter, $g/d$: gas to dust mass ratio, $M_{\mathrm{disk}}$: total disk mass}
\tablenotetext{}{UV excess?: whether stellar UV radiation field has excess emission components, such as Lyman-$\alpha$ line emission and optically thin hydrogenic bremsstrahlung radiation.}
\end{deluxetable*}
\noindent 
In this paper, we adopt the molecular abundance distribution of a T Tauri disk and a Herbig Ae disk which were calculated in paper I \citep{Notsu2016}\footnote[2]{In the remainder of this paper, we define \citet{Notsu2016} and \citet{Notsu2017a} as papers I and II, respectively.} and paper II \citep{Notsu2017a}, using self-consistent disk physical models. Here we briefly explain our disk model.
\\ \\
In our works (see also papers I and II), we used self-consistent physical models of steady, axisymmetric Keplerian disks surrounding a T Tauri star and a Herbig Ae star.
They were calculated on the basis of the methods in \citet{NomuraMillar2005} with X-ray heating \citep{Nomura2007}.
\citet{Walsh2010, Walsh2012, Walsh2014a, Walsh2015}, \citet{Heinzeller2011}, \citet{Furuya2013}, and \citet{Notsu2015} adopted the same physical models to study various physical and chemical effects.
Table 1 shows the central star and disk parameters of our adopted models in detail.
\\ \\
The gas and dust are assumed to be mixed well. In this model, we adopt a size distribution of spherical, compact dust grains that replicates the extinction curve observed in dense clouds \citep{Mathis1977, WeingartnerDraine2001}.
\\ \\
In Figure 1 of papers I and II, we displayed the gas number densities, the gas temperatures $T_{\mathrm{g}}$, the dust-grain temperatures $T_{\mathrm{d}}$, and the wavelength-integrated UV fluxes in a T Tauri disk and a Herbig Ae disk, respectively. 
In Section 2 of paper II, we discussed the differences between the physical structures of the T Tauri disk and the Herbig Ae disk in detail.
\\ \\
The large chemical network we use to calculate the disk molecular abundances includes gas-phase reactions and gas-grain interactions (freeze-out, and thermal and non thermal desorption).
We adopt the set of atomic oxygen-rich abundances \citep{Graedel1982, Woodall2007} as the initial elemental fractional abundances.
Figure 2 of papers I and II showed the fractional abundances (relative to total gaseous hydrogen nuclei density) of $\mathrm{H_2O}$ vapor ($n_{\mathrm{H2O,gas}}/n_{\mathrm{H}}$) and ice ($n_{\mathrm{H2O,ice}}/n_{\mathrm{H}}$) in a T Tauri disk and a Herbig Ae disk.
\subsection{Water emission line profiles from protoplanetary disks}
\noindent 
We calculate the Keplerian profiles of water emission lines from the T Tauri disk and the Herbig Ae disk, and identify those emission lines which are the best candidates for tracing emission from the inner gaseous water within the $\mathrm{H_2O}$ snowline.
In papers I and II, we used the same method to calculate the profiles of water emission lines from a Herbig Ae disk and a T Tauri disk (based on \citealt{Rybicki1986}, \citealt{Hogerheijde2000}, \citealt{NomuraMillar2005}, and \citealt{Schoier2005}, see also Section 2.3 of paper I.).
In this paper, we extend our water line calculations beyond ortho-$\mathrm{H_2}$$^{16}\mathrm{O}$ lines only
(papers I and II), to sub-millimeter para-$\mathrm{H_2}$$^{16}\mathrm{O}$ and ortho- and para-H$_{2}$$^{18}$O lines.
In addition, we focus on the sub-millimeter lines, in order to investigate more thoroughly
the possibility of observing the $\mathrm{H_2O}$ snowline with ALMA.
\\ \\
For the calculation of line profiles, we modified the 1D code RATRAN\footnote[3]{\url{http://home.strw.leidenuniv.nl/~michiel/ratran/}} \citep{Hogerheijde2000}.
The data for the line parameters are adopted from the Leiden Atomic and Molecular Database LAMDA\footnote[4]{\url{http://home.strw.leidenuniv.nl/~moldata/}} 
\citep{Schoier2005} for the $\mathrm{H_2}$$^{16}\mathrm{O}$ lines and from the HITRAN Database\footnote[5]{\url{http://www.hitran.org}} (e.g., \citealt{Rothman2013}) for the $\mathrm{H_2}$$^{18}\mathrm{O}$ lines.
We also cross-referenced the values of line parameters with the Splatalogue database\footnote[6]{\url{http://www.cv.nrao.edu/php/splat/}}.
The value of the ortho-to-para ratio (OPR) of water is set to 3, which is the high-temperature value \citep{Mumma1987, Hama2013, Hama2016}.
Here we mention that \citet{Hama2016} reported from their experiments that water desorbed from the icy dust-grain surface at 10K shows the OPR = 3, which invalidates the assumed relation between OPR and the formation temperature of water \citep{Mumma1987}. They argue that the role of gas-phase processes that convert the OPR to a lower value in low temperature regions is
important, though the detailed mechanism is not yet understood.
We set the isotope ratio of oxygen $^{16}$O/$^{18}$O to 560 throughout the disk, as \citet{Jorgensen2010} and \citet{Persson2012} adopted.
This $^{16}$O/$^{18}$O value is determined by the observation of local interstellar medium \citep{Wilson1994}.
We do not include emission from jet components and disk winds in calculating the line profiles. 
\\ \\
The assumption of local thermal equilibrium (LTE) is adopted in our calculations to obtain the level populations of the water molecule.
In Section 3.2.5 of paper I and Section 4.2 of paper II, we discussed the validity of this assumption, and we concluded that the LTE assumption is valid when we calculate the candidate water lines which probe emission from hot gaseous water inside the $\mathrm{H_2O}$ snowline. 
This is because these lines mainly come from the dense region ($\sim 10^{11}-10^{14}$ cm$^{-3}$) at $z/r < 0.1$ inside the $\mathrm{H_2O}$ snowline, while they have low values of Einstein $A$ coefficients, such that their critical densities $n_{\mathrm{cr}}=A_{ul}/{<\sigma v>}$\footnote[7]{$<\sigma v>$ is the collisional rates for the excitation of $\mathrm{H_2O}$ molecules by electrons and H$_{\mathrm{2}}$ molecules for an adopted value of the collisional temperature of 200K \citep{Faure2008}.} are relatively small (n$_{\mathrm{cr}}\sim 10^{5}-10^{7}$ cm$^{-3}$, see also Table 1 of paper II).
Here we note that $n_{\mathrm{cr}}$ of the para-$\mathrm{H_2}$$^{16}\mathrm{O}$ 183 GHz and 325 GHz lines are $1.4\times 10^{5}$ and $1.2\times 10^{6}$ cm$^{-3}$, respectively. These values are lower than the values of the total gas density not only in the inner disk midplane, but also in the hot surface water layer of the outer disk ($\sim10^{7}-10^{8}$ $\mathrm{cm}^{-3}$) and in the photodesorbed water layer ($\sim10^{8}-10^{10}$ $\mathrm{cm}^{-3}$).
In contrast, non-LTE effects are important for strong water lines which have large $A_{\mathrm{ul}}$ ($\sim$ $10^{-1}-10^{0}$ s$^{-1}$), and trace the inner/outer hot surface layers (e.g., the ortho-$\mathrm{H_2}$$^{16}\mathrm{O}$ 63.32 $\mu$m line) or the cold photodesorbed layer (e.g., the ortho-$\mathrm{H_2}$$^{16}\mathrm{O}$ 557 GHz and the para-$\mathrm{H_2}$$^{16}\mathrm{O}$ 1113 GHz lines, see e.g., \citealt{Meijerink2009, Woitke2009b, Banzatti2012, Antonellini2015, Antonellini2016}).

\begin{deluxetable*}{rrrrrr}
\tablewidth{0pt}
\tablecaption{{The regional classifications in the disk midplane with different water abundances}}\label{tab:T2}
\tablehead{
\colhead{Region name}&
\colhead{Star}&
\colhead{$r$}&
\colhead{$T_{g}$}&
\colhead{$n_{\mathrm{H2O,gas}}/n_{\mathrm{H}}$} &
\colhead{Comments}\\
\colhead{}&
\colhead{}&
\colhead{[au]}&
\colhead{[K]}&
\colhead{}&
\colhead{}}
\startdata      
A$_{\mathrm{TT}}$& T Tauri & 0$-$2 & $>150$ & $\sim 10^{-5}-10^{-4}$ & inside the $\mathrm{H_2O}$ snowline \\
B$_{\mathrm{TT}}$&                   & 2$-$30 & $<150$ & $\sim 10^{-12}$ & outside the $\mathrm{H_2O}$ snowline\\
 &&&&&\\
A$_{\mathrm{HA}}$& Herbig Ae & 0$-$8 & $>170$ & $\sim 10^{-5}-10^{-4}$ & high $\mathrm{H_2O}$ abundance region \\
B$_{\mathrm{HA}}$&                       & 8$-$14 & $120-170$ & $\sim 10^{-8}$ & inside the $\mathrm{H_2O}$ snowline \\  
C$_{\mathrm{HA}}$&                        & 14$-$30 & $<120$ & $\sim 10^{-12}$ & outside the $\mathrm{H_2O}$ snowline \\
\enddata
\tablenotetext{}{Notes: $r$: the disk radius from the central star, $T_{g}$: disk gas temperature, $n_{\mathrm{H2O,gas}}/n_{\mathrm{H}}$: the fractional abundances (relative to total gaseous hydrogen nuclei density) of water vapor}
\end{deluxetable*}
%
%
\section{Results}
\setcounter{figure}{0}
\begin{figure*}[htbp]
\begin{center}
\includegraphics[scale=0.6]{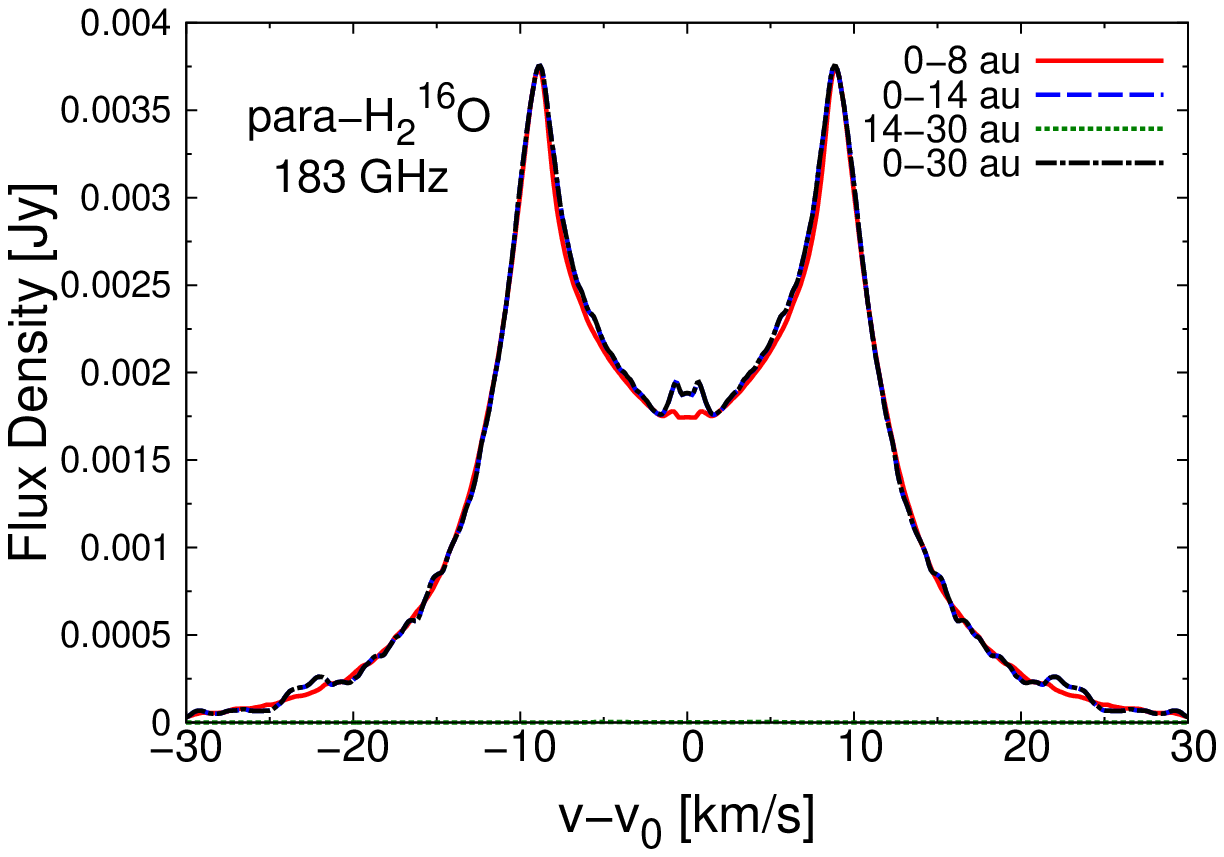}
\includegraphics[scale=0.6]{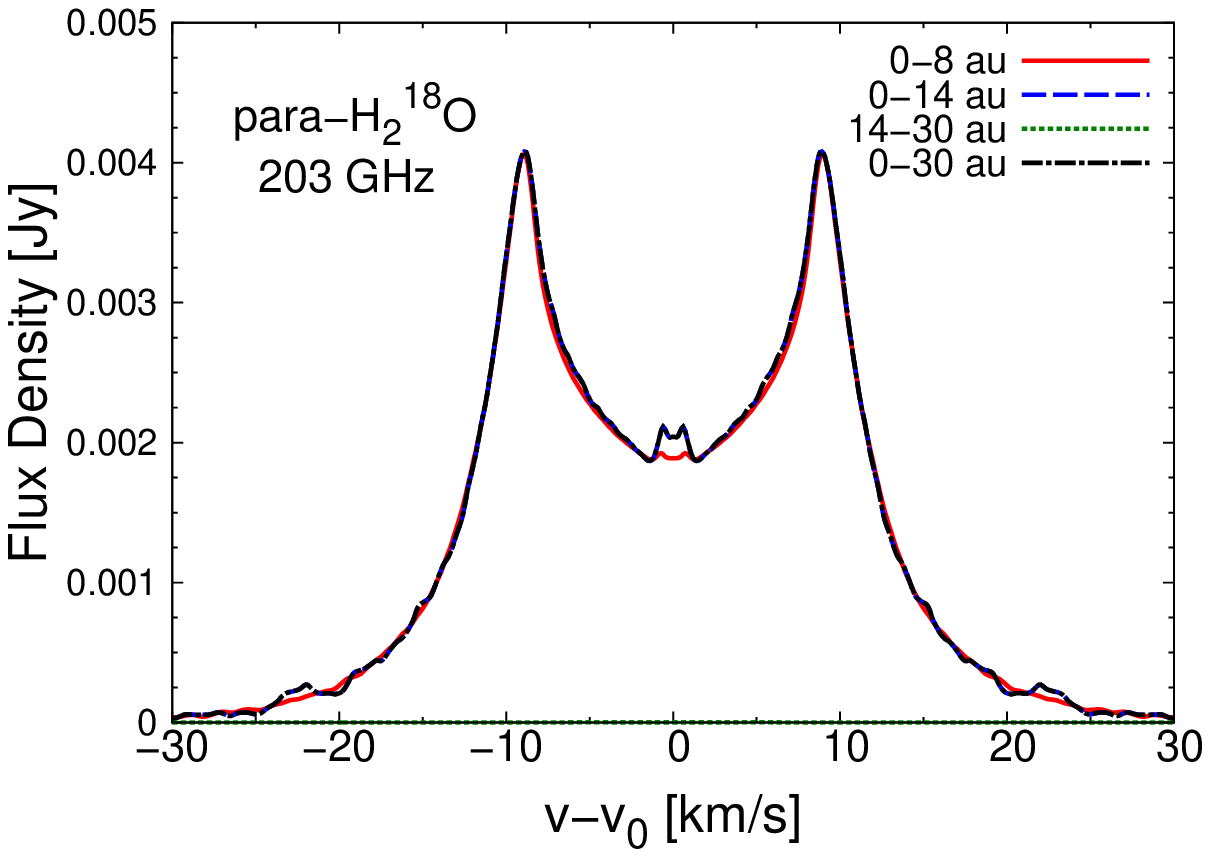}
\includegraphics[scale=0.6]{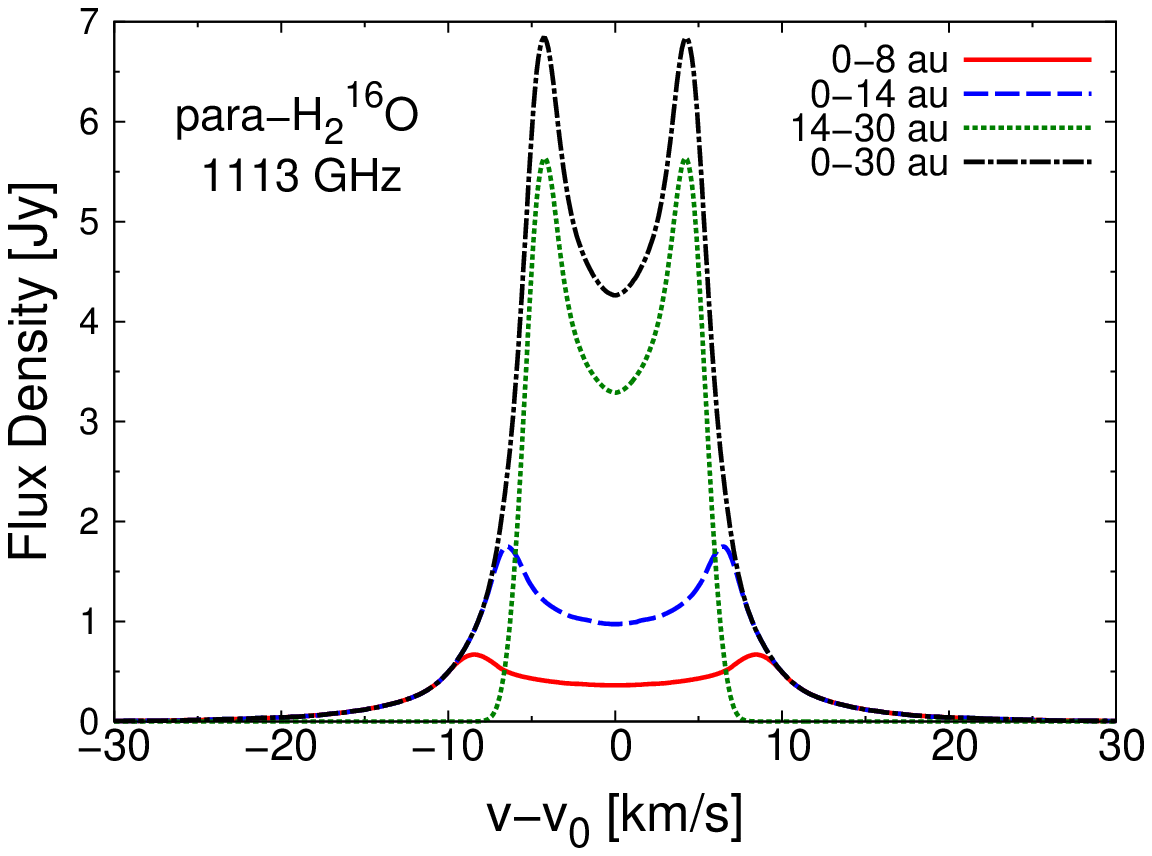}
\includegraphics[scale=0.6]{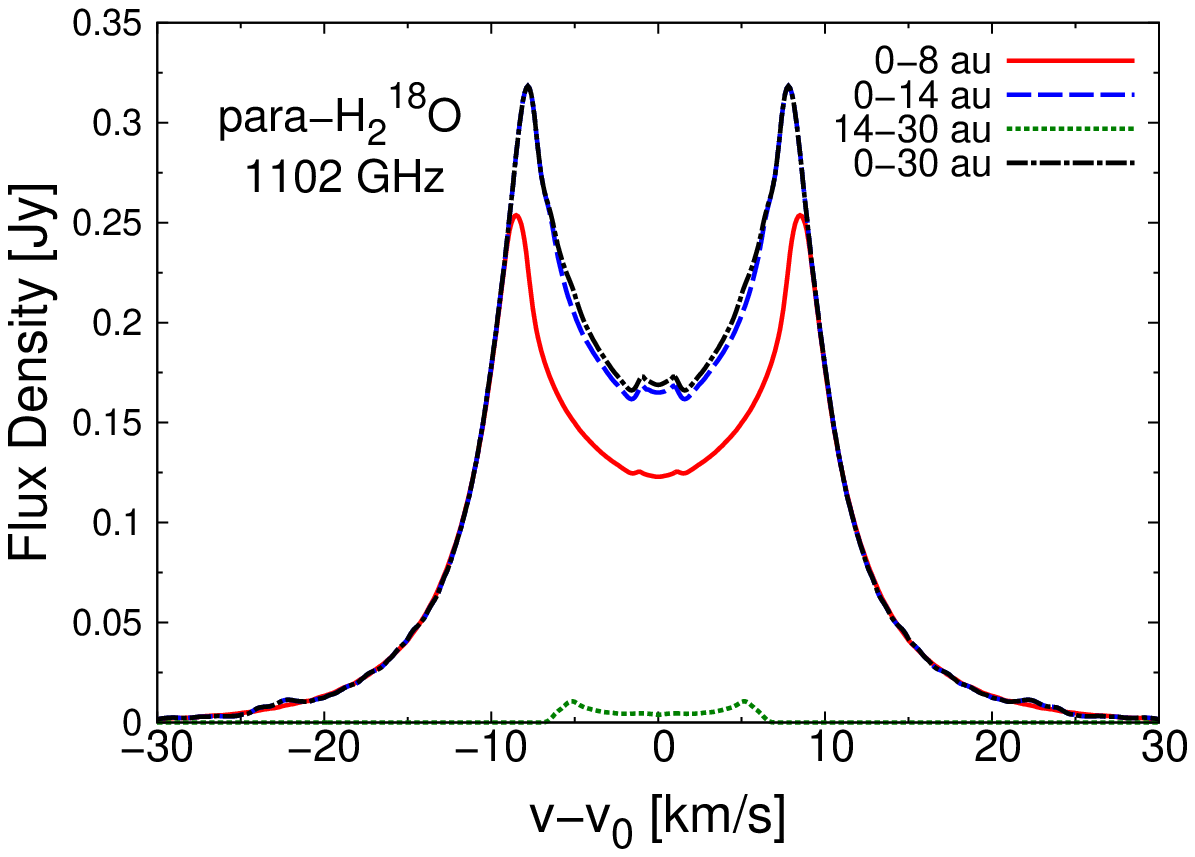}
\end{center}
\vspace{0.6cm}
\caption{\noindent The profiles of para-$\mathrm{H_2}$$^{16}\mathrm{O}$ lines at 183 GHz (top left) and 1113 GHz (bottom left), and para-$\mathrm{H_2}$$^{18}\mathrm{O}$ lines at 203 GHz (top right), and 1102 GHz (bottom right) for the Herbig Ae disk. 
In these line profiles, we ignore dust emission and adopt a disk inclination, $i=30$ deg and the distance to the object, $d=140$ pc.
The line profiles from inside 8 au ($=$the inner high temperature region) are displayed with {\it red solid lines}, those from within 14 au ($\sim$within the $\mathrm{H_2O}$ snowline) are {\it blue dashed lines}, those from 14-30 au ($\sim$outside the $\mathrm{H_2O}$ snowline) are {\it green dotted lines}, and those from the total area inside 30 au are {\it black dashed dotted lines}. 
In the top two panels, the flux densities outside the $\mathrm{H_2O}$ snowline ({\it green dotted lines}, $<$ 10$^{-4}$ Jy) are much smaller than those inside 8 au ({\it red solid lines}).
Therefore, the {\it red solid lines}, {\it blue dashed lines}, and {\it black dashed dotted lines} are almost completely overlapped (see also Figure \ref{Figure8_paperIII}).
\vspace{0.3cm}
}\label{Figure1_paperIII}
\end{figure*} 
\begin{figure*}[htbp]
\begin{center}
\includegraphics[scale=0.6]{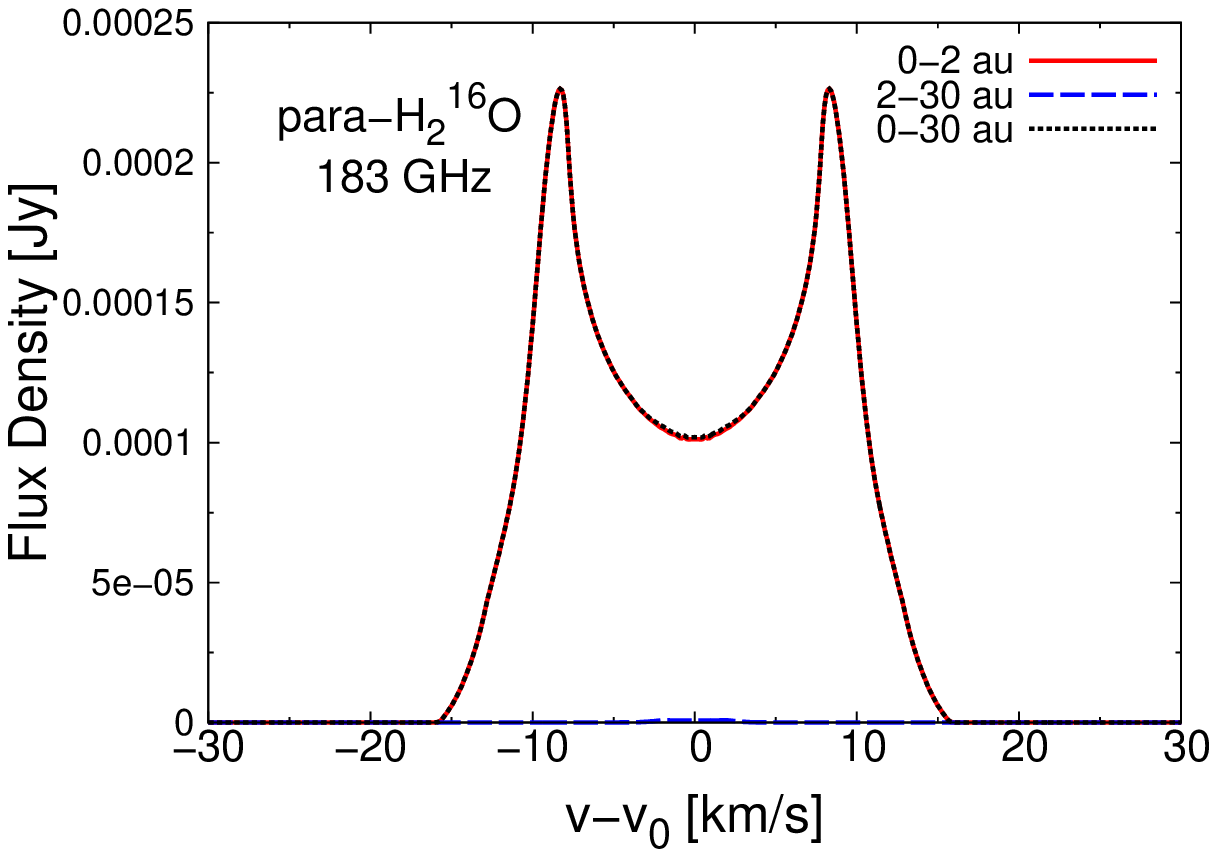}
\includegraphics[scale=0.6]{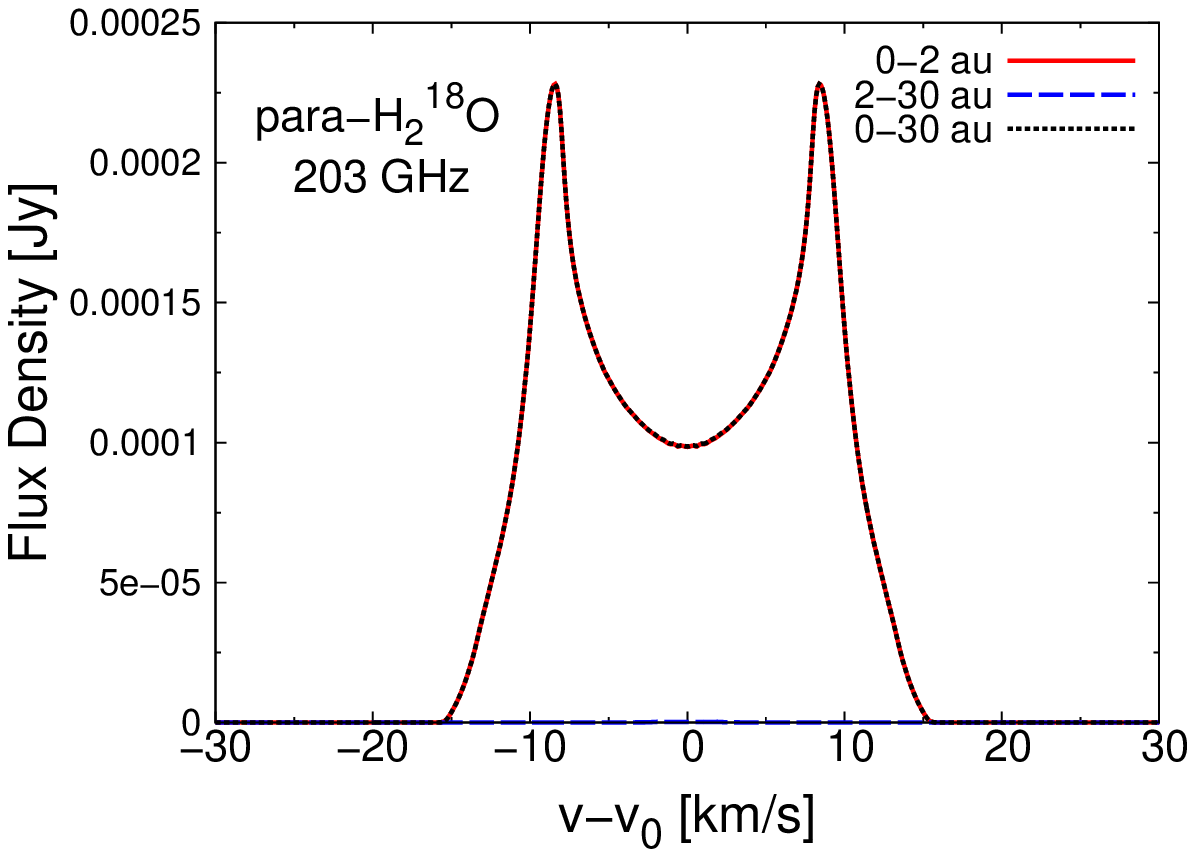}
\includegraphics[scale=0.6]{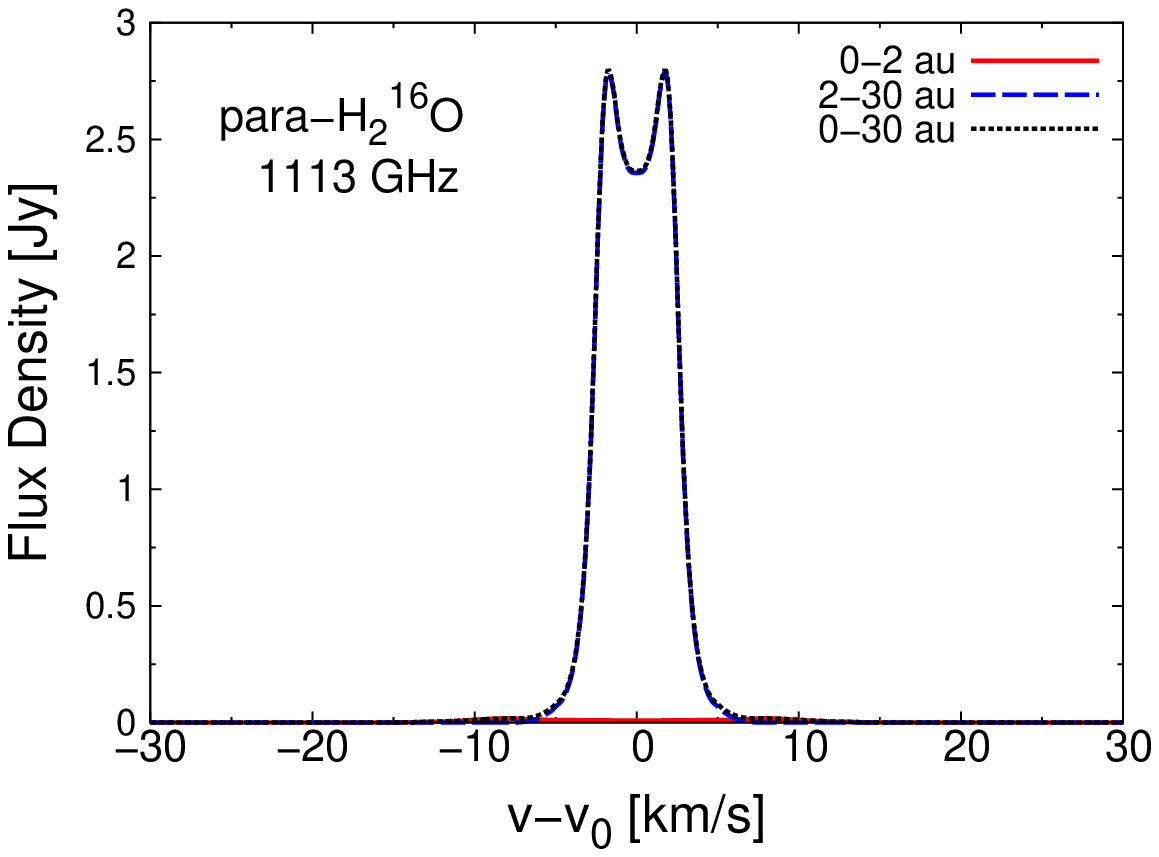}
\includegraphics[scale=0.6]{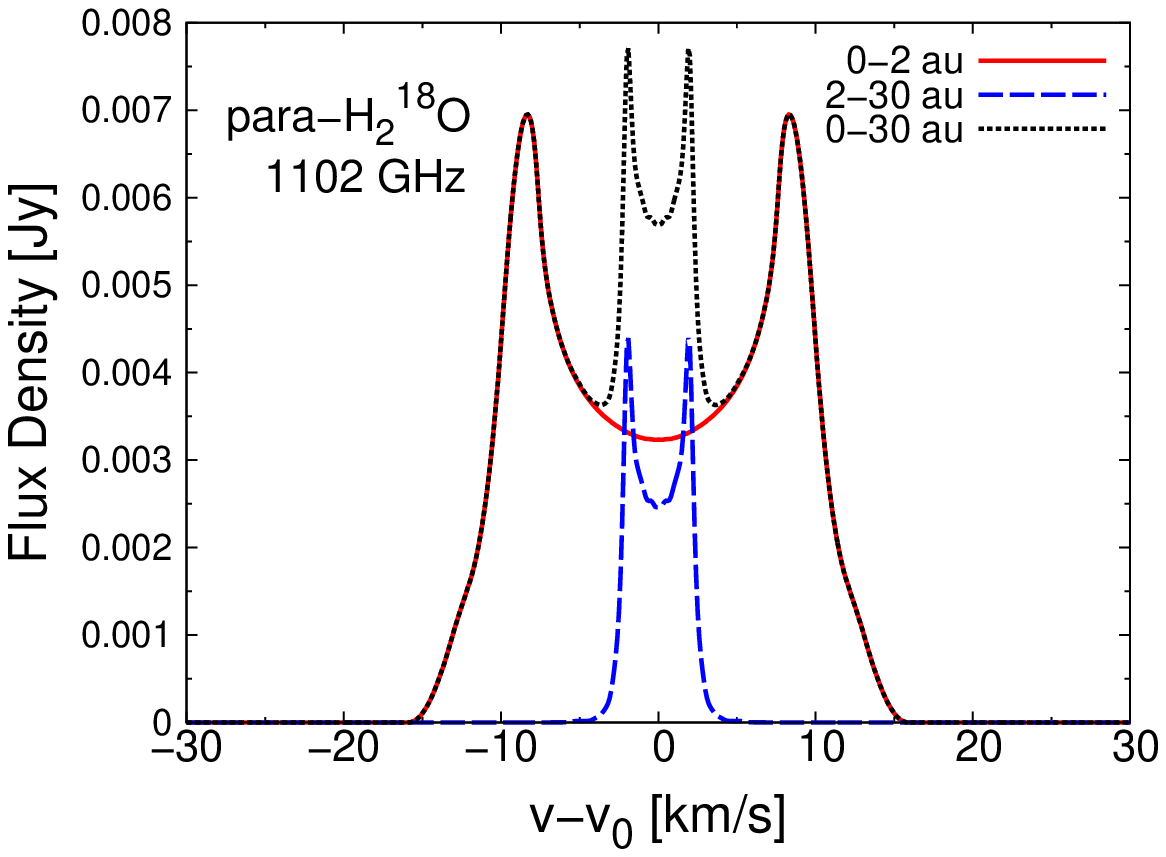}
\end{center}
\vspace{0.6cm}
\caption{\noindent 
The same as Figure \ref{Figure1_paperIII} but for the T Tauri disk.
The line profiles from inside 2 au ($\sim$within the $\mathrm{H_2O}$ snowline) are displayed with {\it red solid lines}, those from $2-30$ au ($\sim$outside the $\mathrm{H_2O}$ snowline) are {\it blue dashed lines}, and those from the total area inside 30 au are {\it black dotted lines}.
In the top two panels, the flux densities outside the $\mathrm{H_2O}$ snowline ({\it blue dashed lines}, $<$ 10$^{-5}$ Jy) are much smaller than those inside the $\mathrm{H_2O}$ snowline ({\it red solid lines}).
Therefore, the {\it red solid lines} and {\it black dotted lines} are almost completely overlapped.
Moreover in the bottom left panels, the flux densities outside the $\mathrm{H_2O}$ snowline ({\it blue dashed lines}) are much larger than those inside the $\mathrm{H_2O}$ snowline ({\it red solid lines}).
Therefore, the {\it blue dashed lines} and {\it black dotted lines} are almost completely overlapped.
\vspace{0.3cm}
}\label{Figure2_paperIII}
\end{figure*} 
\subsection{The velocity profiles of sub-millimeter water emission lines}
\begin{figure*}[htbp]
\begin{center}
\hspace{0.3cm}
\includegraphics[scale=0.7]{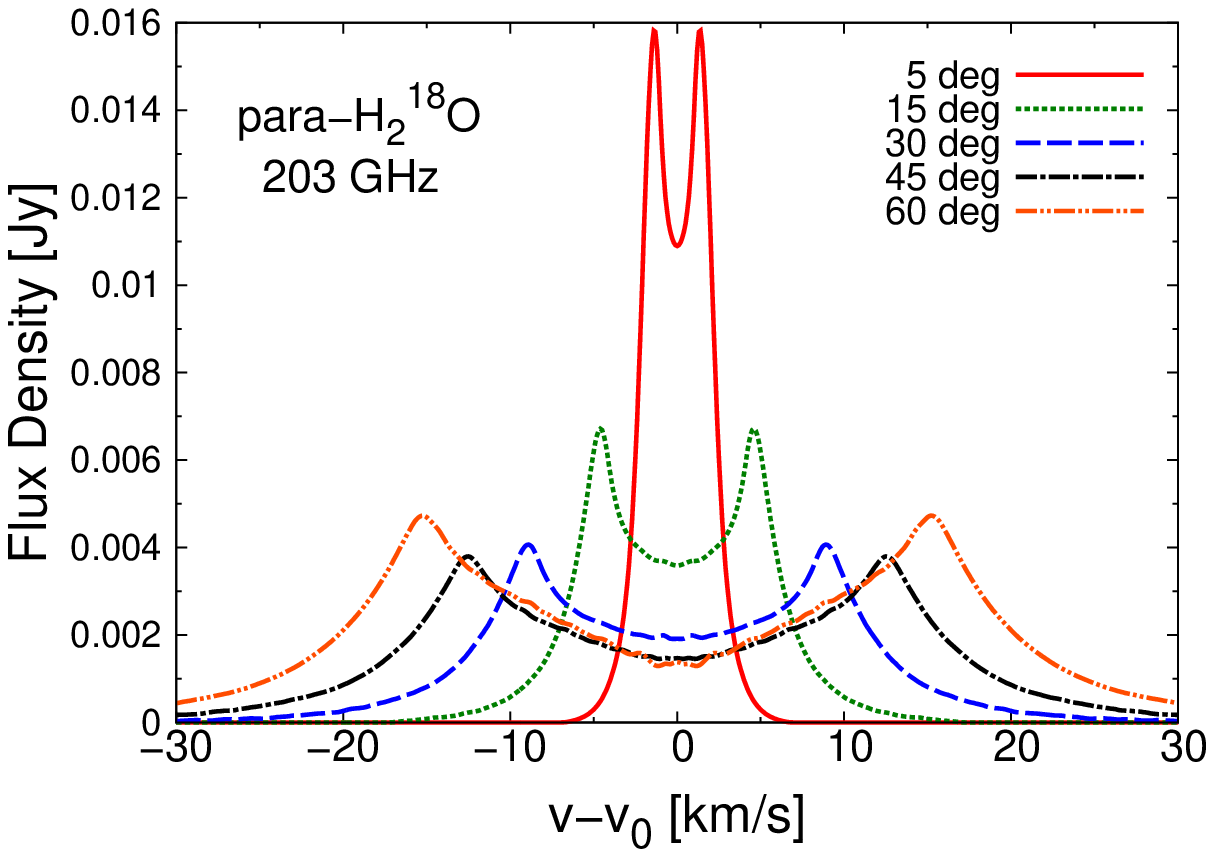}
\hspace{0.3cm}
\includegraphics[scale=0.7]{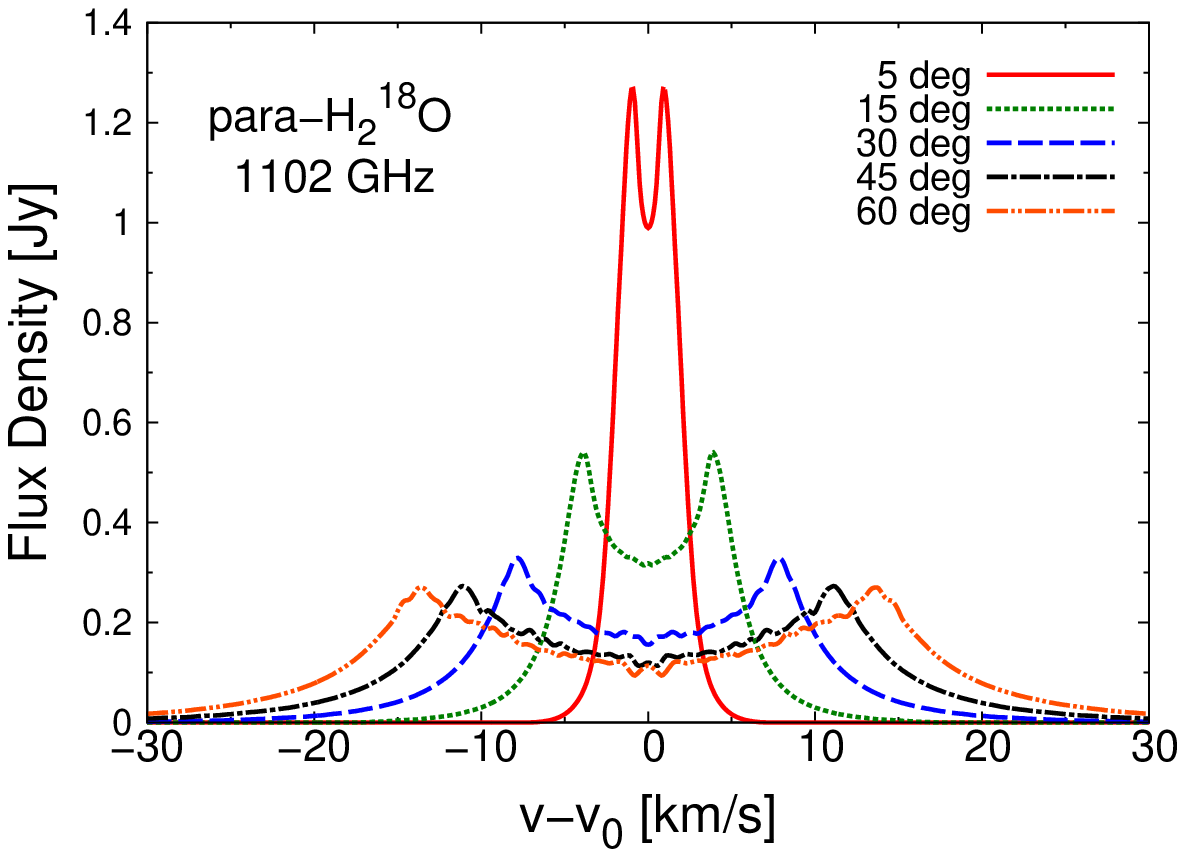}
\end{center}
\vspace{0.6cm}
\caption{\noindent 
The velocity profiles for the para-$\mathrm{H_2}$$^{18}\mathrm{O}$ lines at 203 GHz (top panel) and 1102 GHz (bottom panel) from the Herbig Ae disk inside 30 au with different inclination angles, $i=5 \deg$ ({\it red solid lines}), $15 \deg$ ({\it green dotted lines}), $30 \deg$ ({\it blue dashed lines}), $45 \deg$ ({\it black dashed dotted lines}), and $60 \deg$ ({\it orange dashed double dotted lines}).
\vspace{0.3cm}
}\label{Figure3_paperIII}
\end{figure*} 
\begin{figure*}[htbp]
\begin{center}
\includegraphics[scale=0.7]{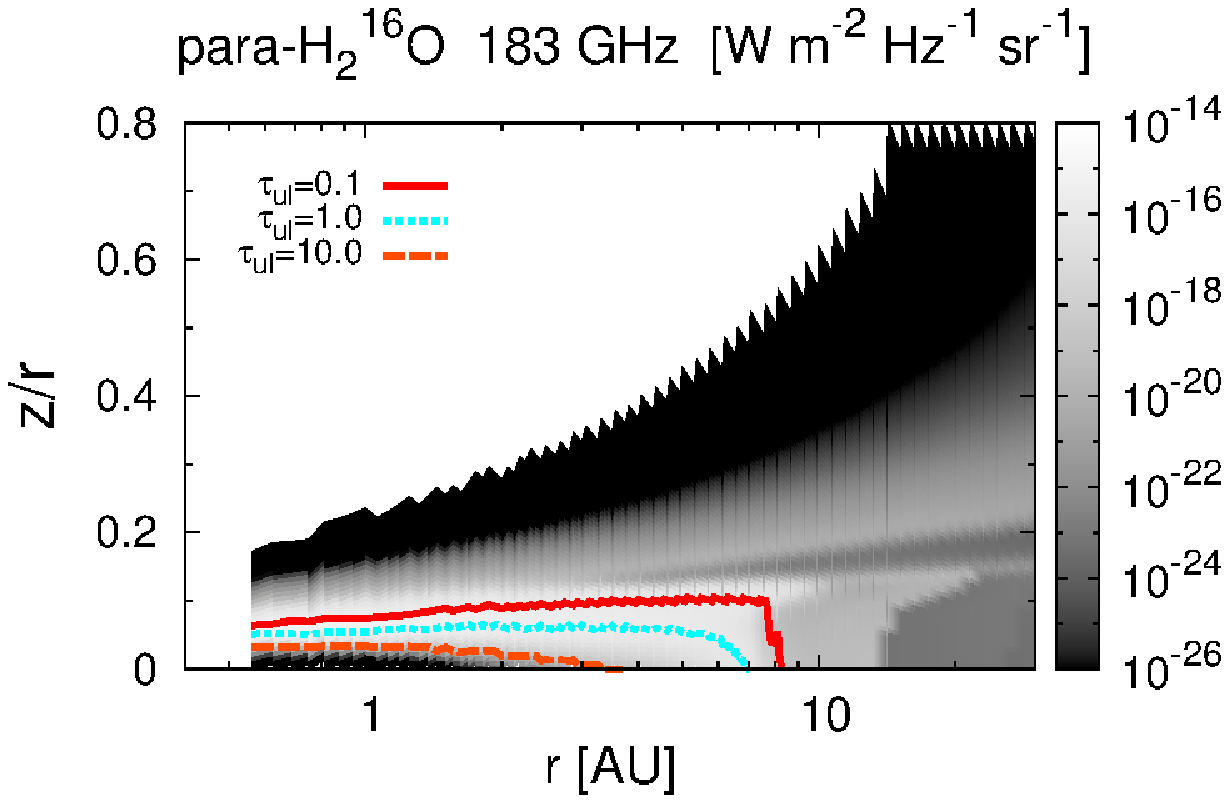}
\includegraphics[scale=0.7]{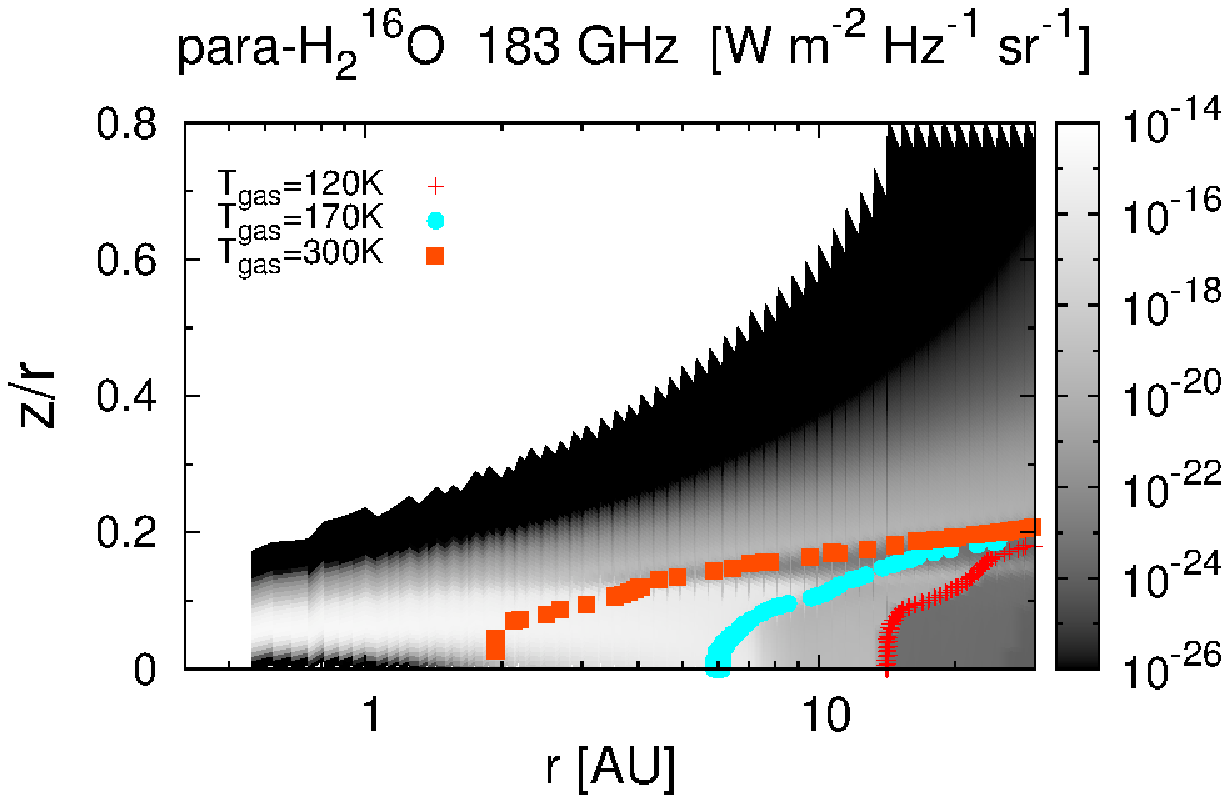}
\includegraphics[scale=0.7]{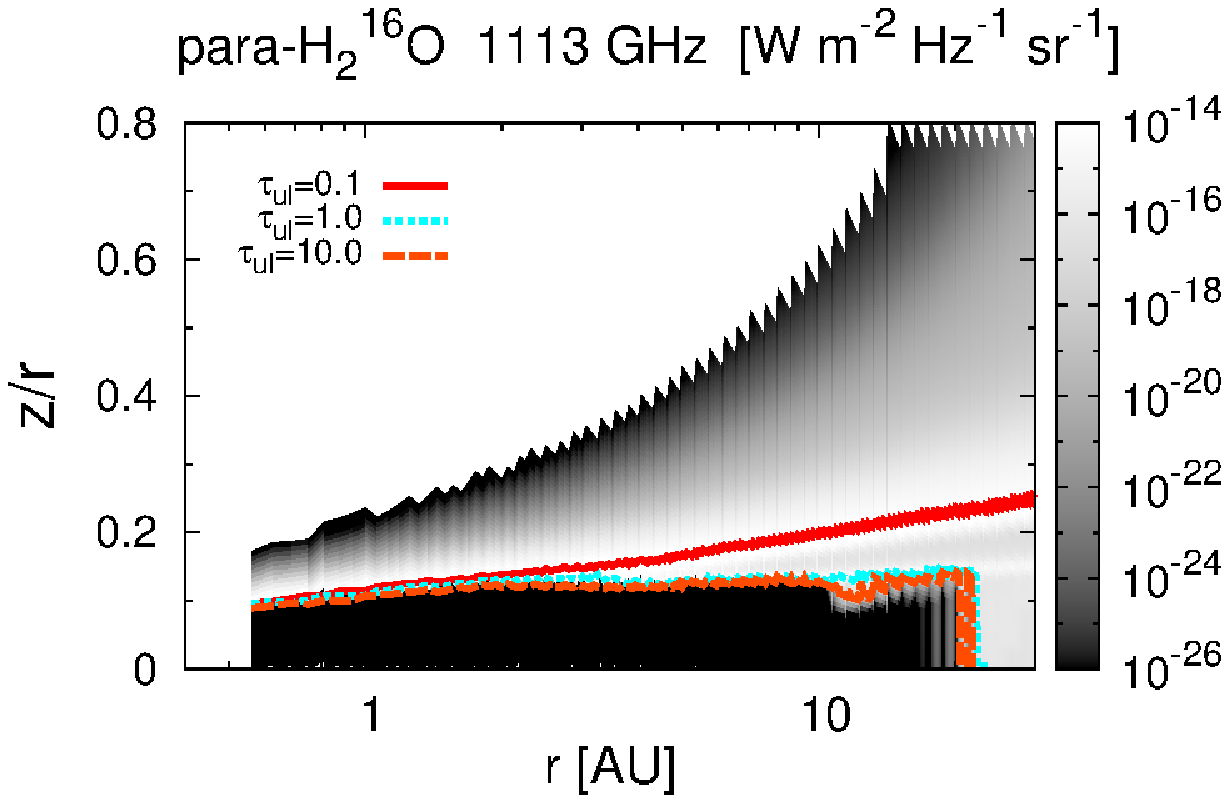}
\includegraphics[scale=0.7]{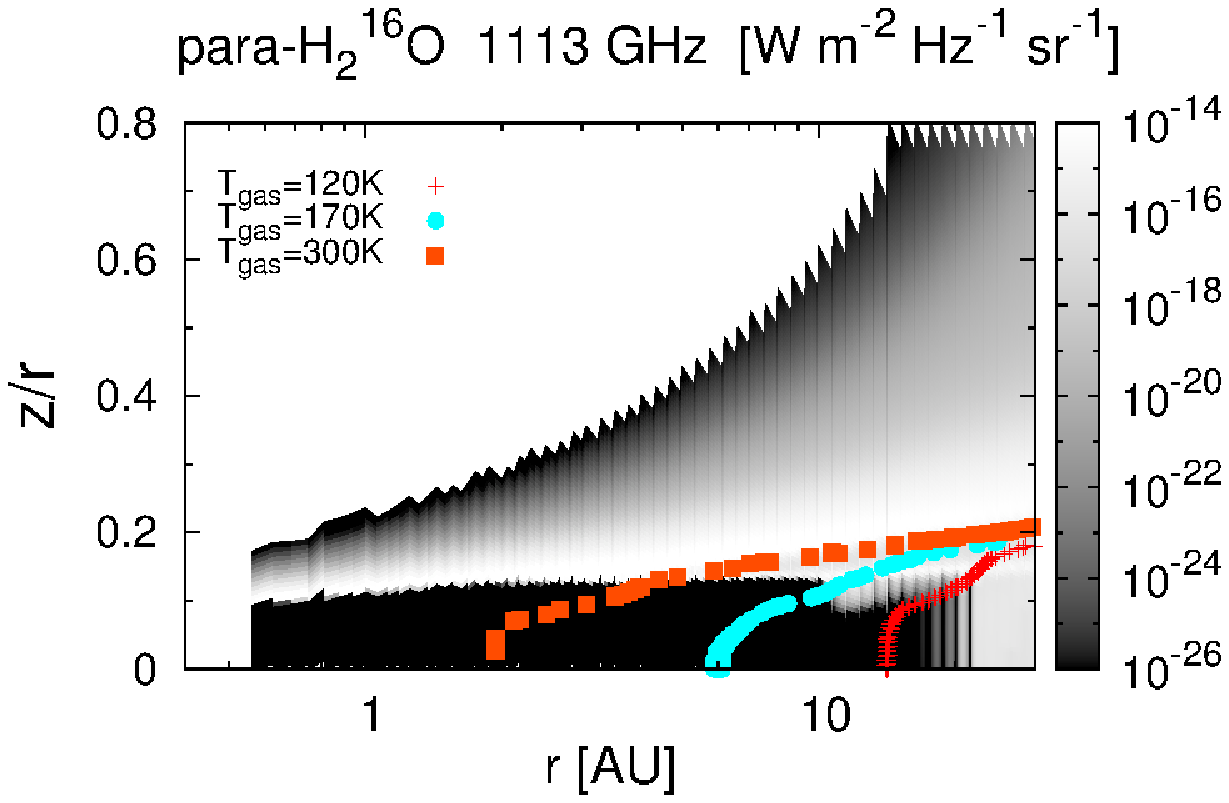}
\end{center}
\vspace{0.5cm}
\caption{
\noindent The local intensity distributions along the line-of sight direction of para-$\mathrm{H_2}$$^{16}\mathrm{O}$ lines at 183 GHz (top panels) and 1113 GHz (bottom panels) for the Herbig Ae disk.
In the left panels, the total (gas and dust) optical depth contours for $\tau_{ul}=$0.1 (red solid lines), 1 (cyan dotted lines), and 10 (orange dashed lines) are plotted on top of the line local intensities. The gas temperature contours for $T_{g}=$120K (red cross points), 170K (cyan circle points), and 300K (orange square points) are plotted in the right panels.
The units of the local intensity are W $\mathrm{m}^{-2}$ $\mathrm{Hz}^{-1}$ ${\mathrm{sr}}^{-1}$.
\vspace{0.3cm}
}\label{Figure4_paperIII}
\end{figure*}   
\setcounter{figure}{3}
\begin{figure*}[htbp]
\begin{center}
\includegraphics[scale=0.7]{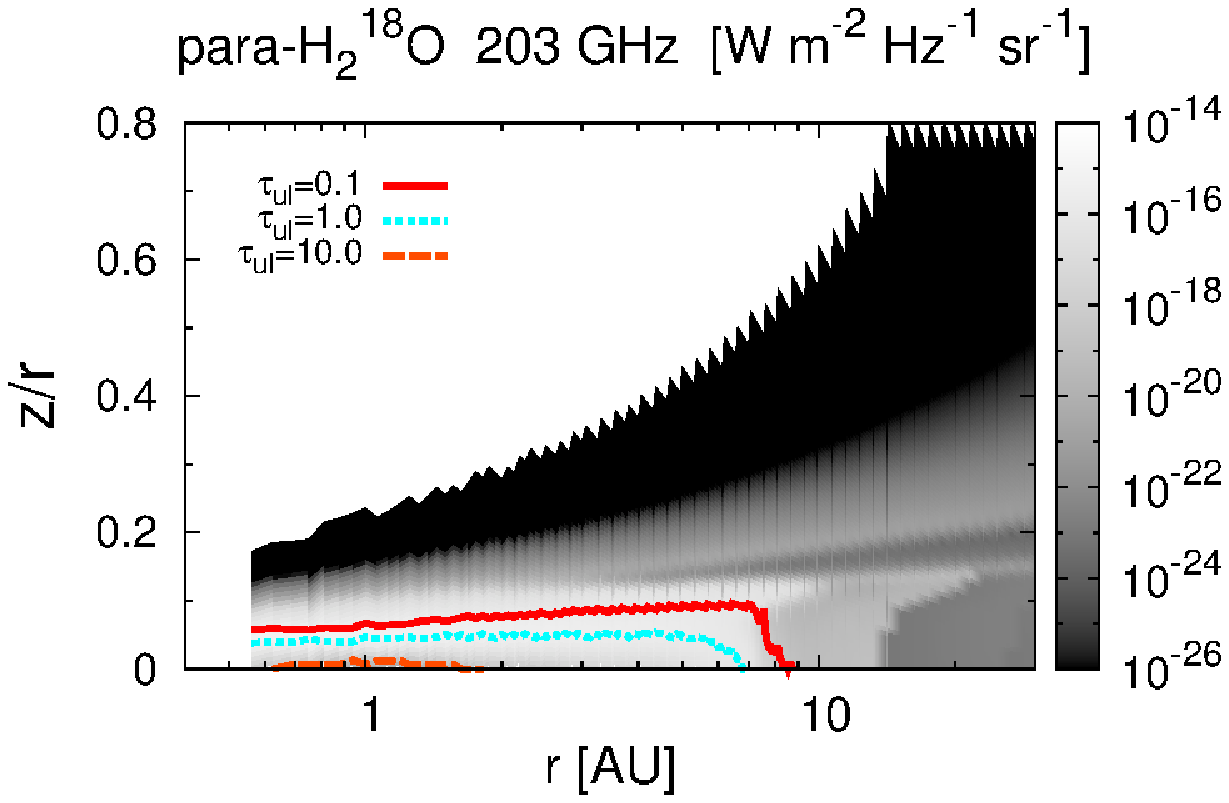}
\includegraphics[scale=0.7]{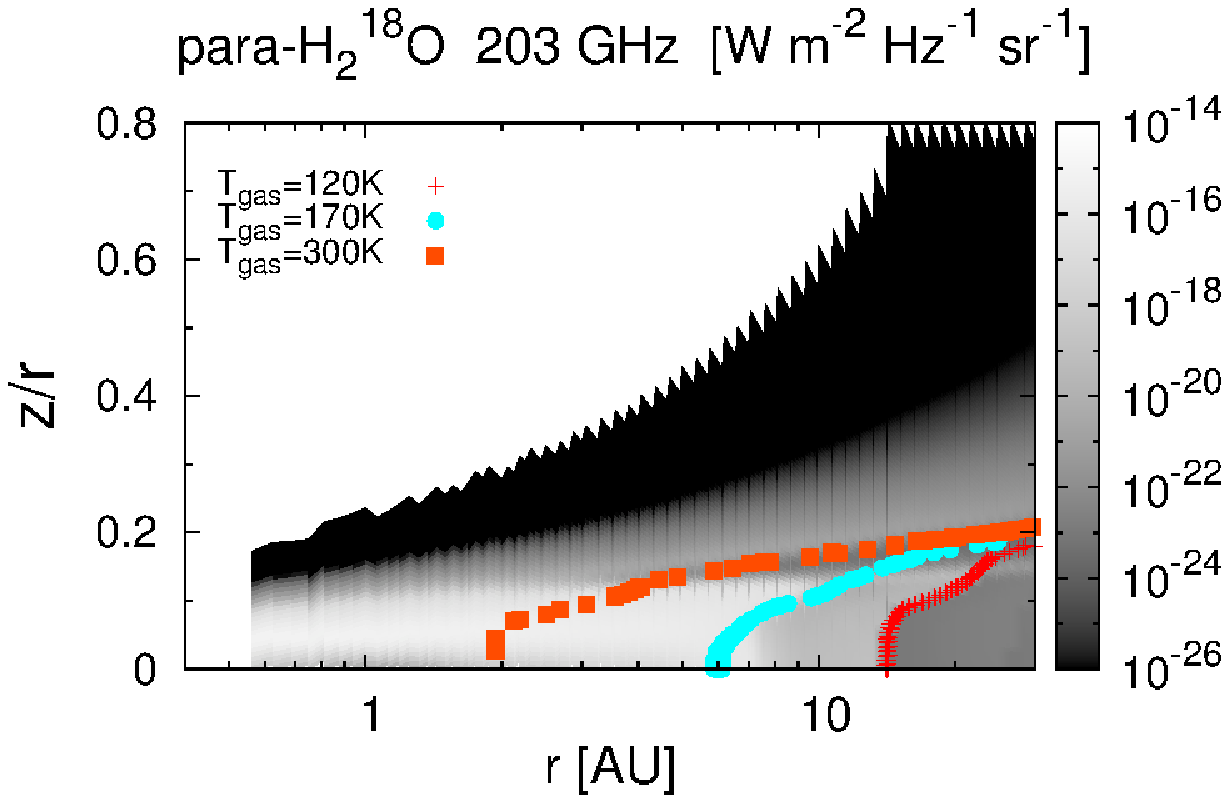}
\includegraphics[scale=0.7]{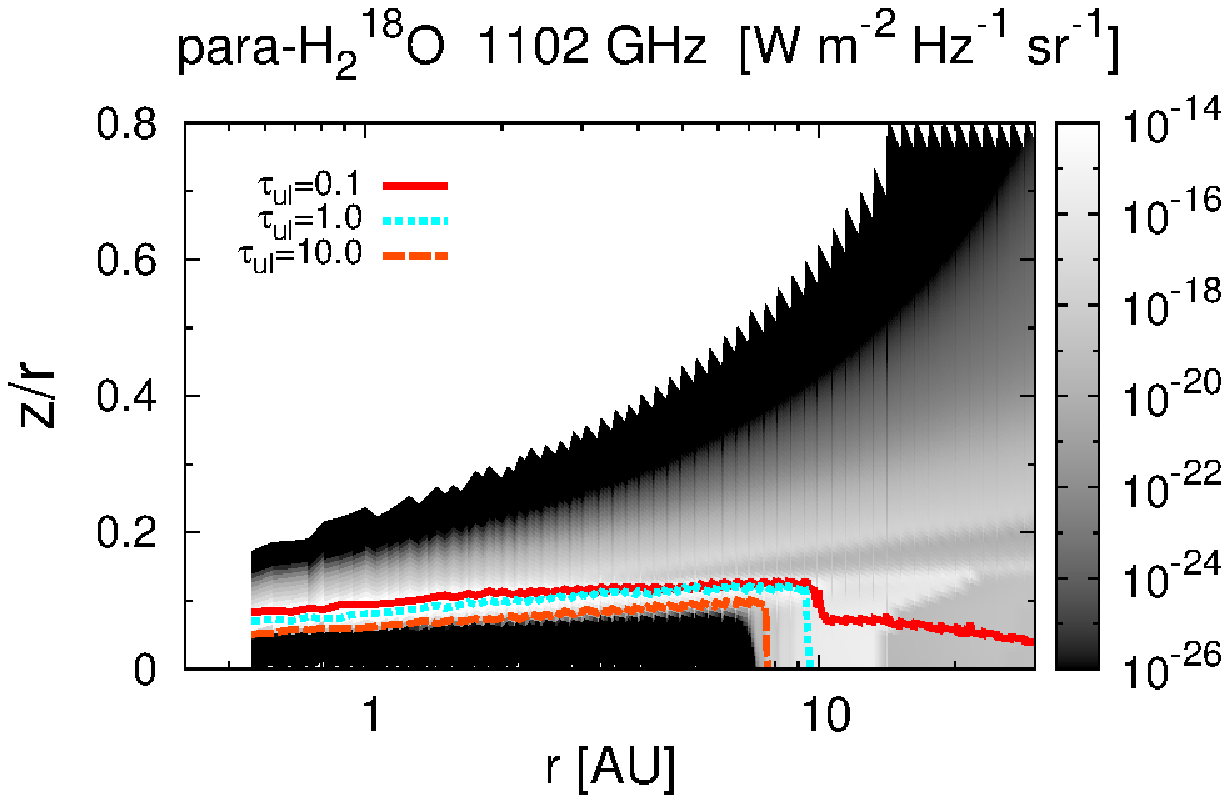}
\includegraphics[scale=0.7]{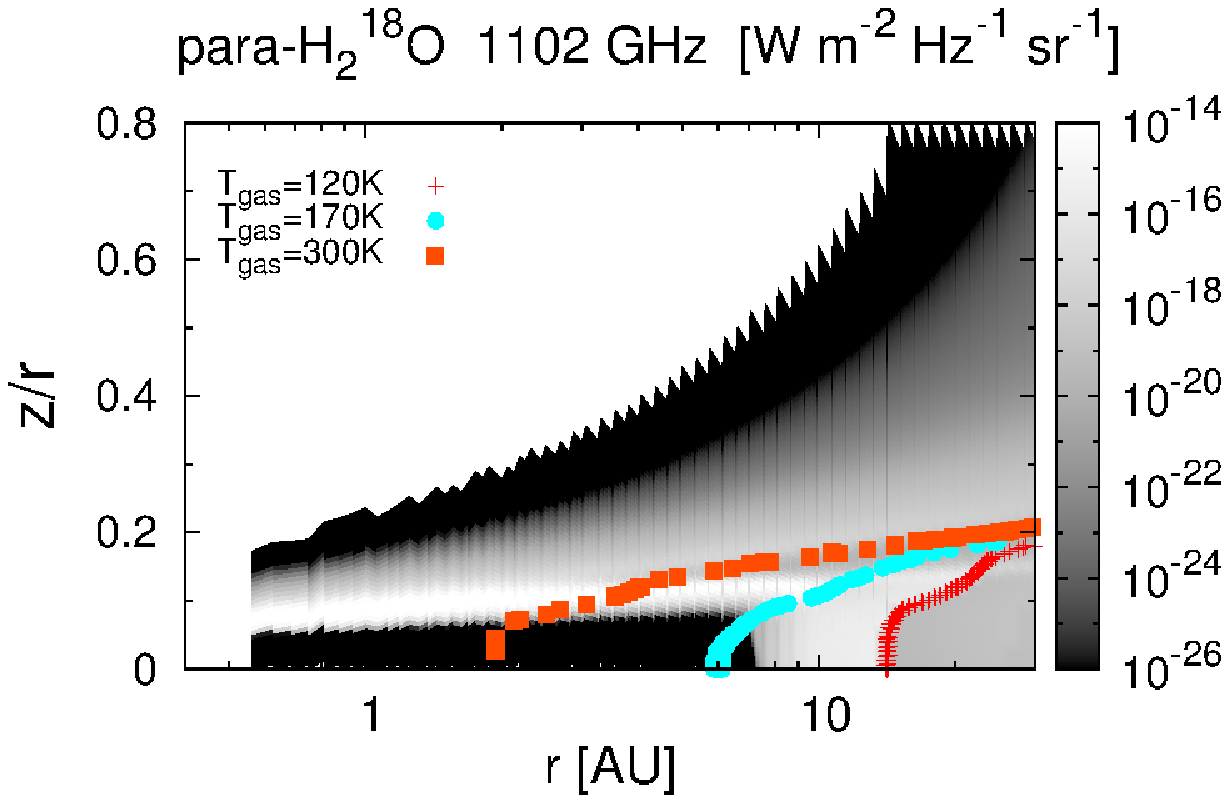}
\end{center}
\vspace{0.5cm}
\caption{\noindent (Continued.) The local intensity distributions along the line-of sight direction of para-$\mathrm{H_2}$$^{18}\mathrm{O}$ lines at 203 GHz (top panels) and 1102 GHz (bottom panels) for the Herbig Ae disk. 
\vspace{0.3cm}
}\label{Figure4_paperIII}
\end{figure*}
\begin{figure*}[htbp]
\begin{center}
\includegraphics[scale=1.8]{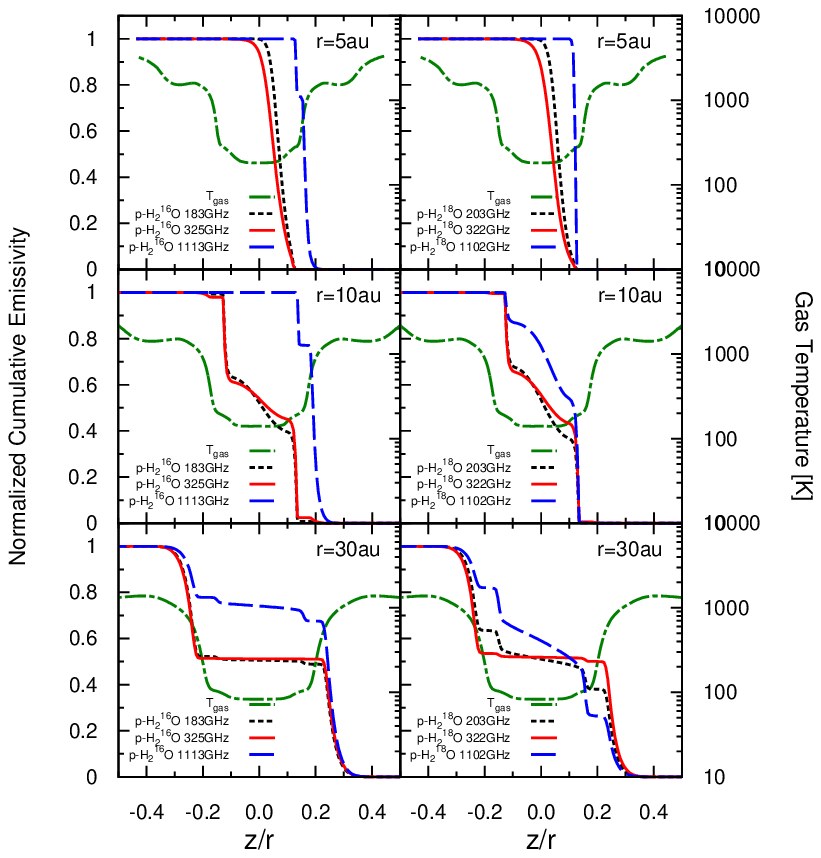}
\end{center}
\vspace{3.0cm}
\caption{\noindent The normalized cumulative line local intensity distributions along the vertical direction at $r=$5 au (top panels), $r=$10 au (middle panels), and $r=$30 au (bottom  panels), and the vertical distribution of the gas temperature $T_{g}$ [Kelvin] ({\it green dashed dotted lines}).
The distributions for three para-$\mathrm{H_2}$$^{16}\mathrm{O}$ lines at 183 GHz ({\it black dotted lines}), 325 GHz ({\it red solid lines}), 1113 GHz ({\it blue dashed lines}) from a Herbig Ae disk are shown in the left three panels.
The distributions for three para-$\mathrm{H_2}$$^{18}\mathrm{O}$ lines at 203 GHz ({\it black dotted lines}), 322 GHz ({\it red solid lines}), 1102 GHz ({\it blue dashed lines}) from a Herbig Ae disk are shown in the right three panels.
The cumulative local intensities are normalized using the values at $z=-\infty$.
\vspace{0.3cm}
}\label{Figure5_paperIII}
\end{figure*}
\begin{figure*}[htbp]
\begin{center}
\includegraphics[scale=0.65]{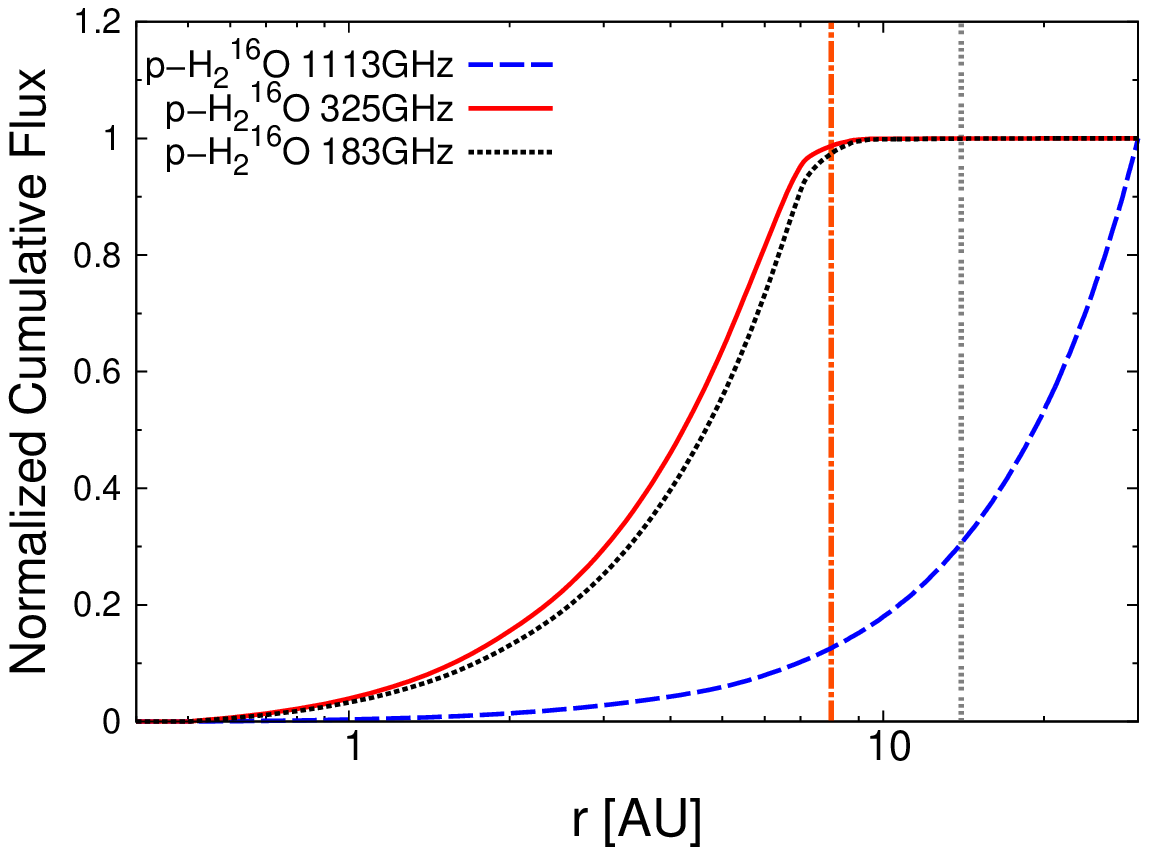}
\includegraphics[scale=0.65]{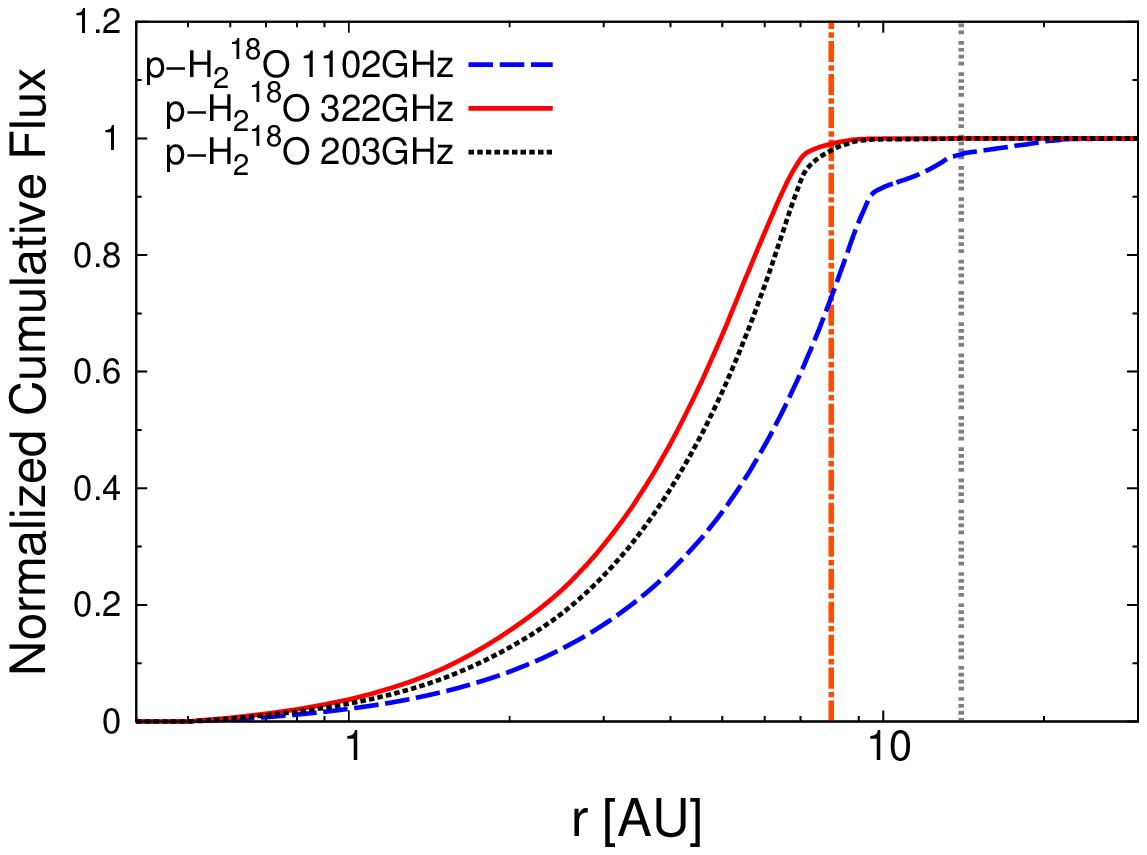}
\includegraphics[scale=0.65]{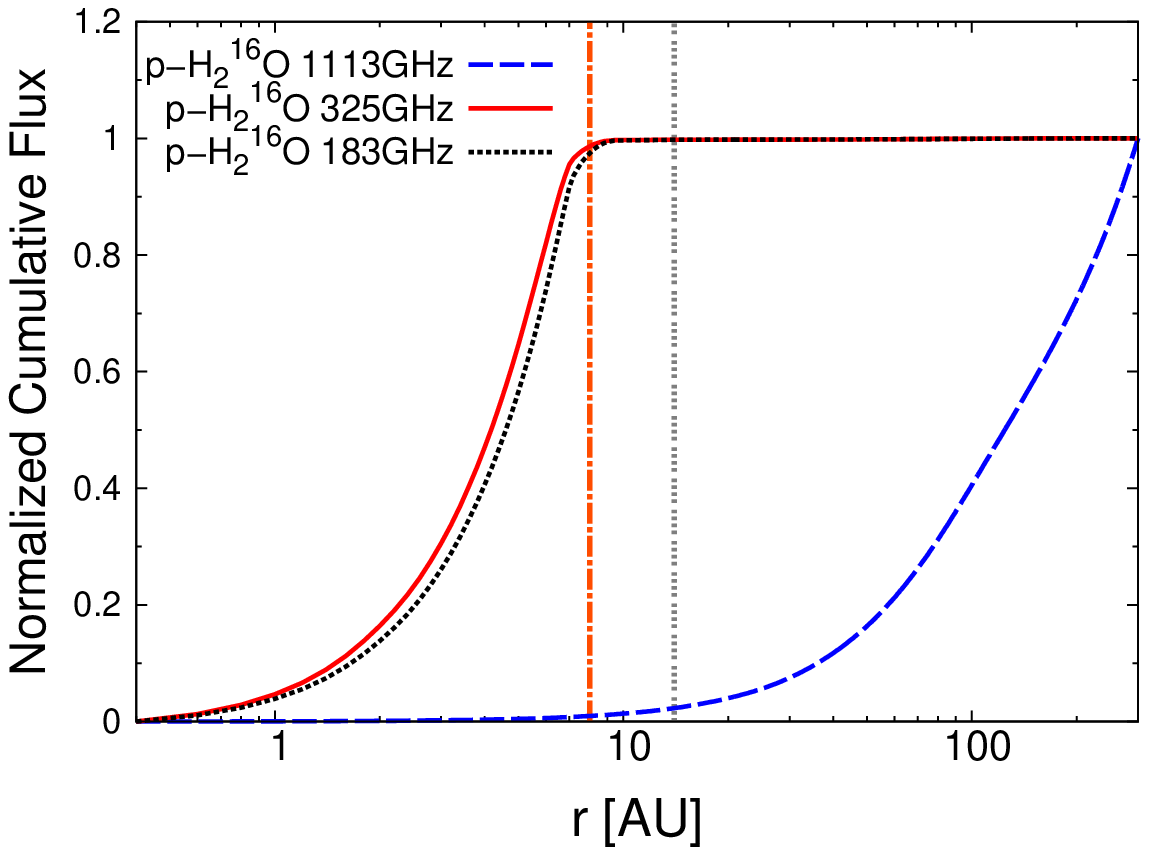}
\includegraphics[scale=0.65]{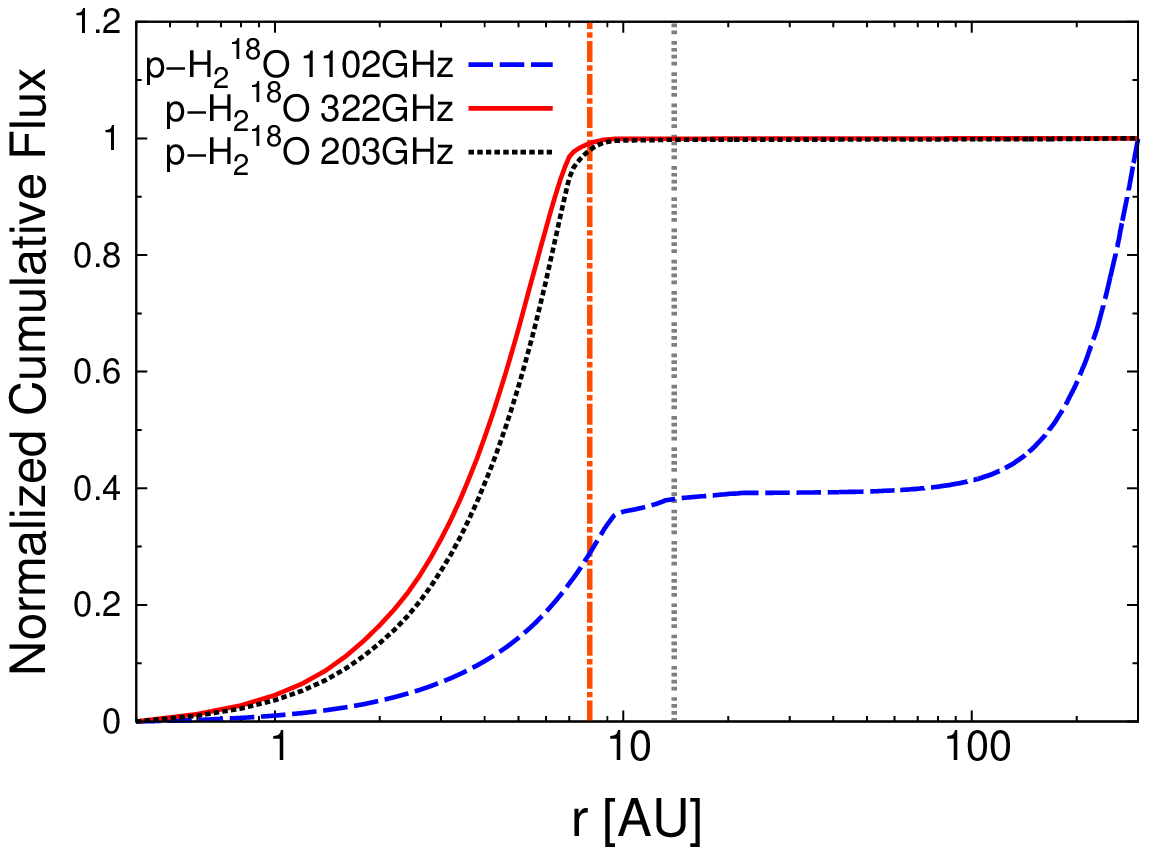}
\end{center}
\vspace{0.6cm}
\caption{\noindent (Left two panels): The distributions of the normalized cumulative fluxes along the radial directions of three para-$\mathrm{H_2}$$^{16}\mathrm{O}$ lines at 183 GHz ({\it black dotted lines}), 325 GHz ({\it red solid lines}), and 1113 GHz ({\it blue dashed lines}) from a Herbig Ae disk.
(Right two panels): The radial distributions of the normalized cumulative fluxes of three para-$\mathrm{H_2}$$^{18}\mathrm{O}$ lines at 203 GHz ({\it black dotted lines}), 322 GHz ({\it red solid lines}), and 1102 GHz ({\it blue dashed lines}) from the Herbig Ae disk.
The vertical straight lines display the positions of $r$=14 au ({\it grey dotted line}) and 8 au ({\it orange dashed dotted line}), respectively.
The cumulative fluxes are normalized using the values at $r=30$ au (top panels) and at $r=300$ au (bottom panels).}
\vspace{0.3cm}
\label{Figure6_paperIII}\end{figure*}
\noindent Figures \ref{Figure1_paperIII} and \ref{Figure2_paperIII} show the velocity profiles for 
representative characteristic para-$\mathrm{H_2}$$^{16}\mathrm{O}$ lines at 183 GHz (top left) and 1113 GHz (bottom left), and para-$\mathrm{H_2}$$^{18}\mathrm{O}$ lines at 203 GHz (top right) and 1102 GHz (bottom right) from the Herbig Ae disk and the T Tauri disk, respectively.
The para-$\mathrm{H_2}$$^{16}\mathrm{O}$ 183 GHz line and the para-$\mathrm{H_2}$$^{18}\mathrm{O}$ 203 GHz line are the same transition levels and fall in ALMA Band 5 \citep{Immer2016, Humphreys2017}.
The detailed parameters, such as transition quantum numbers ($J_{K_{a}K_{c}}$), wavelength $\lambda$, frequency, $A_{\mathrm{ul}}$, $E_{\mathrm{up}}$, 
and total fluxes of lines are listed in Tables 3 and 4 in the Appendix.
In calculating the line profiles in Figures \ref{Figure1_paperIII} and \ref{Figure2_paperIII} (also Figures \ref{Figure7_paperIII} and \ref{Figure8_paperIII}, see Section 3.4), and total fluxes in Tables 3 and 4, we do not include dust emission, although we do include both gas and dust absorption.
We discuss the effects of dust emission in Section 4.1.
In calculating line profiles and total fluxes in this paper (see Figures \ref{Figure1_paperIII}, \ref{Figure2_paperIII}, \ref{Figure7_paperIII}, \ref{Figure8_paperIII}, \ref{Figure9_paperIII} and \ref{Figure10_paperIII}, and Tables 3 and 4), we adopted the inclination angle of the disk, $i=30 \deg$ and the distance to the object $d=$140 pc ($\sim$ the distance of Taurus molecular cloud).
\\ \\
The position of the $\mathrm{H_2O}$ snowline in the T Tauri disk ($r\sim$1.6 au, $T_{\mathrm{g}}\sim$150K, see paper I) is closer to the central star than that in the Herbig Ae disk ($r\sim$14 au, $T_{\mathrm{g}}\sim$120K, see paper II), in agreement with previous studies (e.g., \citealt{Woitke2009b, Walsh2015}).
Table 2 shows the regional classifications in the disk midplane with different water fractional abundances (for more details, see also papers I and II).
Inside the $\mathrm{H_2O}$ snowline (regions A$_{\mathrm{TT}}$, A$_{\mathrm{HA}}$ and B$_{\mathrm{HA}}$, see also Table 2), the temperature exceeds the sublimation temperature under the pressure conditions of the midplane and most of the $\mathrm{H_2O}$ molecules are released into the gas-phase through thermal desorption. Thus the column densities of water vapor become larger ($>10^{16}$ cm$^{-2}$, see Figure 3 of paper II) than those of the outer disk ($\sim10^{14}$ cm$^{-2}$).
In the case of Herbig Ae disks, the gas-phase chemistry to form $\mathrm{H_2O}$ molecules is efficient in the inner region at a higher temperature (region A$_{\mathrm
{HA}}$, see Table 2), and the column densities of gaseous water molecules become much larger ($\sim10^{20}-10^{22}$ cm$^{-2}$, see Figure 3 of paper II and Section 3.2 of this paper).
Since the radial temperature profile in the midplane of the T Tauri disk is steeper than that of the Herbig Ae disk, the T Tauri disk does not have B$_{\mathrm{HA}}$ like transition region with relarively large fractional abundance of water vapor ($\sim 10^{-8}$).
Thus in the cases of candidate water lines with smaller Einstein $A$ coefficients (A$_{\mathrm{ul}}$=$10^{-6} \sim10^{-3}$ s$^{-1}$) and relatively high upper state energies (E$_{\mathrm{up}}$$\sim$1000K), the contribution to the line emission of the optically thick hot midplane inside the H$_{2}$O snowline is much higher than that of the outer optically thin surface layer.
\\ \\
In the top panels of Figures \ref{Figure1_paperIII} and \ref{Figure2_paperIII}, the contributions from the inner disk within the $\mathrm{H_2O}$ snowline 
(regions A$_{\mathrm{TT}}$, A$_{\mathrm{HA}}$ and B$_{\mathrm{HA}}$, see also Table 2)
are very large compared with the contributions from the outer disk
(regions B$_{\mathrm{TT}}$, C$_{\mathrm{HA}}$, see also Table 2), and they have the characteristic Keplerian rotationally broadened double-peaked profiles.
Moreover, in the line profiles from the Herbig Ae disk, most of the emission fluxes come from the high $\mathrm{H_2O}$ gas abundance region (region A$_{\mathrm{HA}}$).
The two peak positions and the rapid flux density drop between the peaks contains information on the hot $\mathrm{H_2O}$ vapor distribution inside the $\mathrm{H_2O}$ snowline.
The spread in the wings of the emission profile (high velocity regions) represents the inner edge of the $\mathrm{H_2O}$ gas distribution in the disk.
This is because emission from each radial region in the disk is Doppler shifted due to the Keplerian rotation (see also Equations (11) and (12) of Paper I).
Comparing sub-millimeter water lines having the same transition quantum numbers ($J_{K_{a}K_{c}}$), the total flux of the para-$\mathrm{H_2}$$^{16}\mathrm{O}$ 183 GHz line tends to be 0.8 times larger than that of the para-$\mathrm{H_2}$$^{18}\mathrm{O}$ 203 GHz line (see Tables 3 and 4).
\\ \\
Figure \ref{Figure3_paperIII} shows the velocity profiles for the para-$\mathrm{H_2}$$^{18}\mathrm{O}$ lines at 203 GHz (top) and 1102 GHz (bottom) from the Herbig Ae disk inside 30 au with different inclination angles ($i=5, 15, 30, 45, 60 \deg$).
If we observe objects with smaller/larger disk inclination angles than those of our original model value ($i=$30 $\deg$), the widths of the line peaks become smaller/larger.
This is because the projected Keplerian velocity is proportional to $\sin i$.
If the object is nearly face on ($i\sim0 \deg$), the shape of the line profile is close to that expected from thermal broadening only, and the flux densities at the line peaks become larger.
In contrast, in objects with larger disk inclination angles, the line emission components from various places in the disks are dispersed at various velocities due to the effect of Doppler shift.
Thus the flux densities at the line peaks tend to be smaller as the disk inclination angles become larger.
Here we note that the flux densities at the line peaks are similar in the cases of much larger inclination angles ($i>45 \deg$).
Since the line width depends on central star mass and inclination angle of the disk, we have to know these values in advance through other observations (e.g., resolved imaging of strong molecular lines like CO lines) in order to locate the $\mathrm{H_2O}$ snowline from the profiles of water line emission.
\\ \\
In the bottom panels of Figures \ref{Figure1_paperIII} and \ref{Figure2_paperIII}, except for the para-$\mathrm{H_2}$$^{18}\mathrm{O}$ 1102 GHz line for the Herbig Ae disk, the outer disk contributions (regions B$_{\mathrm{TT}}$, C$_{\mathrm{HA}}$, see Table 2) are large compared with that from the inner disk within the $\mathrm{H_2O}$ snowline (regions A$_{\mathrm{TT}}$, A$_{\mathrm{HA}}$ and B$_{\mathrm{HA}}$, see Table 2).
In addition, the widths between the line double peaks are about 2-3 times narrower than those of candidate water lines (see top panels of Figures \ref{Figure1_paperIII} and \ref{Figure2_paperIII}), although the values of $A_{\mathrm{ul}}$ are not so high. 
This is because these water lines are the ground-state rotational transitions and have much smaller values of $E_{\mathrm{up}}$ ($\sim$50K) compared to those of other lines. The fluxes of these lines mainly come from the water reservoir in the outer cold photodesorption region.
\subsection{The local intensity and optical depth distributions of sub-millimeter water emission lines}
\noindent Figure \ref{Figure4_paperIII} shows the water line local intensity (emissivity times line-of-sight extinction and local length, $\eta_{ul} e^{-\tau_{ul}}$$\mathrm{ds}$;
see also Equation (14) of Paper I) 
for the Herbig Ae disk.
In the left panels of Figure \ref{Figure4_paperIII}, the total (gas and dust) optical depth contours ($\tau_{ul}=$0.1, 1, and 10) are plotted on top of the line local intensities. The gas temperature contours
($T_{g}=120, 170, 300$K) are plotted in the right-hand panels. 
The line-of-sight direction corresponds to $z=+\infty$ to $-\infty$ at each disk radius, thus the inclination angle is assumed to be 0 $\deg$.
In calculating the values of $\tau_{ul}$, we consider the contributions of both line absorption by the water gas and absorption by dust grains. 
Here we note that the units in Figures \ref{Figure1_paperIII}, \ref{Figure2_paperIII}, and \ref{Figure3_paperIII} are Jy ($=10^{26}$ W m$^{-2}$ Hz$^{-1}$) and the values are those observed on Earth. In contrast, we plot the ``local intensity" (not ``flux density") of each grid in the disk in Figure \ref{Figure4_paperIII}, and the units in this figure are W m$^{-2}$ Hz$^{-1}$ sr$^{-1}$ (see also papers I and II). 
Figure \ref{Figure5_paperIII} shows 
the normalized cumulative line local intensity distributions along the vertical direction at $r=$5 au (top panels), $r=$10 au (middle panels), and $r=$30 au (bottom panels), along with the gas temperature $T_{g}$ in K.
In Figure \ref{Figure5_paperIII}, we also plot the para-$\mathrm{H_2}$$^{16}\mathrm{O}$ 325 GHz line and the para-$\mathrm{H_2}$$^{18}\mathrm{O}$ 322 GHz line, which are the same transition levels and fall in ALMA Band 7. Here we note that the detailed profiles of these two lines are shown in Figure \ref{Figure8_paperIII}.
Looking at the top panels of Figures \ref{Figure4_paperIII} and all panels of Figure \ref{Figure5_paperIII}, the value of local intensities within $r<$14 au ($=$ the position of the $\mathrm{H_2O}$ snowline),  $T_{g}>$120K, and $z/r \lesssim 0.1$ are larger than those from the outer optically thin hot surface layer and the photodesorption region. In particular, the local intensities from the high $\mathrm{H_2O}$ vapor abundance region (region A$_{\mathrm{HA}}$)
and $z/r \lesssim 0.1$ are much larger (see also Section 3.2.1 of paper II).
Here we note that these sub-millimeter lines have smaller $A_{\mathrm{ul}}$ values ($\sim10^{-5}-10^{-6}$ s$^{-1}$), and relatively smaller E$_{\mathrm{up}}$ values ($\sim200-500$K) compared with candidate ortho-$\mathrm{H_2}$$^{16}\mathrm{O}$ lines which were discussed in papers I and II ($A_{\mathrm{ul}}$$\sim10^{-3}-10^{-6}$ s$^{-1}$, E$_{\mathrm{up}}$$\sim700-2100$K).
For the sub-millimeter $\mathrm{H_2}$$^{18}\mathrm{O}$ and para-$\mathrm{H_2}$$^{16}\mathrm{O}$ lines, since the values of $A_{\mathrm{ul}}$ tend to be smaller (typically $<$$10^{-4}$ s$^{-1}$) than those of infrared candidate ortho-$\mathrm{H_2}$$^{16}\mathrm{O}$ lines (see paper II), and the number densities of $\mathrm{H_2}$$^{18}\mathrm{O}$ and para-$\mathrm{H_2}$$^{16}\mathrm{O}$ molecules are smaller (OPR=3, $^{16}$O/$^{18}$O=560, see also Section 2.2) than those of ortho-$\mathrm{H_2}$$^{16}\mathrm{O}$ molecules, contributions from the outer optically thin surface region become much smaller.
In the sub-millimeter candidate water line cases, the local intensity from the outer optically thin disk is around $10^{4}$ times smaller than that for the infrared candidate water line cases (see paper II).
Therefore, we recommend that sub-millimeter $\mathrm{H_2}$$^{18}\mathrm{O}$ and para-$\mathrm{H_2}$$^{16}\mathrm{O}$ lines with relatively small values of E$_{\mathrm{up}}$ ($\sim$200K) can be used for locating the position of the $\mathrm{H_2O}$ snowline. 
\\ \\
The optical depths become larger as $E_{\mathrm{up}}$ becomes smaller (see Figure \ref{Figure5_paperIII}), because the absorption by lines is stronger even in the colder region of the disk (see also paper II).
The dust opacity at sub-millimeter wavelengths is small compared with those at infrared wavelengths. 
Hence, for the case of candidate sub-millimeter lines, line absorption by excited molecules mainly determines the emitting regions of lines and total optical depth profiles in the disk midplane of the inner disk with a high $\mathrm{H_2O}$ gas abundance (region A$_{\mathrm{HA}}$). On the other hand, dust absorption mainly controls the line opacity in the disk surface and colder midplane in the outer disk.
Therefore, using candidate sub-millimeter lines, we can detect H$_{2}$O vapor closer to the midplane ($z=0$) inside the H$_{2}$O snowline, compared with infrared lines.
In the cases of the para-$\mathrm{H_2}$$^{16}\mathrm{O}$ lines (see Figures \ref{Figure4_paperIII} and \ref{Figure5_paperIII}), emission from $z < 0.03$ at $r\lesssim$ 3 au is not detectable. This is because the optical depth of the innermost disk midplane is high due to absorption by excited water molecules and dust grains in the upper disk layer. 
\\ \\
Since the number densities of the para-H$_{2}$$^{16}$O molecules are one third smaller than that of ortho-H$_{2}$$^{16}$O molecules, the former can trace deeper into the disk than the latter.
Furthermore, because the number densities of the H$_{2}$$^{18}$O molecules are around 560 times smaller than that of H$_{2}$$^{16}$O molecules, H$_{2}$$^{18}$O lines can trace much deeper into the disk (down to $z=0$) than H$_{2}$$^{16}$O lines, and thus H$_{2}$$^{18}$O lines are better targets to extract the position of the H$_{2}$O snowline at the disk midplane. 
If the dust opacity of the object is much larger than that of our disk model, however, the H$_{2}$$^{16}$O lines are preferred targets.
We also discuss in detail the impact of dust emission in Section 4.1.
\\ \\
In the Herbig Ae disk (see paper II), there was a difference between the exact $\mathrm{H_2O}$ snowline location ($r\sim$14 au) and the outer edge of the hot water vapor area ($r\sim$ 8 au), although there was no difference between them in a T Tauri disk with the radial steeper temperature profile in the disk midplane (see also paper I).
This is because the water formation rate by gas-phase reactions strongly depends on the gas temperature.
Such a water vapor distribution in the Herbig Ae disk midplane was discussed in \citet{Woitke2009b}. 
Here we point out that this distribution will depend on the adopted chemical model.
\citet{Eistrup2016} calculated the chemical evolution in a disk midplane under both initial atomic and molecular abundances.
They reported that for molecular initial abundances, the water gas and ice fractional abundances around the $\mathrm{H_2O}$ snowline ($\sim10^{-4}$) are larger than that for atomic initial abundances ($\sim 10^{-6}$).
\\ \\
According to our model calculations, some sub-millimeter candidate ortho-$\mathrm{H_2}$$^{16}\mathrm{O}$ lines (e.g., see also paper II) can trace the position of the exact $\mathrm{H_2O}$ snowline location. In contrast, sub-millimeter candidate para-$\mathrm{H_2}$$^{16}\mathrm{O}$ lines and para-$\mathrm{H_2}$$^{18}\mathrm{O}$ lines discussed in this subsection mainly trace the position of the outer edge of the hot water vapor area (see Figures \ref{Figure1_paperIII}, \ref{Figure4_paperIII}, \ref{Figure6_paperIII}).
The differences in the line properties come from the differences in $A_{\mathrm{ul}}$ and number densities among the lines.
\\ \\
Here we note that \cite{vanKempen2008} conducted the calculations of far-infrared $\mathrm{H_2}$$^{16}\mathrm{O}$ and $\mathrm{H_2}$$^{18}\mathrm{O}$ lines within the $Herschel$/HIFI frequency coverage for Class 0 protostar models, and suggested that higher excitation lines ($E_{\mathrm{up}}$$>$150 K) are usually dominated by emission coming from the warm inner region.
\\ \\
According to Figures \ref{Figure4_paperIII} and \ref{Figure5_paperIII}, the local intensities of the para-$\mathrm{H_2}$$^{16}\mathrm{O}$ 1113 GHz line
are similar both outside and inside the $\mathrm{H_2O}$ snowline for the Herbig Ae disk.
However, most disk-integrated line emission comes from the outer disk on account of the larger surface area.
Moreover, the line opacity in the outer disk midplane is about $10^{3}-10^{4}$ times higher than those of the candidate $\mathrm{H_2O}$ lines with similar line wavelengths and thus similar dust opacities.
\\ \\
Here we note that in the case of the para-$\mathrm{H_2}$$^{18}\mathrm{O}$ line at 1102 GHz for the Herbig Ae disk, the contribution from the inner optically thick layer within the $\mathrm{H_2O}$ snowline is still dominant (see Figure \ref{Figure4_paperIII}).
This is because the number densities of para-$\mathrm{H_2}$$^{18}\mathrm{O}$ molecules are much smaller than those of $\mathrm{H_2}$$^{16}\mathrm{O}$ molecules.
However, the emitting region of this line extends farther into the outer disk (see also Section 3.3). 
Thus the contribution of the outer disk is expected to be larger if we integrate emission components within a larger disk radius (e.g., $r\lesssim$ hundreds au).
Therefore, we see that these two lines are not appropriate to detect the hot water emission inside the $\mathrm{H_2O}$ snowline, as concluded also for the ortho-$\mathrm{H_2}$$^{16}\mathrm{O}$ 557 GHz line (see papers I and II).
\subsection{The normalized radial cumulative line fluxes}
\noindent The top two panels of Figure \ref{Figure6_paperIII} show the normalized cumulative fluxes in the radial directions of three para-$\mathrm{H_2}$$^{16}\mathrm{O}$ lines at 183 GHz, 325 GHz, and 1113 GHz, and the bottom two panels show those of para-$\mathrm{H_2}$$^{18}\mathrm{O}$ lines at 203 GHz, 322 GHz, and 1102 GHz, for the Herbig Ae disk.
For the candidate $\mathrm{H_2}$$^{16}\mathrm{O}$ and $\mathrm{H_2}$$^{18}\mathrm{O}$ lines, most line flux is emitted from the region with a high water vapor abundance ($\sim10^{-5}-10^{-4}$, $r<$8 au).
For the para-$\mathrm{H_2}$$^{16}\mathrm{O}$ 1113 GHz line case, the line emitting region is much further out from the position of $\mathrm{H_2O}$ snowline ($r\sim 50-300$ au), and it is similar to that of the ortho-$\mathrm{H_2}$$^{16}\mathrm{O}$ 557 GHz line (see paper II).
\\ \\
Here we point out that line emission from the para-$\mathrm{H_2}$$^{18}\mathrm{O}$ 1102 GHz line is both from the region inside ($r<$14 au) and much further out ($r>$100 au) than the position of $\mathrm{H_2O}$ snowline. We also discussed this line property in the last paragraph of Section 3.2.
\subsection{The properties of all other sub-millimeter water emission lines}
\begin{figure*}[htbp]
\begin{center}
\includegraphics[scale=0.6]{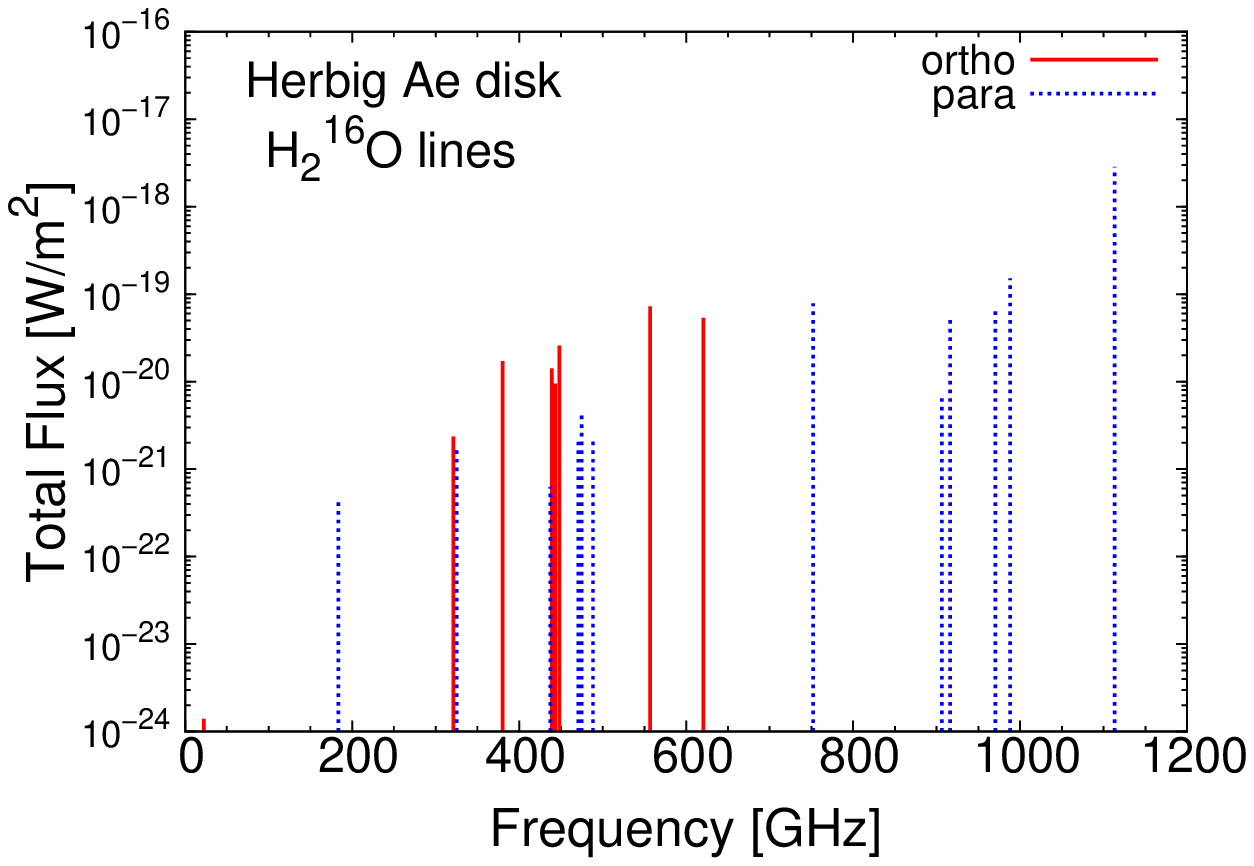}
\includegraphics[scale=0.6]{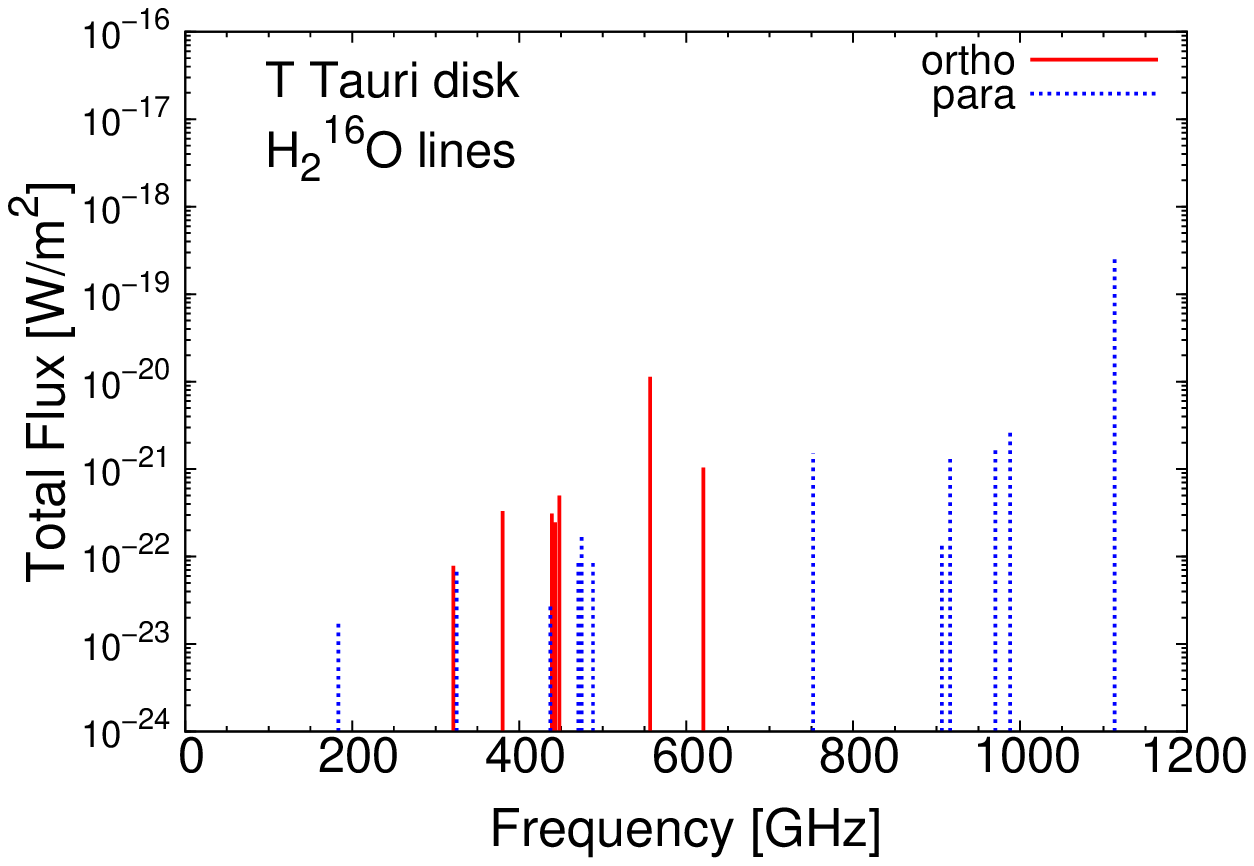}
\includegraphics[scale=0.6]{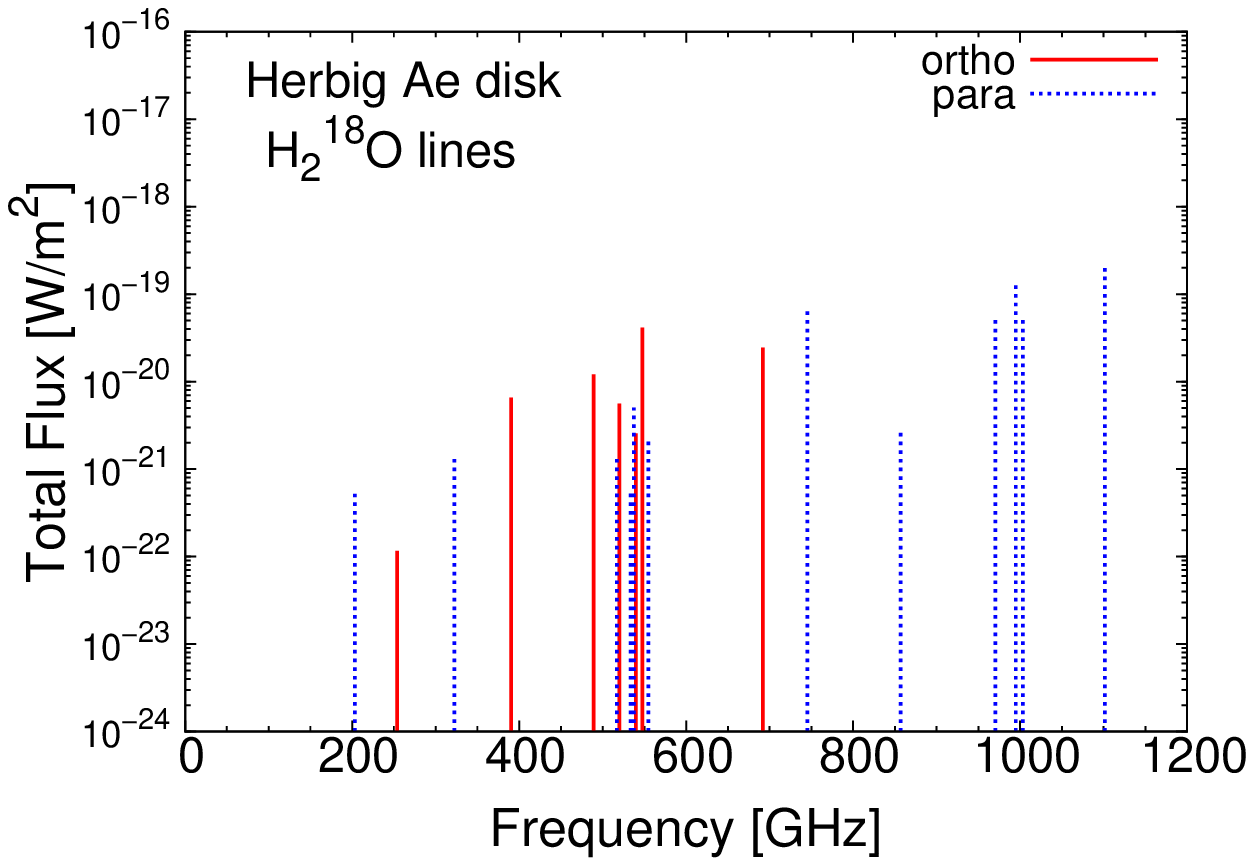}
\includegraphics[scale=0.6]{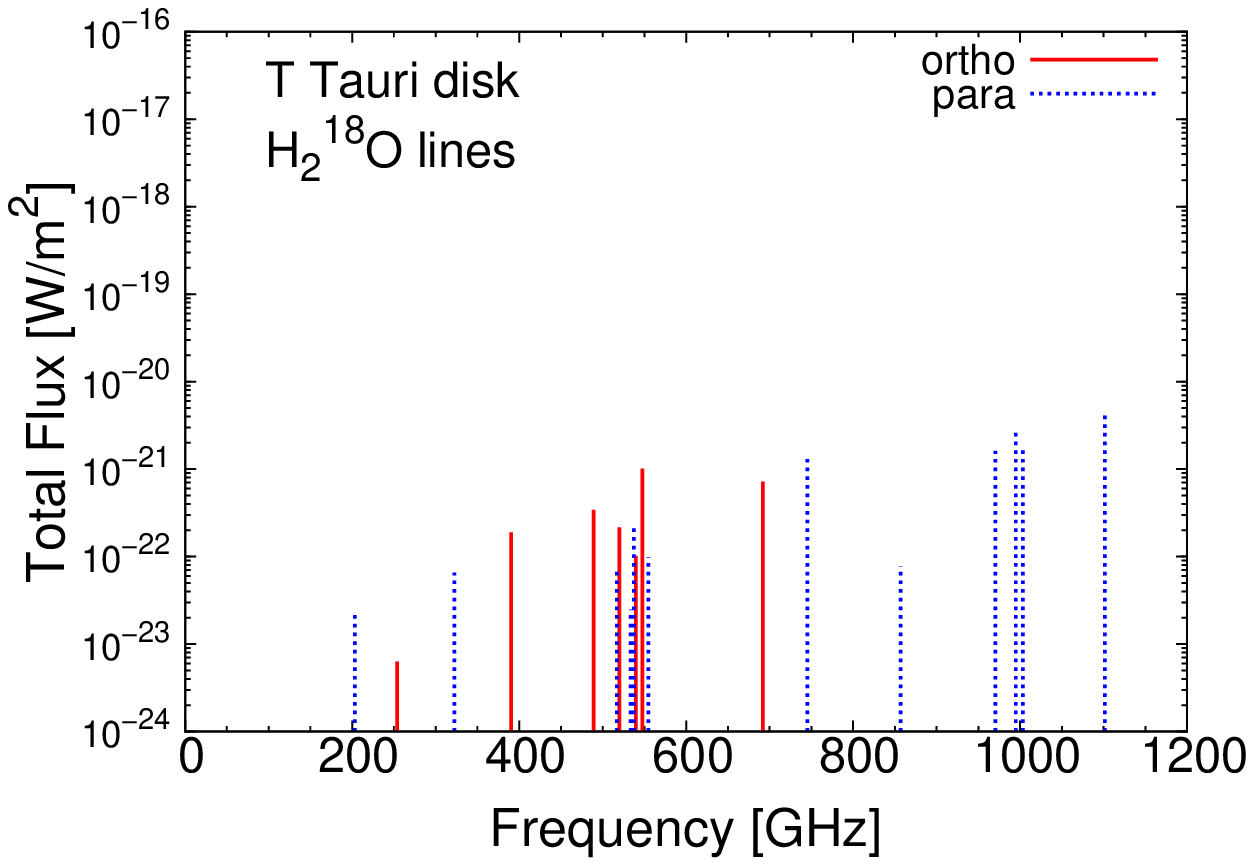}

\end{center}
\vspace{0.6cm}
\caption{\noindent 
The disk integrated emergent fluxes of sub-millimeter ortho- and para-$\mathrm{H_2}$$^{16}\mathrm{O}$ lines (top panels), and ortho- and para-$\mathrm{H_2}$$^{18}\mathrm{O}$ lines (bottom panels), 
from a Herbig Ae disk (left panels) and a T Tauri disk (right panels). The ortho- and para- water lines are plotted with {\it red solid lines} and {\it blue dotted lines}, respectively.
These water lines are selected on the basis of their excitation energies, $0 < E_{\mathrm{up}} < 2000$K, and lower frequencies, $\nu = 0-1003$ GHz ($\lambda > 297 $$\mu$m).
In these panels, we also plot the total fluxes of the ground-state rotational transitions of the para-$\mathrm{H_2}$$^{16}\mathrm{O}$ 1113 GHz line and para-$\mathrm{H_2}$$^{18}\mathrm{O}$ 1102 GHz line.
\vspace{0.5cm}
}\label{Figure7_paperIII}
\end{figure*}
\begin{figure*}[htbp]
\begin{center}
\includegraphics[scale=0.6]{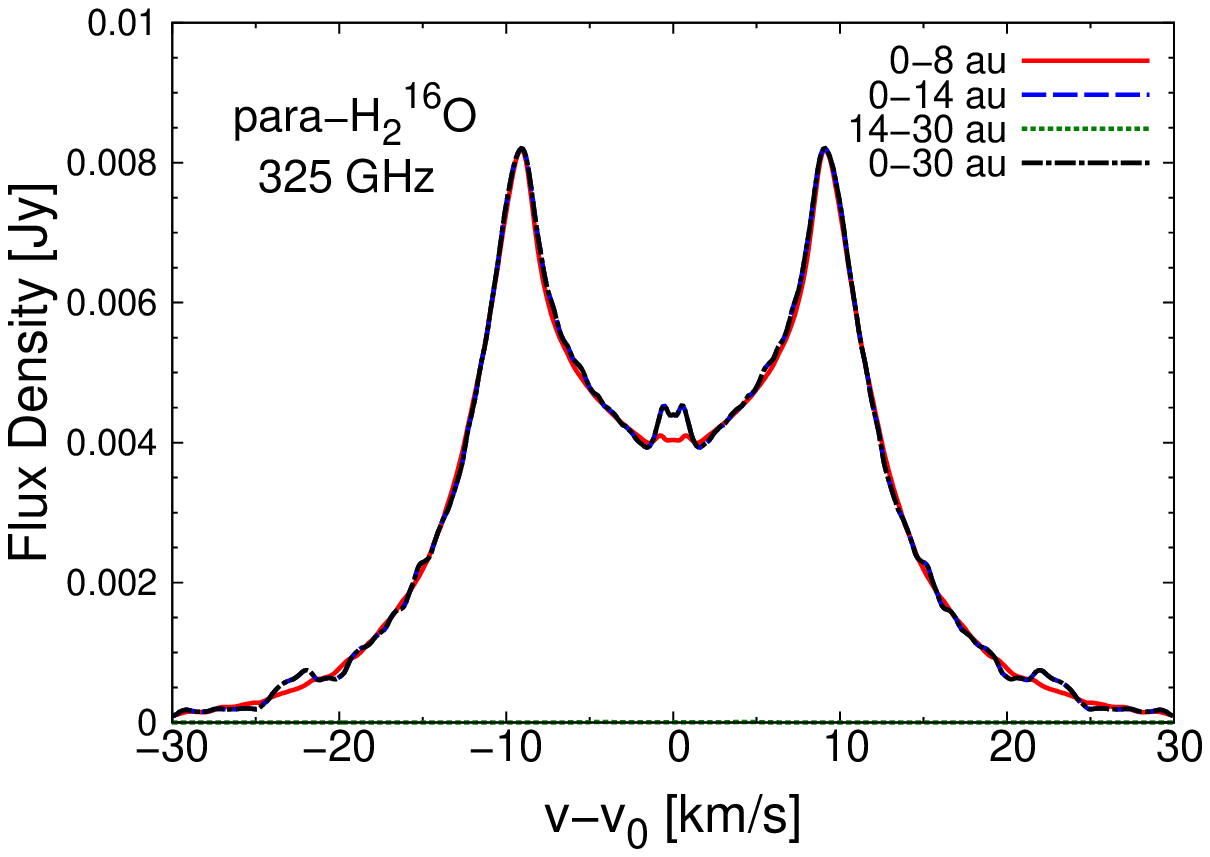}
\includegraphics[scale=0.6]{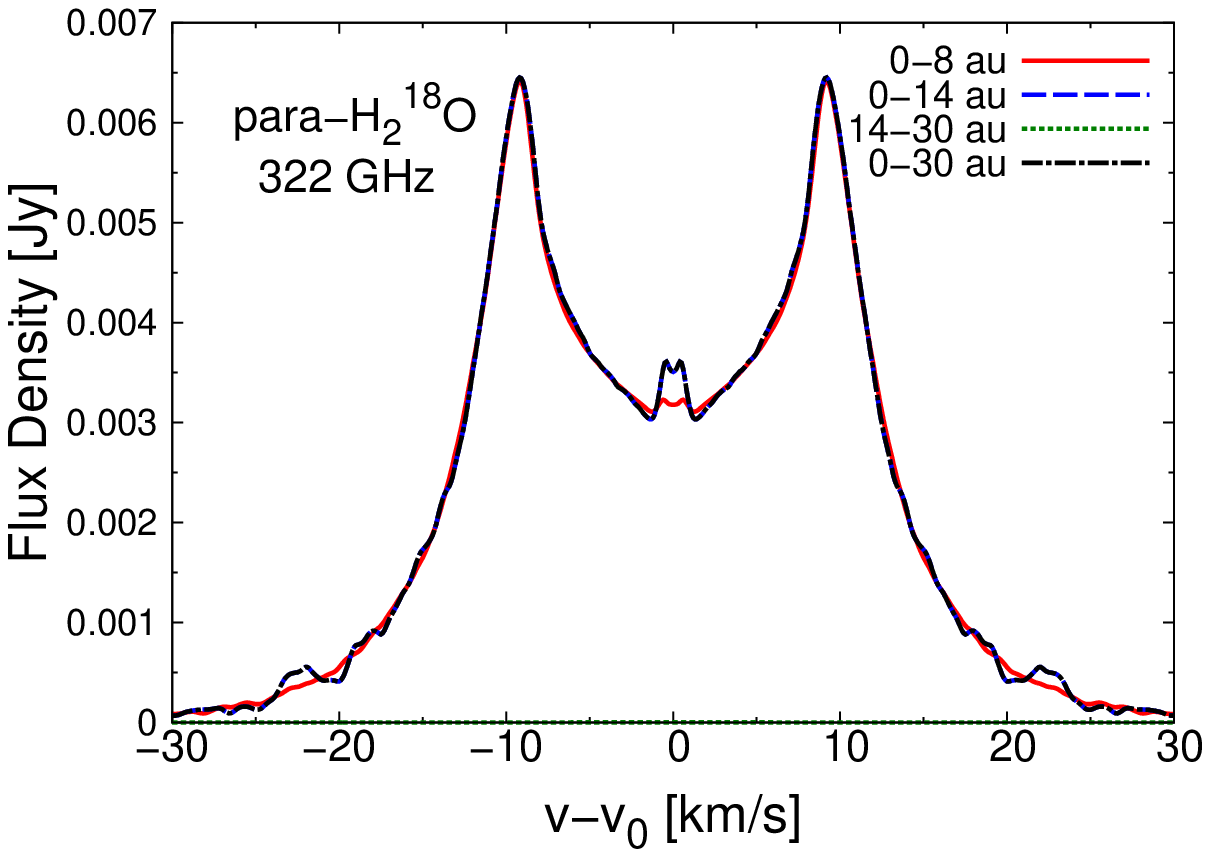}
\includegraphics[scale=0.6]{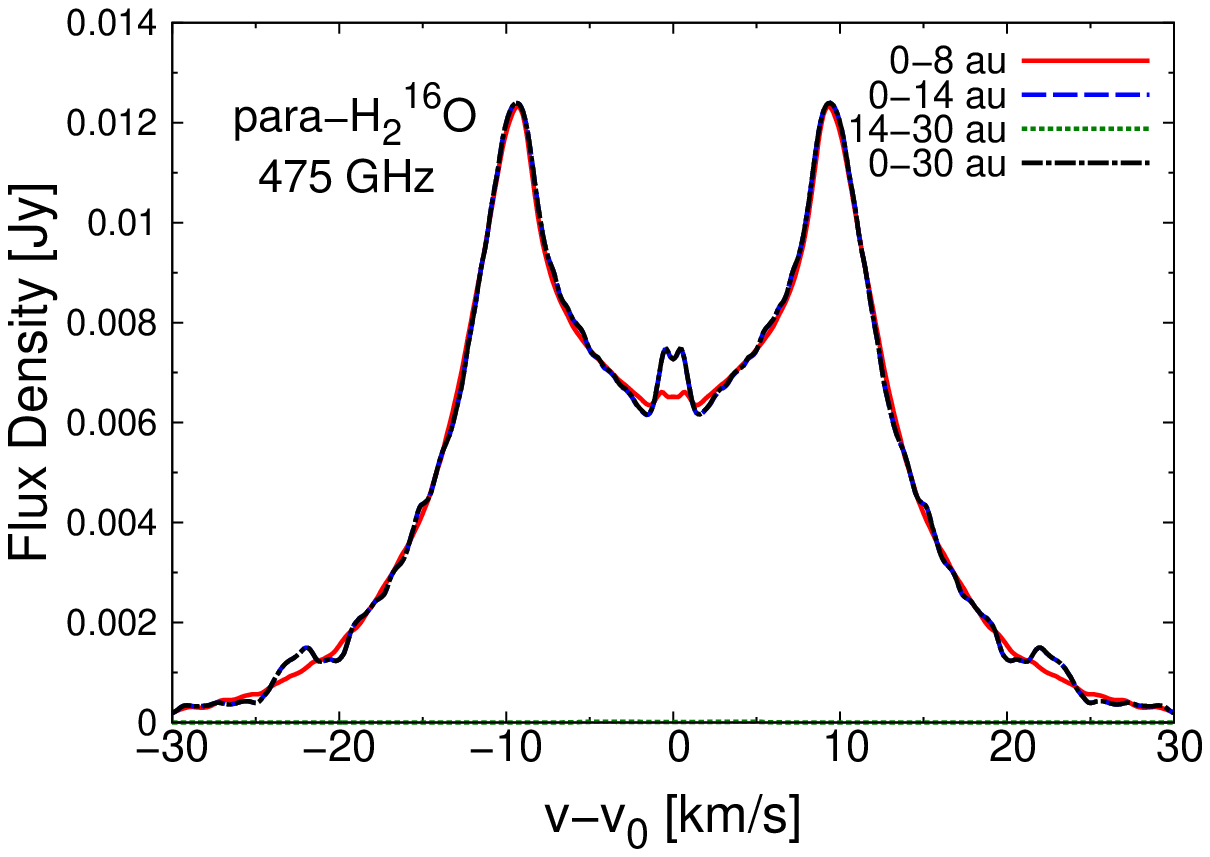}
\includegraphics[scale=0.6]{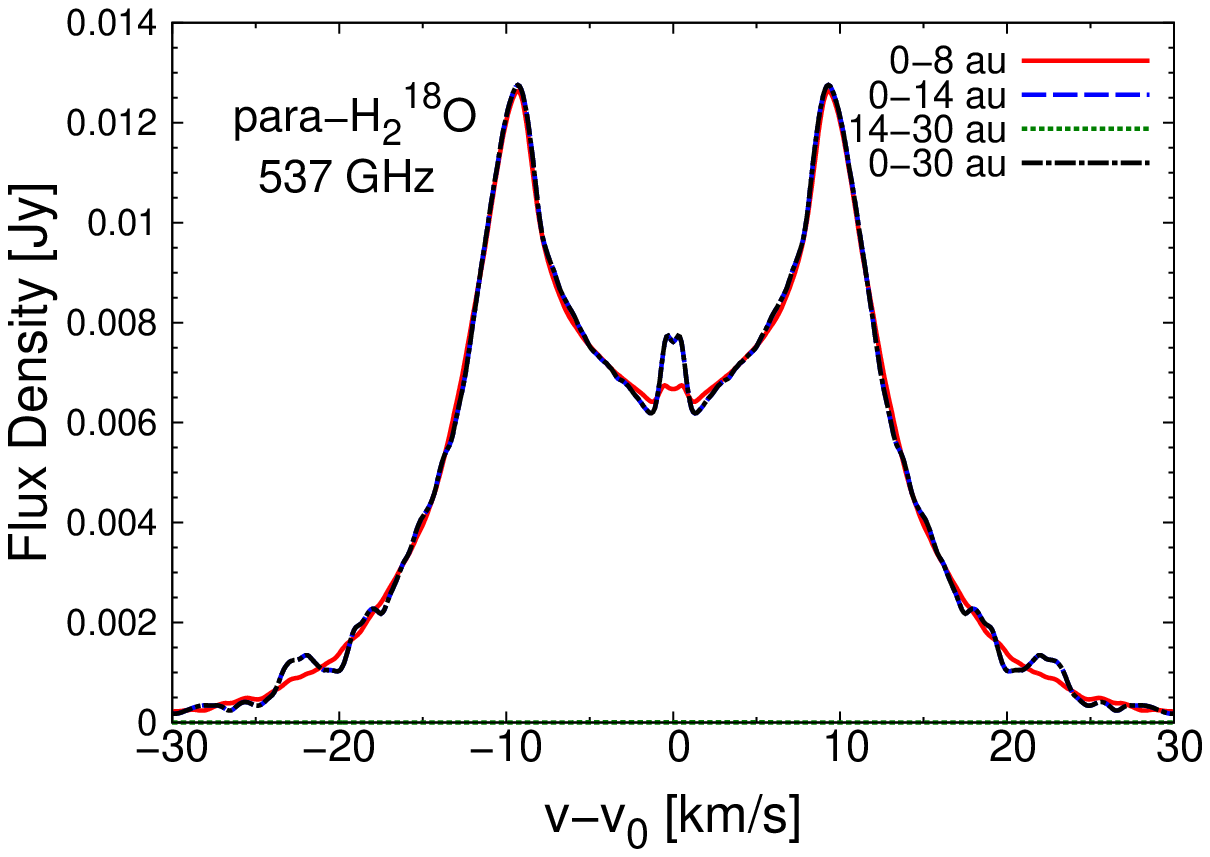}
\end{center}
\vspace{0.6cm}
\caption{\noindent The profiles of para-$\mathrm{H_2}$$^{16}\mathrm{O}$ lines at 325 GHz (top left) and 475 GHz (bottom left), and para-$\mathrm{H_2}$$^{18}\mathrm{O}$ lines at 322 GHz (top right), 
and 537 GHz (bottom right) from the Herbig Ae disk.
The parameters and total fluxes of these $\mathrm{H_2O}$ lines are listed in Tables 3 and 4.
In these profiles, we do not include dust emission.
The line profiles from within 8 au, the inner high temperature region, are displayed with {\it red solid lines}, those from within the $\mathrm{H_2O}$ snowline, 14 au, with {\it blue dashed lines}, those from 14-30 au, outside the $\mathrm{H_2O}$ snowline, with {\it green dotted lines}, and those from the total area inside 30 au, with {\it black dashed dotted lines}. 
In the panels of para-$\mathrm{H_2}$$^{16}\mathrm{O}$ and para-$\mathrm{H_2}$$^{18}\mathrm{O}$ line profiles, the flux densities outside the $\mathrm{H_2O}$ snowline ({\it green dotted lines}, $<$ 10$^{-4}$ Jy) are much smaller than those inside 8 au ({\it red solid lines}).
Therefore, the {\it red solid lines}, {\it blue dashed lines}, and {\it black dashed dotted lines} are almost completely overlapped (see also Figure \ref{Figure1_paperIII}).
\vspace{0.3cm}
}\label{Figure8_paperIII}
\end{figure*} 
\setcounter{figure}{7}
\begin{figure*}[htbp]
\begin{center}
\includegraphics[scale=0.6]{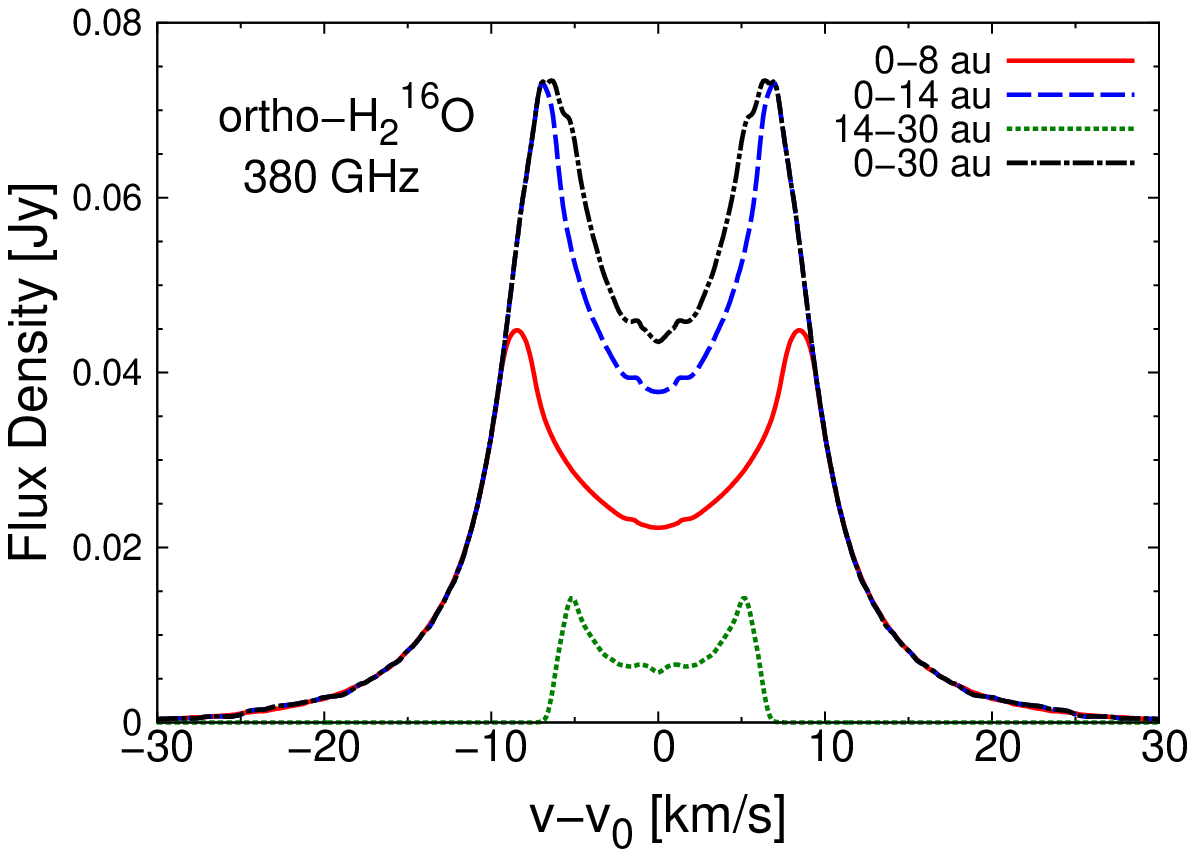}
\includegraphics[scale=0.6]{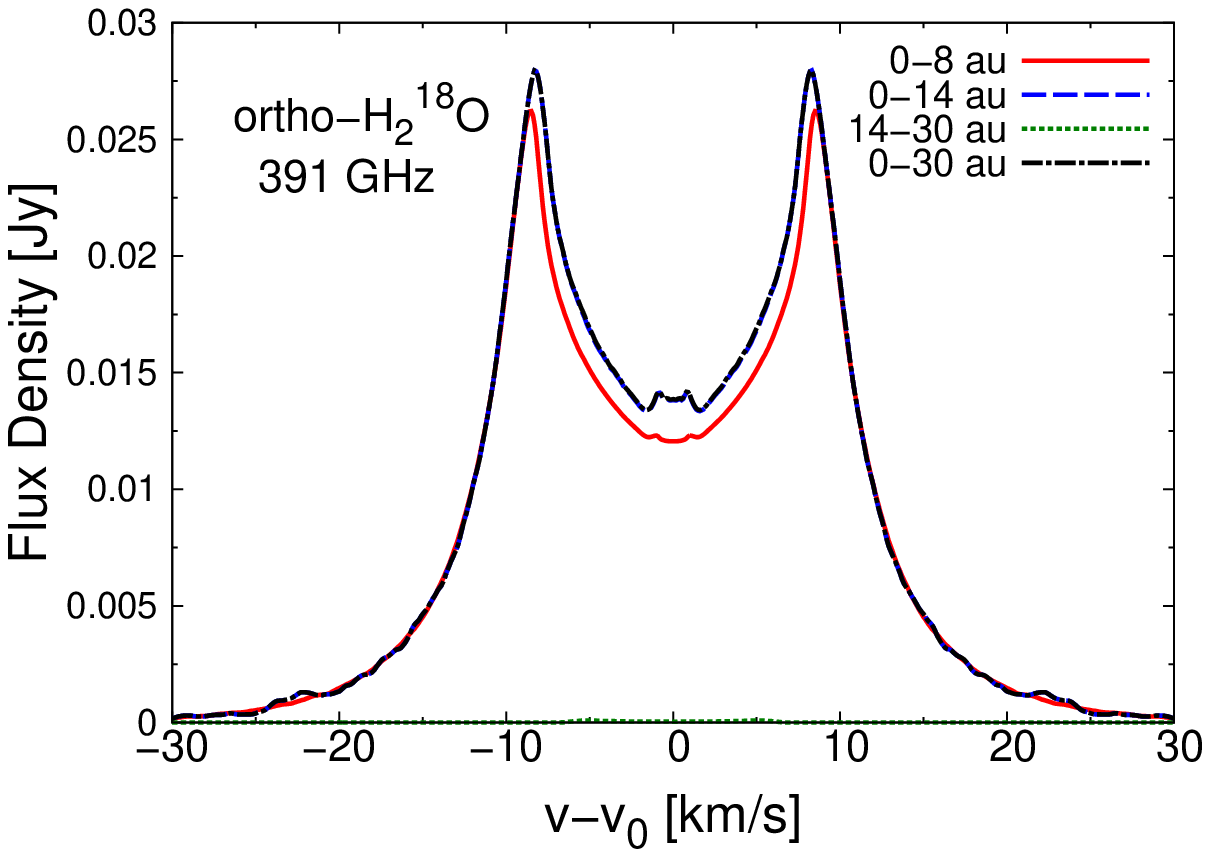}
\includegraphics[scale=0.6]{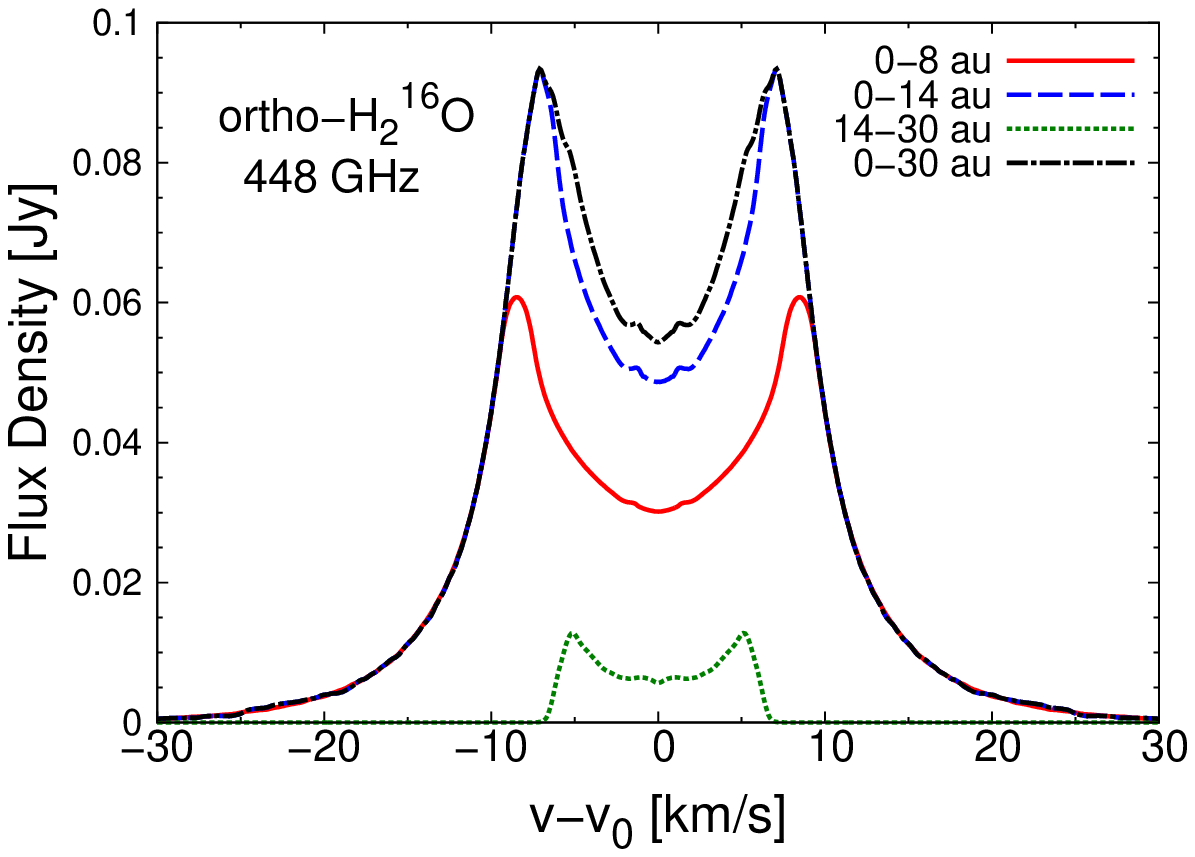}
\includegraphics[scale=0.6]{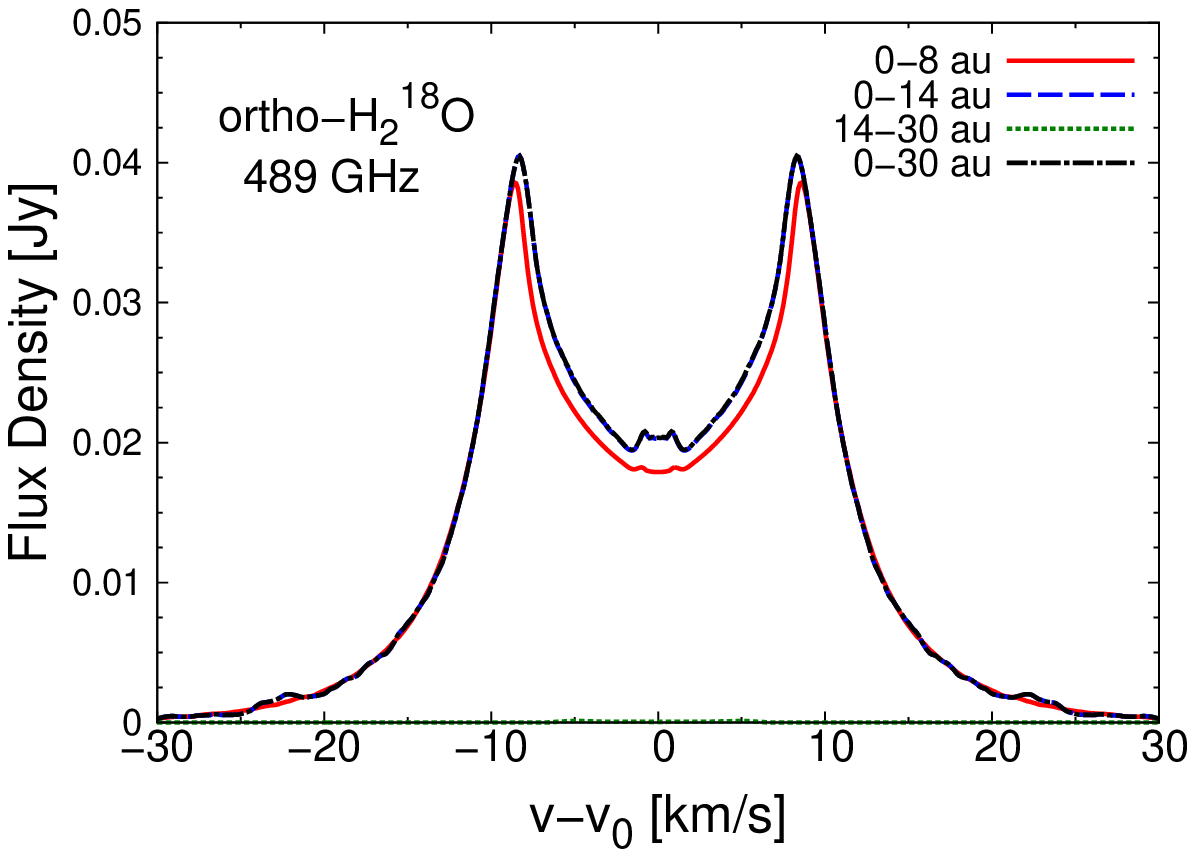}
\includegraphics[scale=0.6]{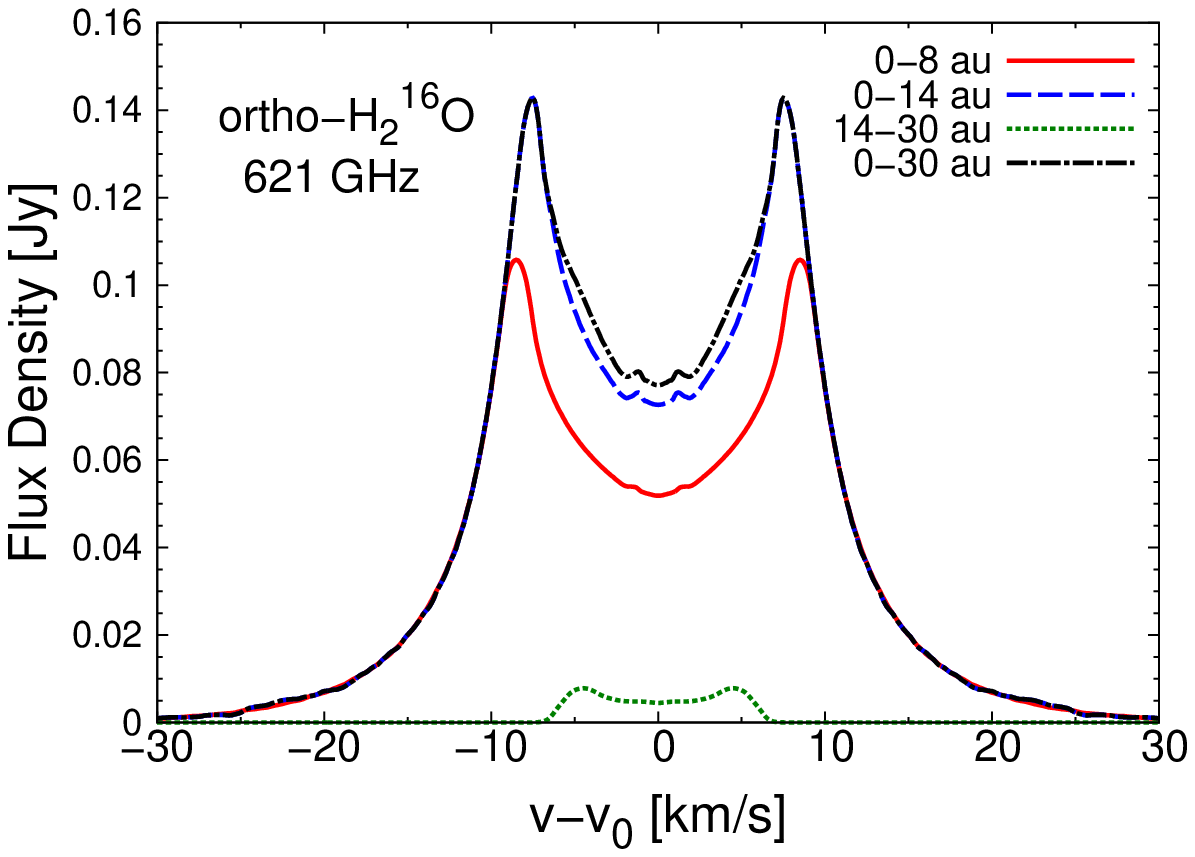}
\includegraphics[scale=0.6]{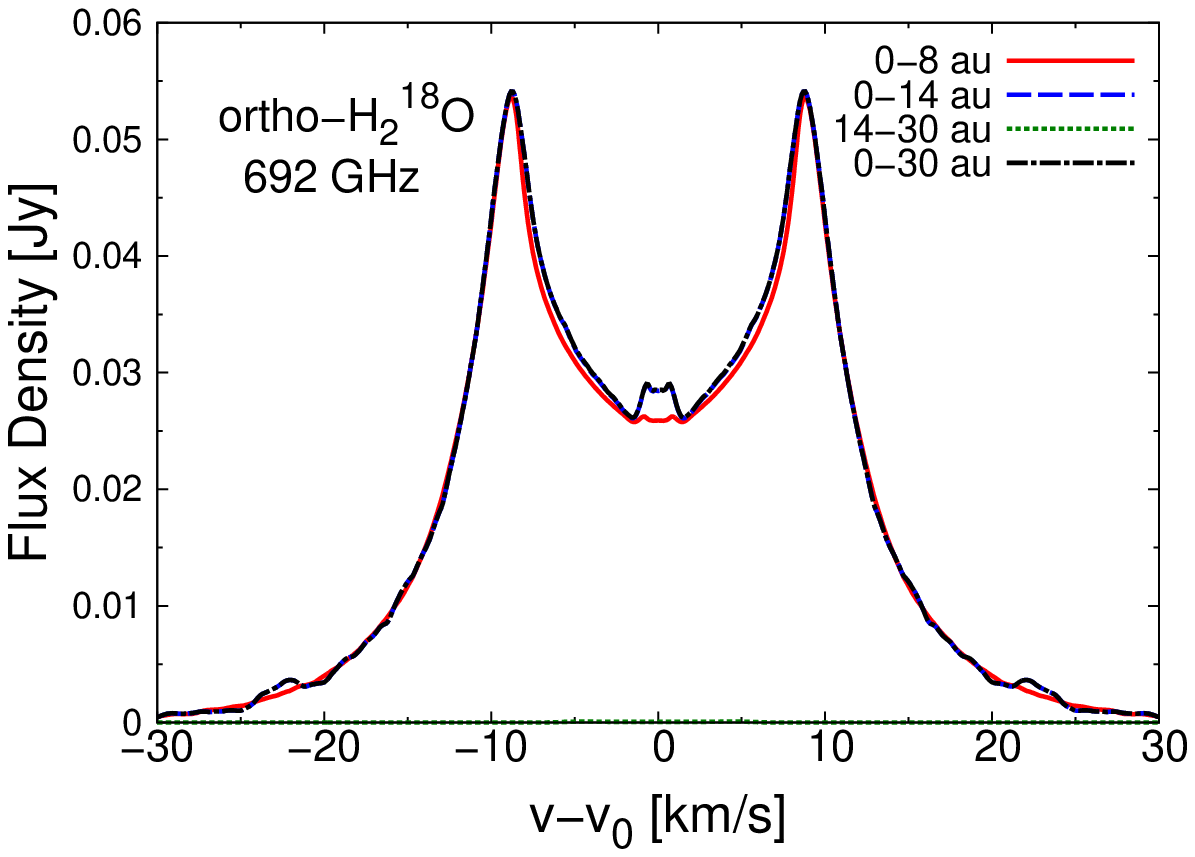}
\end{center}
\vspace{0.6cm}
\caption{\noindent (Continued.) The profiles of ortho-$\mathrm{H_2}$$^{16}\mathrm{O}$ lines at 380 GHz (top left), 448 GHz (middle left), and 621 GHz (bottom left), ortho-$\mathrm{H_2}$$^{18}\mathrm{O}$ lines at 391 GHz (top right), 489 GHz (middle right), and 692 GHz (bottom right) from the Herbig Ae disk. 
The parameters and total fluxes of these $\mathrm{H_2O}$ lines are listed in Tables 3 and 4.
In the panels of ortho-$\mathrm{H_2}$$^{18}\mathrm{O}$ line profiles, the flux densities outside the $\mathrm{H_2O}$ snowline ({\it green dotted lines}, $<$ 10$^{-3}$ Jy) are much smaller than those inside 8 au ({\it red solid lines}).
Therefore, the {\it red solid lines}, {\it blue dashed lines}, and {\it black dashed dotted lines} are almost completely overlapped (see also Figure \ref{Figure1_paperIII}).
\vspace{0.3cm}
}
\label{Figure8_paperIII}
\end{figure*} 
\noindent
Figure \ref{Figure7_paperIII} shows the total fluxes of calculated all sub-millimeter ortho- and para-$\mathrm{H_2}$$^{16}\mathrm{O}$ lines and ortho- and para-$\mathrm{H_2}$$^{18}\mathrm{O}$ lines from a Herbig Ae disk (left panels) and a T Tauri disk (right panels).
The detailed line parameters and total line fluxes of $\mathrm{H_2}$$^{16}\mathrm{O}$ lines and $\mathrm{H_2}$$^{18}\mathrm{O}$ lines in Figure \ref{Figure7_paperIII} are listed in Tables 3 and 4, respectively. 
In calculating the total fluxes in Figure \ref{Figure7_paperIII} and Tables 3 and 4, we do not include dust emission, although we do include both gas and dust absorption.
In Figure \ref{Figure7_paperIII} and Tables 3 and 4, we calculated the two ground-state rotational transitions of the 
para-$\mathrm{H_2}$$^{16}\mathrm{O}$ and para-$\mathrm{H_2}$$^{18}\mathrm{O}$ lines (at 1113 GHz and 1102 GHz, respectively).   
We also include all sub-millimeter ortho- and para-$\mathrm{H_2}$$^{16}\mathrm{O}$ lines and ortho- and 
para-$\mathrm{H_2}$$^{18}\mathrm{O}$ lines with $E_{\mathrm{up}}<2000$~K and with a frequency of $\nu \le 1003$~GHz ($\lambda \ge 297$~$\mu$m).
\\ \\
On the basis of Figure \ref{Figure7_paperIII} and Tables 3 and 4, the values of these line fluxes from the Herbig Ae disk are about $1-5\times10^{2}$ larger than those of the T Tauri disk.
This is because the $\mathrm{H_2O}$ snowline position in the T Tauri disk is at a smaller radial distance than that in the Herbig Ae disk.
The total line fluxes tend to be larger as the values of $E_{\mathrm{up}}$ are smaller and $A_{\mathrm{ul}}$ are larger.
In addition, they tend to be smaller as the wavelengths of these water lines are longer, because mid-infrared wavelengths are the peak wavelengths of the Planck function at the gas temperatures around the $\mathrm{H_2O}$ snowline ($T_{g}$$\sim$ 100-200 K).
Furthermore, comparing water lines with the same transition quantum numbers ($J_{K_{a}K_{c}}$), the total fluxes of $\mathrm{H_2}$$^{16}\mathrm{O}$ lines tend to be $0.5-10^{2}$ times larger than those of $\mathrm{H_2}$$^{18}\mathrm{O}$ lines, and the flux ratios tend to be larger in the cases of ortho-$\mathrm{H_2O}$ lines.
The differences in flux ratios are mainly due to the differences in emitting regions.
\\ \\
\noindent Figure \ref{Figure8_paperIII} shows the velocity profiles of several water lines in the frequency range 300$-$700 GHz for the Herbig Ae disk.
The values of $E_{\mathrm{up}}$ ($\sim300-700$K) of lines in Figure \ref{Figure8_paperIII} are relatively smaller than those of candidate ortho-$\mathrm{H_2}$$^{16}\mathrm{O}$ lines discussed in papers I and II ($E_{\mathrm{up}}\sim$1000K).
In these cases contributions from the relatively high water vapor abundance region ($\sim$$10^{-8}$, $r\sim8-14$ au) dominate.
By contrast, in the case of para-$\mathrm{H_2}$$^{16}\mathrm{O}$ and $\mathrm{H_2}$$^{18}\mathrm{O}$ lines, most of the flux comes from the high $\mathrm{H_2O}$ gas vapor abundance region ($\sim$$10^{-5}-10^{-4}$, $r<$8 au), since the number densities of $\mathrm{H_2}$$^{18}\mathrm{O}$ and para-$\mathrm{H_2}$$^{16}\mathrm{O}$ molecules are smaller than those of ortho-$\mathrm{H_2}$$^{16}\mathrm{O}$ molecules.
\section{Discussion}
\begin{figure*}[htbp]
\begin{center}
\includegraphics[scale=0.6]{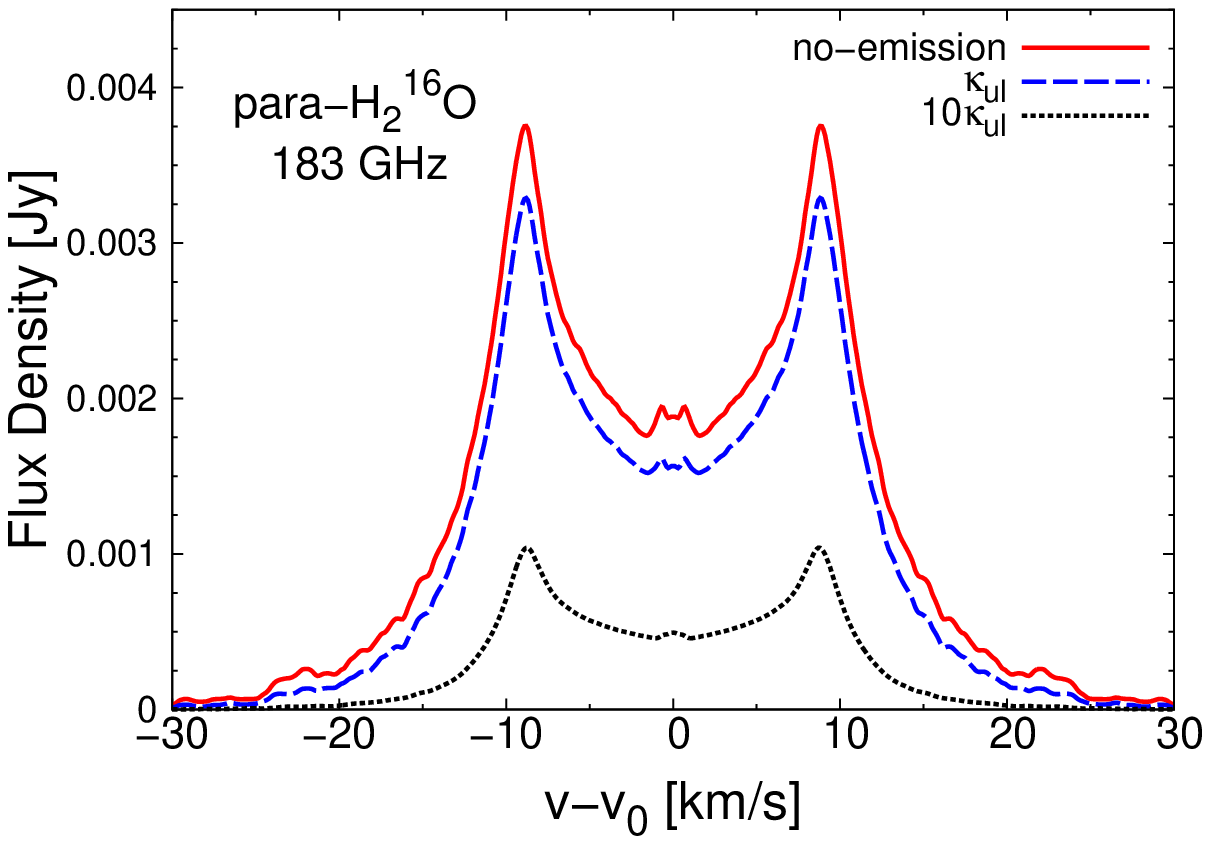}
\includegraphics[scale=0.6]{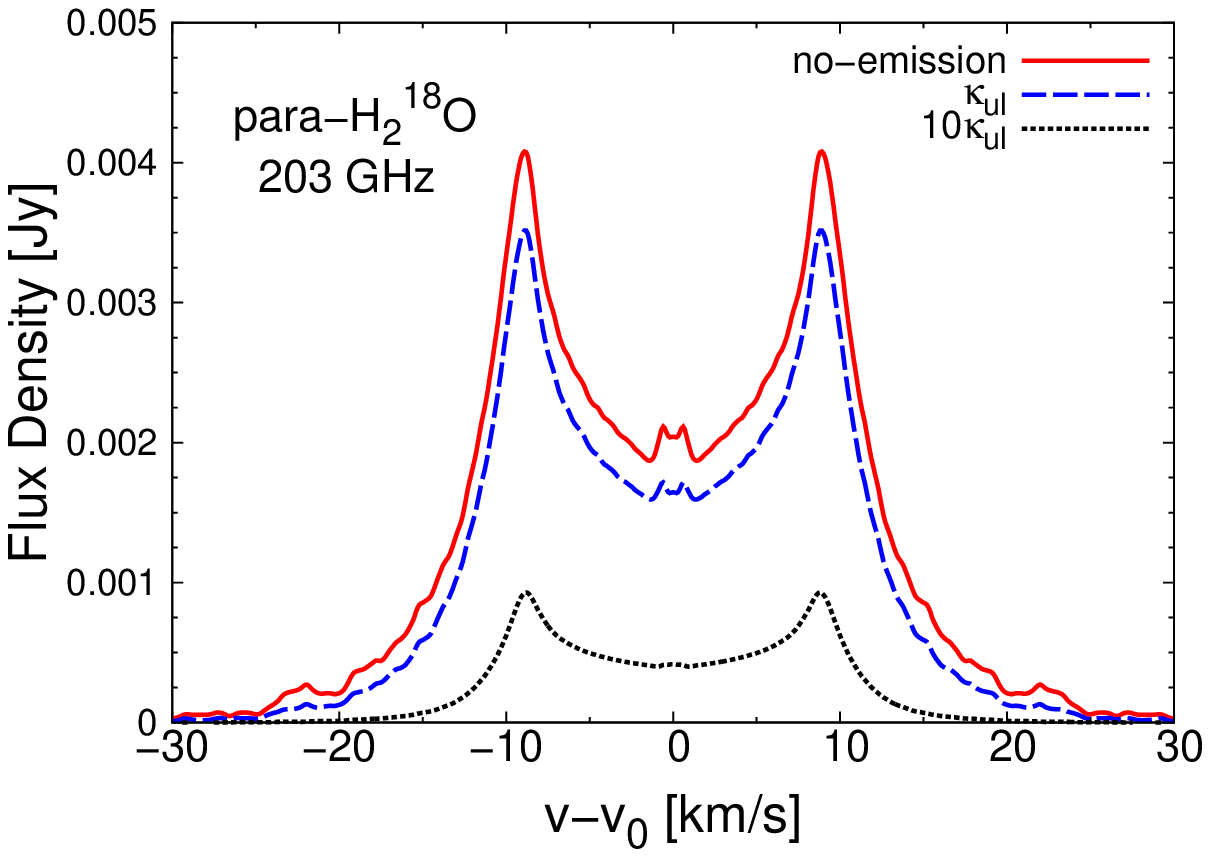}
\includegraphics[scale=0.6]{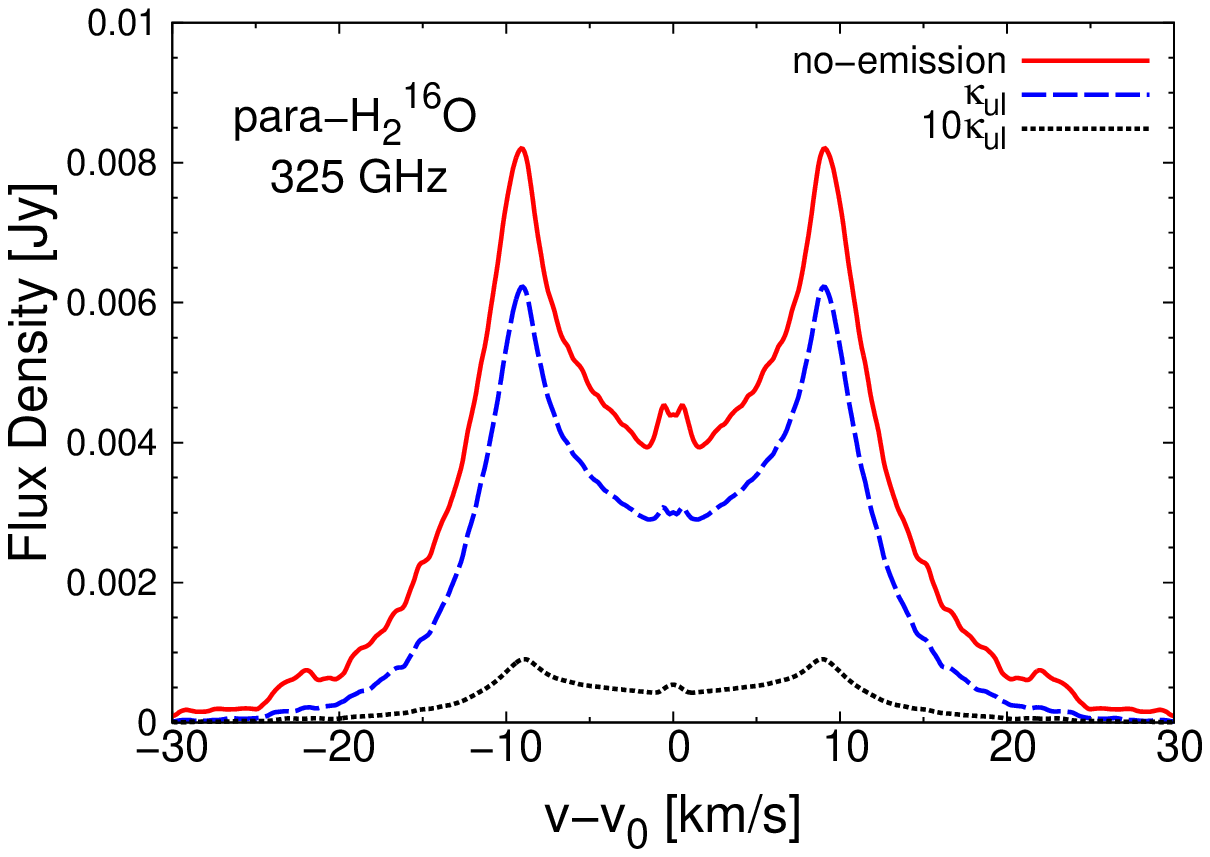}
\includegraphics[scale=0.6]{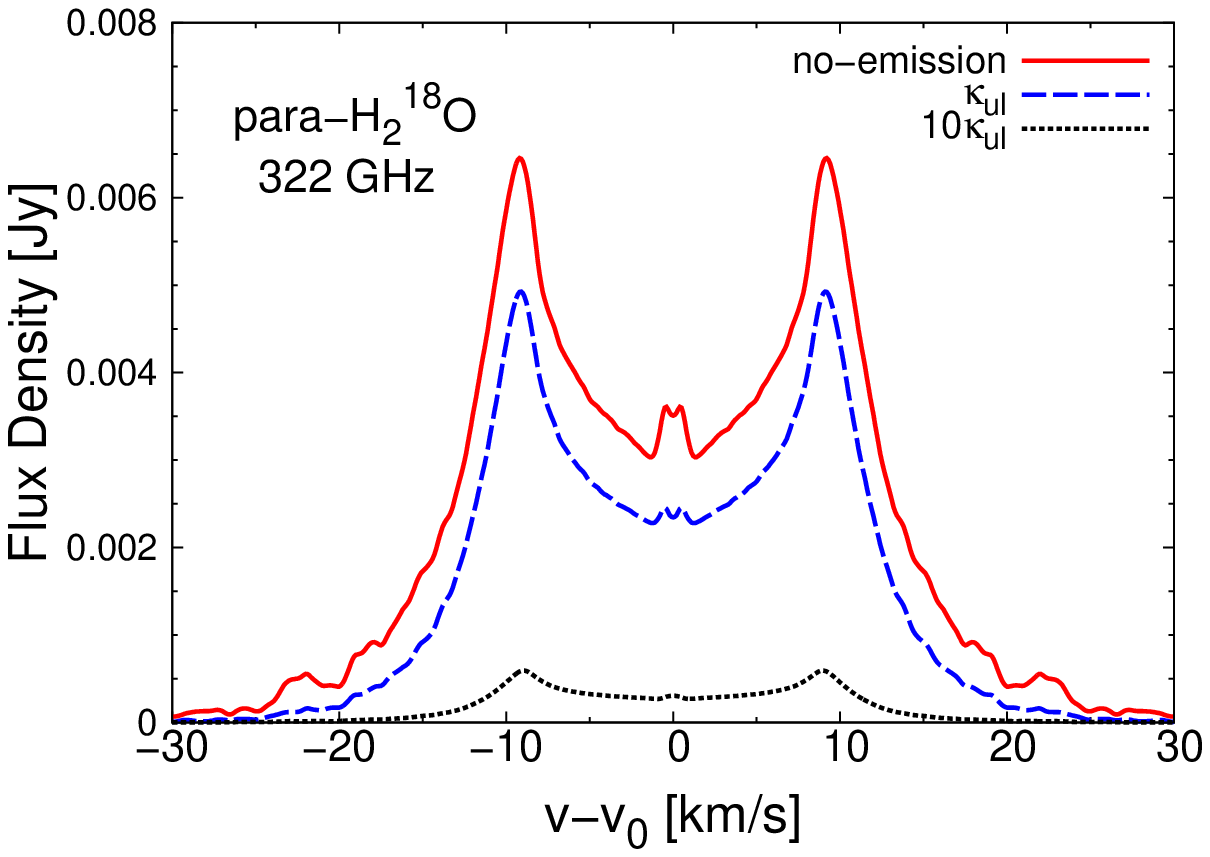}
\includegraphics[scale=0.6]{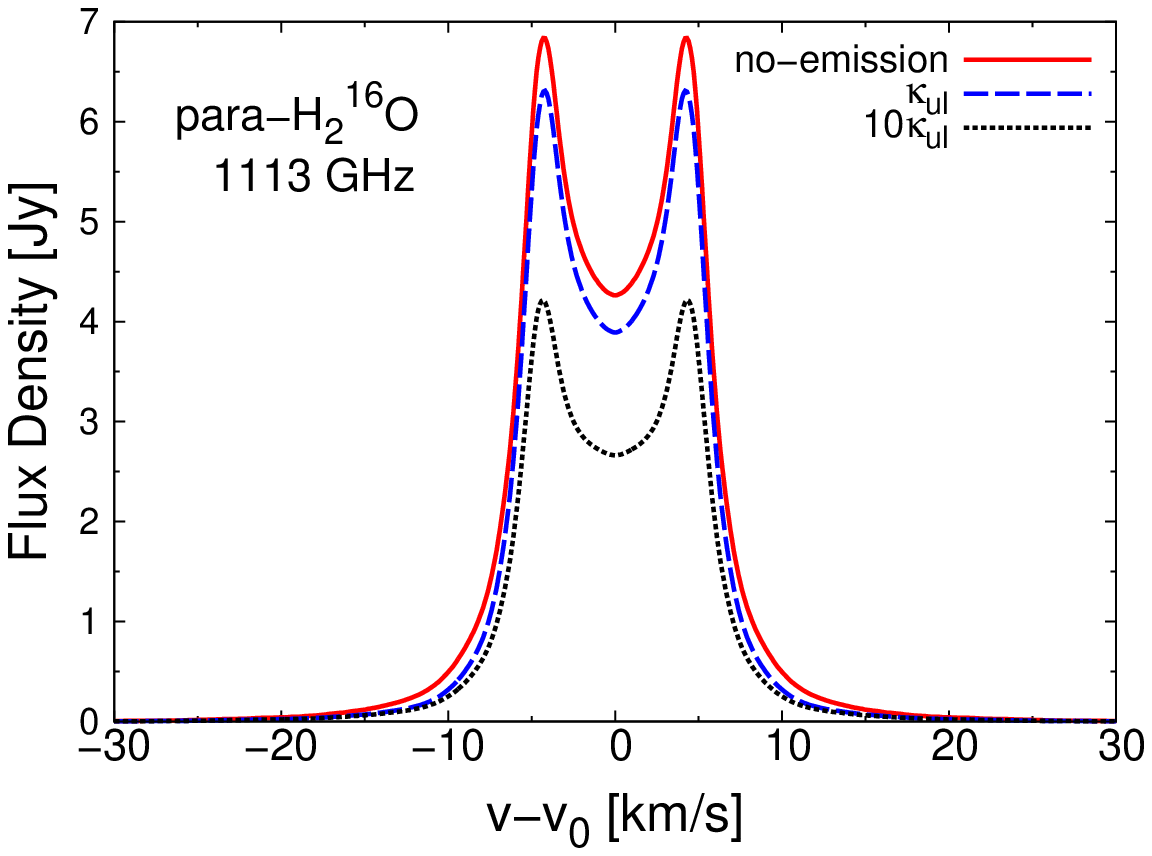}
\includegraphics[scale=0.6]{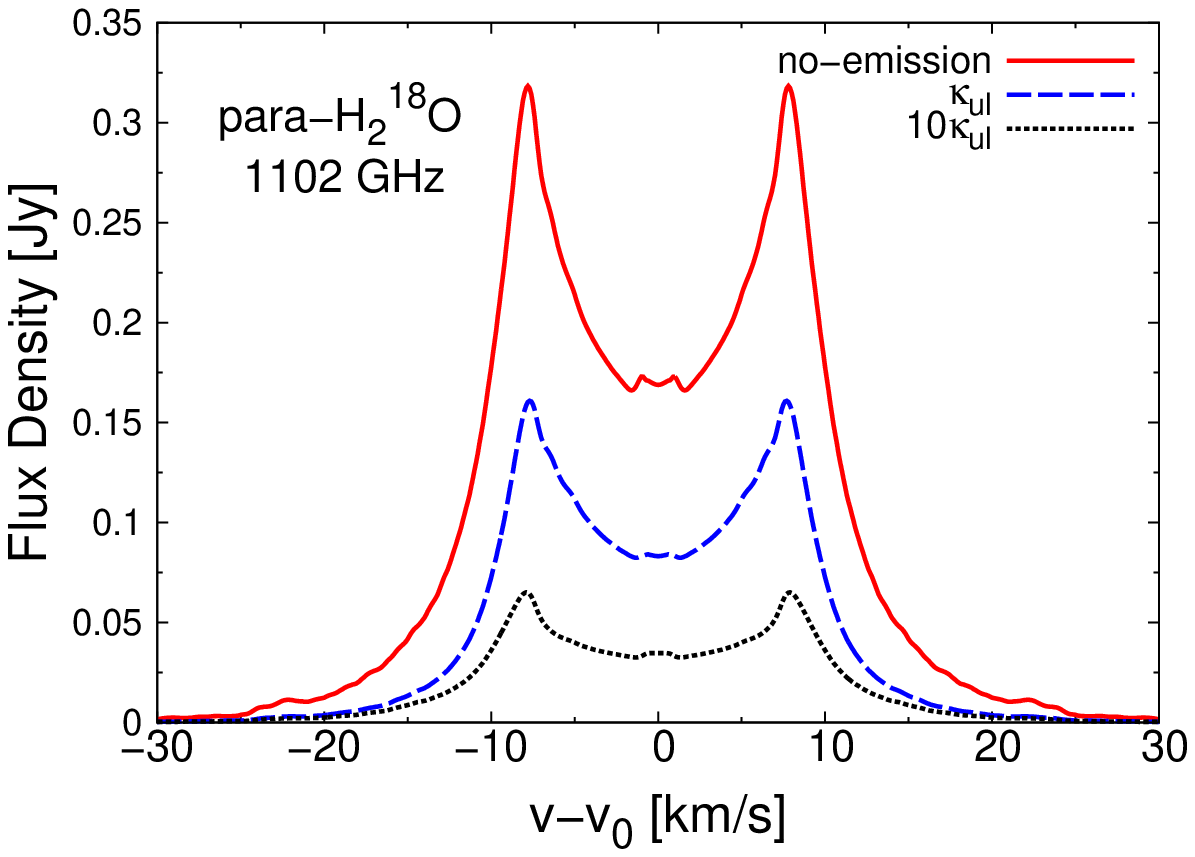}
\end{center}
\vspace{0.6cm}
\caption{\noindent 
The profiles of para-$\mathrm{H_2}$$^{16}\mathrm{O}$ lines at 183 GHz (top left), 325 GHz (middle left), 1113 GHz (bottom left), and
para-$\mathrm{H_2}$$^{18}\mathrm{O}$ lines at 203 GHz (top right), 322 GHz (middle right), 1102 GHz (bottom right) from a Herbig Ae disk inside 30 au.
In the line profiles with {\it red solid lines}, we do not include dust emission components (see also Figures \ref{Figure1_paperIII} and \ref{Figure8_paperIII}).
In the line profiles with {\it blue dashed lines} and {\it black dotted lines}, we include both dust and gas emission components, and we subtract dust emission components (the values of fluxes at $v-v_{0}=\pm\infty$) after the calculation to show the line emission component more clearly.
In the line profiles with {\it black dotted lines}, we set the values of dust opacity $\kappa_{ul}$ ten times larger in order to investigate the influence of dust opacity on line properties.
\vspace{0.3cm}
}\label{Figure9_paperIII}
\end{figure*} 
\subsection{Influence of dust emission on water line properties}
\noindent
Figure \ref{Figure9_paperIII} shows the velocity profiles of several para-$\mathrm{H_2}$$^{16}\mathrm{O}$ and para-$\mathrm{H_2}$$^{18}\mathrm{O}$ lines
for a Herbig Ae disk inside 30 au.
In the line profiles with {\it red solid lines}, we do not include dust emission components (see also Figures \ref{Figure1_paperIII} and \ref{Figure8_paperIII}).
In the line profiles with {\it blue dashed lines}, we include both dust and gas emission components, and then we subtract dust emission components (the values of fluxes at $v-v_{0}=\pm\infty$) after the calculation to show the line emission more clearly.
\\ \\
According to the {\it red solid lines} and {\it blue dashed lines} in Figure \ref{Figure9_paperIII}, if we do not include dust emission, the values of line peak flux densities become $1.1-2$ times larger.
This is because the column density of the inner disk is very high so that the line becomes optically thick.
The position of $\tau_{ul}$=1 at the line center in the $z$ direction is higher than those at line-free frequencies, and the total intensity of dust emission is smaller at the line center than those at line-free frequencies. 
Therefore, the line flux density relative to the flux density of dust emission becomes lower, compared with the case in which both the line and dust continuum emission are optically thin, such as found for molecular clouds.
Here we note that in the inner hot disk midplane at $r<7-8$ au, the vertically integrated column densities of gaseous water molecules are $\sim 10^{20}-10^{22}$ cm$^{-2}$ (see also Figure 3 of paper II), and the values of dust optical depth at $z\sim0$ are around $0.05-0.4$ at 183 GHz, and around $0.1-0.8$ at 325 GHz. The shapes of line profiles are similar among Figures \ref{Figure1_paperIII}, \ref{Figure8_paperIII}, and \ref{Figure9_paperIII}.
\\ \\
In the line profiles with {\it black dotted lines} in Figure \ref{Figure9_paperIII}, we also include both dust and gas components in the same way as those with {\it blue dashed lines}, but we artificially increase the values of dust opacity $\kappa_{ul}$ by a factor of ten in order to investigate the influence of dust opacity on line properties.
We note that the dust opacity at sub-millimeter wavelengths changes by a factor of around ten, depending on the properties of the dust grains (e.g., \citealt{Miyake1993, Draine2006}).
In our fiducial disk model, dust opacities appropriate for the dark cloud model are used and they are relatively small at sub-millimeter wavelengths, compared with the model with grain growth (see e.g., \citealt{NomuraMillar2005, Aikawa2006}, and paper I).
We again subtract dust emission components to show the line emission more clearly.
The disk physical structure is the same as the original reference model.
In these cases with larger values of dust opacity (see {\it black dotted lines} in Figure \ref{Figure9_paperIII}), 
the effect of the dust emission becomes stronger and
the values of peak line flux densities become around $0.1-0.6$ times smaller.
\\ \\
The differences in line flux densities are larger in the cases of Band 7 water lines when compared with Band 5 water lines (see Figure \ref{Figure9_paperIII}).
This is because the dust opacity becomes larger as the line frequency increases (e.g., \citealt{Miyake1993, NomuraMillar2005, Draine2006}).
In addition, the differences in line flux densities are also larger in the cases of H$_{2}$$^{18}$O lines, compared with H$_{2}$$^{16}$O lines.
The emitting regions of $\mathrm{H_2}$$^{18}\mathrm{O}$ lines are closer to the disk midplane (down to $z=0$) than those of $\mathrm{H_2}$$^{16}\mathrm{O}$ lines. 
The temperature is higher as the disk height is higher (see Figure \ref{Figure5_paperIII}) and thus the temperature around $\tau_{ul}\sim1$ of H$_{2}$$^{16}$O lines is higher than those of H$_{2}$$^{18}$O lines. Therefore, the line intensities are larger in the cases of H$_{2}$$^{16}$O lines, compared with the cases of H$_{2}$$^{18}$O lines.
Thus, if the dust opacity of the disk is much larger than that of our disk model, the H$_{2}$$^{16}$O lines and Band 5 lines are better candidates for sub-millimeter detection of the inner water reservoir.
Moreover, the differences in line flux densities of para-H$_{2}$$^{16}$O 1113 GHz and para-H$_{2}$$^{18}$O 1102 GHz are smaller than other water lines.
This is because the main line emitting region of these lines is the cold water vapor of the outer photodesorption region where the dust opacity is much smaller than that in the inner disk, and where also the line intensity is greater (see also Section 3.1).
Since all H$_{2}$$^{16}$O and H$_{2}$$^{18}$O lines treated here are optically thick (see Figure \ref{Figure4_paperIII}) and the temperature is higher as the disk height is higher (see Figure \ref{Figure5_paperIII}), line emission is stronger than dust emission, even for the case with an assumed large dust opacity.
\\ \\
We point out that previous mid- and near-infrared observations for Herbig Ae disks have not yet detected water line emission from the inner disk surface \citep{Pontoppidan2010a, Fedele2011, Salyk2011}, although far-infrared water lines have been detected for a few bright Herbig Ae disks \citep{Fedele2012, Fedele2013, Meeus2012}, which are classified as group II Herbig Ae stars (e.g., \citealt{Honda2015}). Several explanations (e.g., inner holes/gaps, strong UV radiation fields in the disk surface, and strong infrared dust emission in the infrared wavelengths, see Section 4.3 of paper II and \citealt{Antonellini2015, Antonellini2016}) have been considered responsible for this non-detection toward Herbig Ae disks.
In contrast, since the dust continuum fluxes in the sub-millimeter wavelengths are low, the possibility of water line detections from protoplanetary disks is expected to be higher.
\\ \\
Here we note that the para-$\mathrm{H_2}$$^{16}\mathrm{O}$ 1113 GHz line and ortho-$\mathrm{H_2}$$^{16}\mathrm{O}$ 557 GHz line were detected from disks around two T Tauri stars, TW Hya and DG Tau, and a Herbig Ae star, HD~100546 through space spectroscopic observations with $Herschel$/HIFI (e.g., \citealt{Hogerheijde2011, Podio2013, Zhang2013, vanDishoeck2014, Salinas2016, Du2017}).
\citet{Du2017} reported the upper limit flux values of these ground level $\mathrm{H_2}$$^{16}\mathrm{O}$ lines for disks around two other T Tauri stars (AA Tau, DM Tau) and a Herbig Ae star (HD163296).
From the properties of lines detected by previous observations and other modeling works (e.g., \citealt{Meijerink2008, Woitke2009b, Antonellini2015}), the main line emitting region is the cold outer disk, consistent with our results.
The number of detections is smaller than those expected from previous modeling works (e.g., \citealt{Antonellini2015}).
\citet{Du2017} discussed the reasons for the low detection rate, and proposed that gas-phase oxygen bearing molecules (such as water and CO) were removed from the line emitting layers of the outer disk by freeze-out of onto dust grains followed by grain growth and settling/migration.
\begin{figure*}[htbp]
\begin{center}
\includegraphics[scale=0.6]{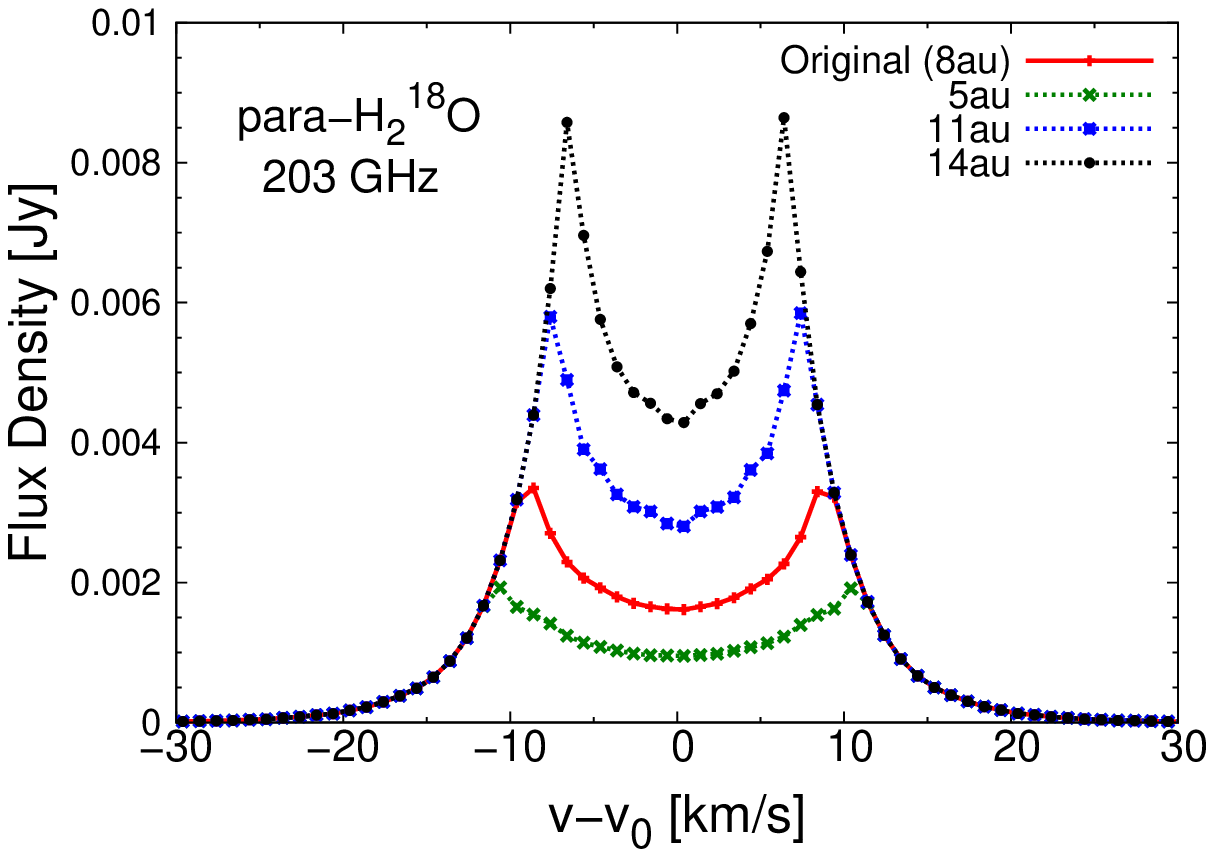}
\end{center}
\vspace{0.6cm}
\caption{
\noindent The profiles of the para-$\mathrm{H_2}$$^{18}\mathrm{O}$ line at 203 GHz from a Herbig Ae disk inside 30 au with a reduced velocity resolution (d$v=$1.0 km s$^{-1}$).
When we calculate these line profiles, we include both dust and gas emission components, and we subtract dust emission components (the values of fluxes at $v-v_{0}=\pm\infty$) after calculations to show the line emission component more clearly.
The red solid line with cross symbols is the line profile of our original Herbig Ae disk model (see also Figure \ref{Figure9_paperIII}).
In other plots, we artificially change the outer edge of the region with a high $\mathrm{H_2O}$ water vapor abundance ($\sim10^{-5}-10^{-4}$) region to 5 au (the green dotted line with cross symbols), 11 au (the blue dotted line with filled square and cross symbols), and 14 au (the black dotted line with circle symbols).
\vspace{0.3cm}
}\label{Figure10_paperIII}
\end{figure*} 
\subsection{Influence of different snowline positions and velocity resolution of the line profiles}
\noindent Our model, which has a continuous disk in the radial direction and no inner gap and/or hole, is applicable to group II Herbig Ae stars (e.g. \citealt{Honda2015}). For a model having an inner hole (i.e., group I Herbig Ae stars), the fluxes of the $\mathrm{H_2O}$ lines, especially from the hot water gas within the $\mathrm{H_2O}$ snowline, are expected to decrease.
In addition, if we adopt different values of dust opacities (due to dust evolution), mass accretion rates (which determine the effects of viscous heating), and central star temperatures, the position of $\mathrm{H_2O}$ snowline is also changed (e.g., \citealt{Oka2011, Harsono2015, Piso2015}).
In Section 3.2.4 of paper I and Sections 4.1 and 4.3 of paper II, the uncertainties in predictions of our model were presented in detail, and the water line properties for some cases in which we artificially changed of $\mathrm{H_2O}$ vapor distribution were discussed (see Figures 8 and 9 of paper I).
If the $\mathrm{H_2O}$ snowline is farther from the star, and the fractional water vapor abundance in the surface of the outer disk is lower than that in the original disk model, the line fluxes from the hot region within the $\mathrm{H_2O}$ snowline are larger than those from the cooler outer disk.
\\ \\
In the cases of sub-millimeter para-$\mathrm{H_2}$$^{16}\mathrm{O}$ lines and $\mathrm{H_2}$$^{18}\mathrm{O}$ lines, the emission fluxes from the optically thin disk surface and photodesorption region are smaller than those of ortho-$\mathrm{H_2}$$^{16}\mathrm{O}$ lines (see Section 3.1). Therefore, the variations in fractional water abundances in those regions will have little impact on the profiles of the para-$\mathrm{H_2}$$^{16}\mathrm{O}$ lines and $\mathrm{H_2}$$^{18}\mathrm{O}$ lines with smaller values of $A_{\mathrm{ul}}$ ($<10^{-4}$ s$^{-1}$) and relatively higher values of $E_{\mathrm{up}}$ ($\gtrsim$200K).
Here we discuss the behavior of these lines in which we reduce the velocity resolution and artificially change the distribution of $\mathrm{H_2O}$ vapor in the disk midplane, and test the validity of our model predictions.
\\ \\
In Figure \ref{Figure10_paperIII}, we display the profiles of the para-$\mathrm{H_2}$$^{18}\mathrm{O}$ line at 203 GHz from a Herbig Ae disk inside 30 au with a reduced velocity resolution (d$v=$1.0 km s$^{-1}$).
This velocity resolution we adopted in this figure is ten times larger than those of other line profiles in this paper and our previous papers (d$v=$0.1 km s$^{-1}$, see also papers I and II), and a velocity resolution of $0.1-1.0$ km s$^{-1}$ is often adopted for ALMA observations.
The red solid line with cross symbols is the line profile of our original Herbig Ae disk model (see also Figure \ref{Figure9_paperIII}).
In other plots, we artificially change the outer edge of the region with high $\mathrm{H_2O}$ water vapor abundance ($=10^{-5}$) to 5 au ($T_{g}\sim180$K), 11 au ($T_{g}\sim135$K), and 14 au ($T_{g}\sim120$K).
Since this line is mainly emitted from the region with a high water vapor abundance ($\sim10^{-5}-10^{-4}$, see also Figures \ref{Figure4_paperIII}, \ref{Figure6_paperIII}, \ref{Figure8_paperIII}, and \ref{Figure9_paperIII}) in the Herbig Ae disk, we changed the outer edge of the region with a high $\mathrm{H_2O}$ water vapor abundance (originally 8 au, see also Table 2). 
\\ \\
According to Figure \ref{Figure10_paperIII}, as the region with a high water vapor abundance becomes larger, the flux density of the line peaks becomes larger, and the line width, especially the width between the two peaks becomes narrower. 
In the outer disk the changes in local velocity widths are smaller, since the velocity widths are inversely proportional to the square root of the radial distance of the emitting region from the central star (see also Equations (11) and (12) of Paper I).
The figure shows that we will be able to distinguish the differences of
the outer edge positions (or the positions of the $\mathrm{H_2O}$ snowline) with the resolution of of a few au through
observing the separation between the two peaks in the line profile
with the velocity resolution of $dv=$1.0 km s$^{-1}$ for the Herbig Ae disk with the inclination angle of 30 $\deg$.
Future detailed calculations, which consider the noise expected in actual observations and the errors of physical parameters such as central star masses and disk inclination angles (obtained by other previous observations), are also important to investigate the observational possibilities with ALMA.
\\ \\
Here we note that the disk physical structure is the same as the original reference model (see Section 2.1), because calculating several different disk physical structures and chemical structures self-consistently is computationally demanding and beyond the scope of this work. Even if we adopt self-consistent models, we expect that the line
widths will not be affected; however, we do expect that line fluxes will be affected by about 0.5-4 times since the temperature of line emitting regions will be about 0.7$-$1.2 times different, on the basis of the differences in gas temperatures around the outer edges of 
high water vapor abundance regions (see also Section 3.2.4 of paper I).
\subsection{Requirement for the future observations}
\noindent In order to trace the hot water gas inside the $\mathrm{H_2O}$ snowline,
high-dispersion spectroscopic observations (R=$\lambda$/$\delta \lambda$ $>$ tens of thousands) of the water candidate lines in Tables 3 and 4 are needed. This is because the velocity width between the peaks is $\sim15-20$ km s$^{-1}$.
The para-$\mathrm{H_2}$$^{16}\mathrm{O}$ 183 GHz line and para-$\mathrm{H_2}$$^{18}\mathrm{O}$ 203 GHz line, which have the same transition levels, are in the frequency coverage of ALMA Band 5 \citep{Immer2016, Humphreys2017}, and the para-$\mathrm{H_2}$$^{16}\mathrm{O}$ 325 GHz line and the para-$\mathrm{H_2}$$^{18}\mathrm{O}$ 322 GHz line, which have the same transition levels, are in ALMA Band 7.
The candidate ortho-$\mathrm{H_2}$$^{16}\mathrm{O}$ 321 GHz line (see paper II) is also in ALMA Band 7.
Since their line fluxes are larger than those of other water lines at similar wavelengths and the atmospheric conditions of ALMA Bands 5 and 7 are usually better than Bands at shorter wavelengths, they are suitable for tracing the hot water vapor inside the $\mathrm{H_2O}$ snowline.
Because of the small values of $E_{\mathrm{up}}$, the para-$\mathrm{H_2}$$^{16}\mathrm{O}$ lines at 183 GHz and 325 GHz are strongly affected by atmospheric absorption.
By contrast, the effects of atmospheric absorption are less in the cases of $\mathrm{H_2}$$^{18}\mathrm{O}$ lines, and $\mathrm{H_2}$$^{16}\mathrm{O}$ lines with larger values of $E_{\mathrm{up}}$ ($\sim$1000-2000K).
\\ \\
Other sub-millimeter ortho- and para-$\mathrm{H_2O}$ lines that trace the $\mathrm{H_2O}$ snowline exist in ALMA Bands 8, 9 and 10 ($\sim$ $275-950$GHz, see also Figures 7 and 8, Tables 3 and 4).
The number of candidate water lines is largest in ALMA Band 8, and which has three ortho-$\mathrm{H_2}$$^{16}\mathrm{O}$ lines, four para-$\mathrm{H_2}$$^{16}\mathrm{O}$ lines, and one ortho-$\mathrm{H_2}$$^{18}\mathrm{O}$ lines.
The values of $A_{\mathrm{ul}}$ are around $1-7\times10^{-5}$ s$^{-1}$, and the values of $E_{\mathrm{up}}$ range from 300K to 1600K.
The ortho-$\mathrm{H_2}$$^{16}\mathrm{O}$ 621 GHz and the ortho-$\mathrm{H_2}$$^{18}\mathrm{O}$ 692 GHz lines are transitions from the same energy levels and both fall in Band 9.
Two candidate para-$\mathrm{H_2}$$^{16}\mathrm{O}$ lines and one para-$\mathrm{H_2}$$^{18}\mathrm{O}$ line fall in ALMA Band 10.
Here we note that the $\mathrm{H_2}$$^{16}\mathrm{O}$ lines with small values of $E_{\mathrm{up}}$ ($<$ 1000K) are strongly affected by atmospheric absorption.
\\ \\
There are no candidate $\mathrm{H_2}$$^{16}\mathrm{O}$ and $\mathrm{H_2}$$^{18}\mathrm{O}$ lines within the frequency coverage of ALMA Bands 3 and 4.
The ortho-$\mathrm{H_2}$$^{18}\mathrm{O}$ 254 GHz line is in Band 6 and it has the same transition energy level of ortho-$\mathrm{H_2}$$^{16}\mathrm{O}$ 321 GHz line (see paper II), although it has a relatively larger value of $E_{\mathrm{up}}$ ($=1853.5$K) and its fluxes are relatively smaller than those of Bands 5 and 7 lines.
\\ \\
Although we predict that the fluxes of the sub-millimeter candidate lines are too small in T Tauri disks ($\sim 10^{-23}-10^{-22}$ W $\mathrm{m}^{-2}$) to detect with a reasonable integration time with current ALMA sensitivity ($\sim 10^{-21}-10^{-20}$ W $\mathrm{m}^{-2}$; 5$\sigma$, 1 hour), the fluxes are relatively strong and a greater possibility of successful detections are expected in Herbig Ae disks, in T Tauri disks with younger ages (e.g., HL Tau, \citealt{ALMA2015, Banzatti2015, Harsono2015, Zhang2015, Okuzumi2016}), and in disks around FU Orionis type stars (e.g., V883 Ori, \citealt{Cieza2016, Schoonenberg2017}).
The predicted fluxes of the para-$\mathrm{H_2}$$^{16}\mathrm{O}$ 183 GHz (Band 5), para-$\mathrm{H_2}$$^{18}\mathrm{O}$ 203 GHz (Band 7), para-$\mathrm{H_2}$$^{16}\mathrm{O}$ 325 GHz (Band 7), and para-$\mathrm{H_2}$$^{18}\mathrm{O}$ 322 GHz (Band 7) lines are around $4\times 10^{-22}-10^{-21}$ W $\mathrm{m}^{-2}$ (see Tables 3 and 4).
In addition, the fluxes of some water lines in Bands $8-9$ are also estimated to be around $10^{-21}-10^{-20}$  W $\mathrm{m}^{-2}$ for a Herbig Ae disk (see also Tables 3 and 4). 
\\ \\
Here we mention that the ortho-$\mathrm{H_2}$$^{16}\mathrm{O}$ 321 GHz line has been detected in the disk and outflow around the massive protostar candidate, Source I in Orion KL \citep{Hirota2014ApJL}, using ALMA, and around the embedded low mass Class I protostar, HL Tau, using SMA \citep{Kristensen2016}.
The interstellar para-$\mathrm{H_2}$$^{16}\mathrm{O}$ 183 GHz and the para-$\mathrm{H_2}$$^{18}\mathrm{O}$ 203 GHz lines were detected for the first time in the Orion and DR21(OH) molecular clouds (e.g., \citealt{Phillips1978}).
Some previous observations reported the detections of the thermal para-$\mathrm{H_2}$$^{18}\mathrm{O}$ 203 GHz line towards hot molecular cloud cores and high mass protostars \citep{Jacq1988, vanderTak2006}.
Resolved detections of the thermal para-$\mathrm{H_2}$$^{18}\mathrm{O}$ 203 GHz line were reported towards the deeply embedded low mass Class 0 protostars NGC 1333-IRAS4B \citep{Jorgensen2010}, NGC 1333-IRAS2A, and NGC 1333-IRAS4A \citep{Persson2012}, using the IRAM Plateau de Bure Interferometer. 
They suggested that the water emission comes from the inner disk.
\section{Conclusions}
\noindent In this paper, we extended our previous work (papers I and II, \citealt{Notsu2016, Notsu2017a}) on using the profiles of ortho-$\mathrm{H_2}$$^{16}\mathrm{O}$ lines for tracing the location of the $\mathrm{H_2O}$ snowline in a Herbig Ae disk and a T Tauri disk, to inclide sub-millimeter para-$\mathrm{H_2}$$^{16}\mathrm{O}$ and ortho- and para-$\mathrm{H_2}$$^{18}\mathrm{O}$ lines.
\\ \\
The number densities of the para-H$_{2}$$^{16}$O molecules are around one third smaller than that of ortho-H$_{2}$$^{16}$O, thus the para-H$_{2}$$^{16}$O line can trace deeper into the disk than the ortho-H$_{2}$$^{16}$O lines.
Since the number densities of H$_{2}$$^{18}$O molecules are around 560 times smaller than those of H$_{2}$$^{16}$O, they can probe deeper into the disk than the H$_{2}$$^{16}$O lines (down to $z=0$) and thus they are better candidates for detecting water emission within the $\mathrm{H_2O}$ snowline at the disk midplane.
If the dust opacity of the disk is much larger than that adopted in our disk model, the H$_{2}$$^{16}$O lines and lines with longer wavelengths are better candidates for sub-millimeter detection of the inner water reservoir.
This is because the dust opacity becomes larger as the line frequency increases.
In addition, the temperature is higher as the disk height is higher and thus the temperature around $\tau_{ul}\sim1$ of H$_{2}$$^{16}$O lines is higher than those of H$_{2}$$^{18}$O lines. Therefore, the line intensities are larger in the case of H$_{2}$$^{16}$O lines, compared with the case of H$_{2}$$^{18}$O lines.
The values of the Einstein $A$ coefficients of sub-millimeter candidate water lines tend to be smaller (typically $<$$10^{-4}$ s$^{-1}$) than infrared candidate water lines (see paper II).
Thus, in the case of sub-millimeter candidate water lines, the local intensity from the outer optically thin region in the disk is around $10^{4}$ times smaller than that in infrared candidate water line cases (see paper II).
Therefore, in the case of sub-millimeter lines, especially for H$_{2}$$^{18}$O and para-H$_{2}$$^{16}$O, lines with relatively smaller upper state energies ($\sim$ a few 100 K) can also be used to trace the location of the water snowline.
The values of candidate water line fluxes of the T Tauri disk are around $1-5\times10^{2}$ smaller than those of the Herbig Ae disk, because the location of the $\mathrm{H_2O}$ snowline in the T Tauri disk exists at a smaller radius from the star than that in the Herbig Ae disk.
\\ \\
There are several candidate water lines that trace the hot water gas inside $\mathrm{H_2O}$ snowline in ALMA Bands $5-10$.
The successful detection of candidate water lines in Herbig Ae disks and younger T Tauri disks could be achieved with current ALMA capabilities.
\\ \\
\acknowledgments
\noindent We are grateful to Professor Hiroshi Shibai and Professor Inga Kamp for their useful comments.
We thank the referee for many important suggestions and comments.
Our numerical studies were carried out on SR16000 at Yukawa Institute for Theoretical Physics (YITP) and computer systems at Kwasan and Hida Observatory (KIPS) in 
Kyoto University, and PC cluster at Center for Computational Astrophysics, National Astronomical Observatory of Japan.
This work is supported by JSPS (Japan Society for the Promotion of Science) Grants-in-Aid for Scientific Research (Grant Number; 25108004, 25108005, 25400229, 15H03646, 15K17750),
by Grants-in-Aid for JSPS fellows (Grant Number; 16J06887), and by the Astrobiology Center Program of National Institutes of Natural Sciences (NINS) (Grant Number; AB281013).
S. N. is grateful for the support from the educational program organized by Unit of Synergetic Studies for Space, Kyoto University.
C. W. acknowledges support from the Netherlands Organization for Scientific Research (NWO, program number 639.041.335) and start-up funds from the University of Leeds.
Astrophysics at Queen's University Belfast is supported by a grant from the STFC (ST/P000312/1).
\noindent
\software{RATRAN \citep{Hogerheijde2000}}
\\ \\
\begin{deluxetable*}{rrrrrrrrr}
\tablewidth{0pt}
\tablecaption{{Calculated parameters and total fluxes for all sub-millimeter ortho- and para-$\mathrm{H_2}$$^{16}\mathrm{O}$ lines}}\label{tab:T2}
\tablehead{
\colhead{$J_{K_{a}K_{c}}$}&
\colhead{\ $\lambda$\tablenotemark{a}}&
\colhead{\ Frequency}& \colhead{\ $A_{\mathrm{ul}}$}&
\colhead{$E_{\mathrm{up}}$} & \colhead{HAe flux\tablenotemark{b}} & \colhead{TT flux\tablenotemark{c}} & \colhead{Comments\tablenotemark{d}} \\
\colhead{}&\colhead{[$\mu$m]}&
\colhead{[GHz]}& \colhead{[s$^{-1}$]}&
\colhead{[K]}& \colhead{[W $\mathrm{m}^{-2}$]}& \colhead{[W $\mathrm{m}^{-2}$]}& \colhead{}}
\startdata      
       &&& ortho-$\mathrm{H_2}$$^{16}\mathrm{O}$ lines&&&& \\    
      6$_{16}$-5$_{23}$ & 13482.8594 & 22.23508 & 1.835$\times10^{-9}$ & 643.5 & $1.4\times10^{-24}$ &$4.5\times10^{-26}$& \\ 
      10$_{29}$-9$_{36}$\tablenotemark{e} & 933.2767 & 321.22568 & 6.165$\times10^{-6}$ & 1861.2 & $2.3\times10^{-21}$ &$7.8\times10^{-23}$& ALMA Band 7\\  
      4$_{14}$-3$_{21}$ &788.5180 & 380.19736 & 3.083$\times10^{-5}$ & 323.5 & $1.7\times10^{-20}$ &$3.3\times10^{-22}$& \\  
             6$_{43}$-5$_{50}$\tablenotemark{e} & 682.6641 & 439.15079 & 2.816$\times10^{-5}$ & 1088.7 & $1.4\times10^{-20}$  &$3.1\times10^{-22}$& ALMA Band 8\\
       7$_{52}$-6$_{61}$\tablenotemark{e} & 676.7044 & 443.01835 & 2.231$\times10^{-5}$ & 1524.8 & $9.5\times10^{-21}$  &$2.5\times10^{-22}$& ALMA Band 8\\                
       4$_{23}$-3$_{30}$ & 669.1780 & 448.00108 & 5.462$\times10^{-5}$ & 432.1 & $2.6\times10^{-20}$  &$5.0\times10^{-22}$& ALMA Band 8\\  
      1$_{10}$-1$_{01}$\tablenotemark{e} & 538.2889 & 556.93599 & 3.497$\times 10^{-3}$& 61.0 & $7.2\times10^{-20}$ &$1.1\times10^{-20}$& $Herschel$/HIFI\\
       5$_{32}$-4$_{41}$\tablenotemark{e} & 482.9902 &  620.70095 & 1.106$\times10^{-4}$ & 732.1 & $5.4\times10^{-20}$ &$1.0\times10^{-21}$& ALMA Band 9\\
       &&& para-$\mathrm{H_2}$$^{16}\mathrm{O}$ lines&&&& \\    
       3$_{13}$-2$_{20}$ & 1635.4389 & 183.31009 & 3.653$\times10^{-6}$ & 204.7 & $4.4\times10^{-22}$  &$1.8\times10^{-23}$& ALMA Band 5\\
       5$_{15}$-4$_{22}$ & 922.0046 & 325.15290 & 1.168$\times10^{-5}$ & 469.9 & $1.9\times10^{-21}$  &$7.6\times10^{-23}$& ALMA Band 7\\
       7$_{53}$-6$_{60}$ & 685.4802 & 437.34666 & 2.146$\times10^{-5}$ & 1524.6 & $6.3\times10^{-22}$  &$2.8\times10^{-23}$& ALMA Band 8\\
       6$_{42}$-5$_{51}$ & 636.6522 & 470.88890 & 3.483$\times10^{-5}$ & 1090.3 & $2.0\times10^{-21}$  &$8.8\times10^{-23}$& ALMA Band 8\\       
       5$_{33}$-4$_{40}$ & 631.5554 & 474.68911 & 4.815$\times10^{-5}$ & 725.1 & $4.5\times10^{-21}$  &$1.8\times10^{-22}$& ALMA Band 8\\
       6$_{24}$-7$_{17}$ & 613.7112 & 488.49113 & 1.382$\times10^{-5}$ & 867.3 & $2.1\times10^{-21}$  &$9.5\times10^{-23}$& ALMA Band 8\\ 
       2$_{11}$-2$_{02}$ & 398.6426 & 752.03314 & 7.130$\times10^{-3}$ & 136.9 & $7.8\times10^{-20}$  &$1.5\times10^{-21}$& \\
       9$_{28}$-8$_{35}$ & 330.8216 & 906.20590 & 2.221$\times10^{-4}$ & 1554.4 & $6.9\times10^{-21}$  &$1.5\times10^{-22}$& ALMA Band 10\\ 
       4$_{22}$-3$_{31}$ & 327.2230 & 916.17158 & 5.743$\times10^{-4}$ & 454.3 & $5.7\times10^{-20}$  &$1.5\times10^{-21}$& ALMA Band 10\\ 
       5$_{24}$-4$_{31}$ & 308.9640 & 970.31505 & 9.071$\times10^{-4}$ & 598.8 & $6.5\times10^{-20}$  &$1.7\times10^{-21}$& \\ 
       2$_{02}$-1$_{11}$ & 303.4562 & 987.92676 & 5.912$\times10^{-3}$ & 100.8 & $1.5\times10^{-19}$  &$2.9\times10^{-21}$& \\                      
       1$_{11}$-0$_{00}$ & 269.2723 & 1113.34301 & 1.863$\times10^{-2}$ & 53.4 & $2.8\times10^{-18}$  &$2.6\times10^{-19}$& $Herschel$/HIFI\\
\enddata
\tablenotetext{a}{In calculating the line wavelength values $\lambda$ from the values of line frequencies, we adopt the speed of light value $c$$=$$2.99792458\times 10^{8}$ m s$^{-1}$.}
\tablenotetext{b}{The total line flux for a Herbig Ae disk. When we calculate these line fluxes, we do not include dust emission.}
\tablenotetext{c}{The total line flux for a T Tauri disk. When we calculate these line fluxes, we do not include dust emission.}
\tablenotetext{d}{``ALMA Band" means that the line is within the current ALMA Band coverage. ``$Herschel$/HIFI" means that the line was detected from protoplanetary disks by previous $Herschel$/HIFI observations (e.g., \citealt{Hogerheijde2011, Podio2013, vanDishoeck2014, Salinas2016, Du2017}).}
\tablenotetext{e}{We have already reported the values for these ortho-$\mathrm{H_2}$$^{16}\mathrm{O}$ lines in paper II.}
\end{deluxetable*}
\begin{deluxetable*}{rrrrrrrrr}
\tablewidth{0pt}
\tablecaption{{Calculated parameters and total fluxes for all sub-millimeter ortho- and para-$\mathrm{H_2}$$^{18}\mathrm{O}$ lines}}\label{tab:T3}
\tablehead{
\colhead{$J_{K_{a}K_{c}}$}&
\colhead{\ $\lambda$\tablenotemark{a}}&
\colhead{\ Frequency}& \colhead{\ $A_{\mathrm{ul}}$}&
\colhead{$E_{\mathrm{up}}$} & \colhead{HAe flux\tablenotemark{b}} & \colhead{TT flux\tablenotemark{c}} & \colhead{Comments\tablenotemark{d}} \\
\colhead{}&\colhead{[$\mu$m]}&
\colhead{[GHz]}& \colhead{[s$^{-1}$]}&
\colhead{[K]}& \colhead{[W $\mathrm{m}^{-2}$]}& \colhead{[W $\mathrm{m}^{-2}$]}& \colhead{}}
\startdata      
       &&& ortho-$\mathrm{H_2}$$^{18}\mathrm{O}$ lines&&&& \\  
      6$_{16}$-5$_{23}$ & 53295.0158 & 5.62515 & 2.988$\times10^{-11}$ & 640.7 & $1.6\times10^{-27}$ &$1.3\times10^{-28}$& \\ 
      10$_{29}$-9$_{36}$ & 1180.7587 & 253.89816 & 2.806$\times10^{-6}$ & 1853.5 & $1.2\times10^{-22}$ &$6.3\times10^{-24}$& ALMA Band 6\\  
      4$_{14}$-3$_{21}$ & 767.5026 & 390.60776 & 3.160$\times10^{-5}$ & 322.0 & $6.6\times10^{-21}$ &$1.9\times10^{-22}$& ALMA Band 8\\  
      4$_{23}$-3$_{30}$ & 613.0045 & 489.05426 & 6.951$\times10^{-5}$ & 429.6 & $1.2\times10^{-20}$  &$3.4\times10^{-22}$& ALMA Band 8\\ 
       6$_{43}$-5$_{50}$ & 576.3718 & 520.13732 & 4.609$\times10^{-5}$ & 1080.6 & $5.6\times10^{-21}$ &$2.2\times10^{-22}$& \\   
       7$_{52}$-6$_{61}$ & 555.2782 & 539.89596 & 3.982$\times10^{-5}$ & 1512.4 & $2.6\times10^{-21}$  &$1.0\times10^{-22}$&\\                
       1$_{10}$-1$_{01}$ & 547.3897 & 547.67644 & 3.257$\times 10^{-3}$& 60.5 & $4.2\times10^{-20}$ &$1.0\times10^{-21}$&\\       
       5$_{32}$-4$_{41}$ & 433.1766 & 692.07914 & 1.511$\times10^{-4}$ & 727.6 & $2.4\times10^{-20}$ &$7.2\times10^{-22}$& ALMA Band 9\\       
       &&& para-$\mathrm{H_2}$$^{18}\mathrm{O}$ lines&&&& \\             
       3$_{13}$-2$_{20}$ & 1473.8514 & 203.40752 & 4.813$\times10^{-6}$ & 203.7 & $5.2\times10^{-22}$ &$2.3\times10^{-23}$& ALMA Band 5\\
       5$_{15}$-4$_{22}$ & 929.6894 & 322.46517 & 1.060$\times10^{-5}$ & 467.9 & $1.4\times10^{-21}$ &$6.6\times10^{-23}$& ALMA Band 7\\
       6$_{24}$-7$_{17}$ & 579.6653 & 517.18196 & 1.491$\times10^{-5}$ & 865.0 & $1.5\times10^{-21}$ &$7.4\times10^{-23}$& \\
       7$_{53}$-6$_{60}$ & 561.9372 & 533.49817 & 3.841$\times10^{-5}$ & 1512.1 & $5.9\times10^{-22}$ &$2.5\times10^{-23}$& \\       
       5$_{33}$-4$_{40}$ & 557.9220 & 537.33757 & 6.870$\times10^{-5}$ & 720.2 & $5.0\times10^{-21}$ &$2.1\times10^{-22}$& \\
       6$_{42}$-5$_{51}$ & 540.3030 & 554.85987 & 5.616$\times10^{-5}$ & 1082.2 & $2.2\times10^{-21}$ &$9.8\times10^{-23}$& \\       
       2$_{11}$-2$_{02}$ & 402.2331 & 745.32020 & 6.759$\times10^{-3}$ & 136.4 & $6.8\times10^{-20}$ &$1.4\times10^{-21}$&\\       
       9$_{28}$-8$_{35}$ & 349.8796 & 856.84463 & 1.711$\times10^{-4}$ & 1548.0 & $2.8\times10^{-21}$ &$7.7\times10^{-23}$& ALMA Band 10\\
       4$_{22}$-3$_{31}$ & 308.9777 & 970.27204 & 6.759$\times10^{-4}$ & 452.4 & $5.5\times10^{-20}$ &$1.6\times10^{-21}$& \\       
       2$_{02}$-1$_{11}$ & 301.3974 & 994.67513 & 5.970$\times10^{-3}$ & 100.6 & $1.4\times10^{-19}$ &$2.9\times10^{-21}$&\\
       5$_{24}$-4$_{31}$ & 298.8131 & 1003.27760 & 9.714$\times10^{-4}$ & 595.9 & $5.7\times10^{-20}$ &$1.7\times10^{-21}$& \\         
       1$_{11}$-0$_{00}$ & 272.1185 & 1101.69826 & 1.769$\times10^{-2}$ & 52.9 & $2.1\times10^{-19}$ &$4.4\times10^{-21}$& \\ 
\enddata
\tablenotetext{a}{In calculating the line wavelength values $\lambda$ from the values of line frequencies, we adopt the speed of light value $c$$=$$2.99792458\times 10^{8}$ m s$^{-1}$.}
\tablenotetext{b}{The total line flux for a Herbig Ae disk. When we calculate these line fluxes, we do not include dust emission.}
\tablenotetext{c}{The total line flux for a T Tauri disk. When we calculate these line fluxes, we do not include dust emission.}
\tablenotetext{d}{``ALMA Band" means that the line is within the current ALMA Band coverage. 
$\mathrm{H_2}$$^{18}\mathrm{O}$ lines have not been detected from protoplanetary disks by previous $Herschel$/HIFI observations (e.g., \citealt{Hogerheijde2011, Podio2013, vanDishoeck2014, Salinas2016, Du2017}).}
\end{deluxetable*}

%
%
\end{document}